\documentclass[a4paper,11pt]{article}
\usepackage{jcappub}
\usepackage{amsmath}
\usepackage{amssymb}
\usepackage{soul}

\DeclareRobustCommand{\ion}[2]{%
\relax\ifmmode
\ifx\testbx\f@series
{\mathbf{#1\,\mathsc{#2}}}\else
{\mathrm{#1\,\mathsc{#2}}}\fi
\else\textup{#1\,{\mdseries\textsc{#2}}}%
\fi}

\usepackage{enumitem}
\setlist[enumerate]{itemsep=1mm}

\usepackage{xcolor}
\definecolor{darkred}{rgb}{0.5, 0.0, 0.0}
\usepackage{graphicx}
\usepackage{hyperref}

\newcommand{\lya}{Lyman-$\alpha$~}
\newcommand{\Lya}{Lyman-$\alpha$~}

\newcommand{\add}[1]{#1}                                          

\newcommand{\UM}{Department of Physics, University of Michigan, 450 Church St, Ann Arbor, MI 48109, USA}
\newcommand{\LCTP}{Leinweber Institute for Theoretical Physics, 450 Church St, Ann Arbor, MI 48109, USA}

\title{\boldmath 
Small-scale Lyman alpha forest cosmology with PRIYA: Constraints from XQ100 and KODIAQ-SQUAD one-dimensional flux power spectra
}

\author{Ming-Feng Ho$^{1,2,3}$,}\emailAdd{mfho@umich.edu}
\author{Mahdi Qezlou$^{1,4}$,}
\author{Simeon Bird$^1$,}
\author{Yanhui Yang$^1$,}
\author{Camille Avestruz$^{2,3}$,}
\author{M.A. Fernandez$^{1,5}$,}
\author{and Vid Ir{\v{s}}i{\v{c}}$^{6}$,}
\affiliation{$^1$Department of Physics and Astronomy, University of California Riverside, 900 University Ave, Riverside, CA 92521}
\affiliation{$^2$\UM}
\affiliation{$^3$\LCTP}
\affiliation{$^4$Department of Astronomy, University of Texas at Austin, 2515 Speedway Boulevard, Stop C1400, Austin, TX 78712, USA}
\affiliation{$^5$Department of Atmospheric Science, Colorado State University, Fort Collins, CO, USA}
\affiliation{$^6$Center for Astrophysics Research, Department of Physics, Astronomy and Mathematics, University of Hertfordshire, College Lane, Hatfield AL10 9AB, UK}

\abstract{We present a new cosmological analysis of the small-scale \lya\ forest 1D flux power spectrum (P1D) using high-resolution quasar spectra from  XQ100 and KODIAQ-SQUAD, interpreted through the PRIYA emulator. PRIYA is a suite of galaxy formation simulations spanning a range of cosmological and inhomogeneous He\,{\sc ii} reionization parameters, enabling few-percent-level predictions of the P1D. These datasets, probing down to $k \sim 6\,h\,\mathrm{Mpc}^{-1}$ at $z = 2-5$, offer access to non-linear scales inaccessible to large-volume surveys like eBOSS.
We find that the XQ100 P1D yields constraints on the primordial power spectrum parameters $(A_P, n_P)$ at pivot scale $k_0 = 0.78\,\mathrm{Mpc}^{-1}$ that are consistent with PRIYA results from eBOSS DR14 and Planck CMB, albeit with broader uncertainties.
Notably, this is achieved without external IGM temperature data, showing that XQ100 alone provides stronger constraints on thermal history than eBOSS DR14.
In contrast, the KODIAQ-SQUAD P1D favors a significantly higher $A_P$ value,
driven by the selection bias toward high-column density absorbers (HCDs).
We also find that the P1D at $k > 0.045\,\mathrm{s/km}$ is more sensitive to Lyman limit system contamination and thermal history. When imposing a prior on $(A_P, n_P)$, the reduced $\chi^2$ remains unchanged and the inferred mean IGM temperature is unaffected, suggesting that cosmological and thermal parameters are largely sensitive to different scales. The XQ100 P1D therefore provides complementary information on thermal nuisance parameters, which can be jointly fit with eBOSS or DESI P1D measurements to improve cosmological constraints.
}

\begin{document}
\maketitle
\flushbottom
%

\section{Introduction}\label{sec:intro}

The \lya forest traces the distribution of neutral gas at low densities in the intergalactic medium (IGM) at redshifts $z=2-5$ \citep{1965ApJ...142.1633G, 1998ApJ...495...44C, 1998MNRAS.301..478T, 2000ApJ...543....1M, 2001ApJ...552...15H, 2002MNRAS.329..848V, 2006AJ....132..117F, 2006MNRAS.365..231V, 2006ApJS..163...80M}. This neutral gas traces cosmic structure formation, making the \lya forest an exceptionally powerful cosmological probe, sensitive to the distribution of dark matter. \lya forest observations play a crucial role in cosmological spectroscopic surveys, such as the Sloan Digital Sky Survey (SDSS) \citep{McDonald:2006ApJS..163...80M,Palanque-Delabrouille:2013A&A...559A..85P,Slosar:2013JCAP...04..026S,Delubac:2015A&A...574A..59D,2019JCAP...07..017C} and the Dark Energy Spectroscopic Instrument (DESI) \citep{2022AJ....164..207A,2023arXiv230606311R,2023arXiv230606316G,Naim:2025,Ravoux:2025arXiv250509493R,DESI:2025JCAP...01..124A,DESI:2025arXiv250314739D}, enabling these surveys to detect Baryonic Acoustic Oscillations (BAO) and probe the growth of cosmic structure on scales, $k \sim 0.01 - 10$ h/Mpc \citep{2004MNRAS.354..684V, 2005ApJ...635..761M, 2006MNRAS.370L..51V, 2005PhRvD..71j3515S, 2006JCAP...10..014S, 2017JCAP...06..047Y, 2020JCAP...04..038P, 2021JCAP...03..049G}, and at higher redshifts ($z = 2 - 5$) than most other large scale structure probes.

The \lya forest not only serves as a probe of the cosmic web but also provides a unique window into the thermal and ionization history of the  IGM \citep{2008MNRAS.386.1131B, 2014MNRAS.438.2499B, 2016MNRAS.463.2335N, 2019ApJ...872...13W, 2019ApJ...872..101B, 2019MNRAS.490.3177W, 2021MNRAS.506.4389G, 2022ApJ...933...59V}.
This sensitivity to the IGM thermal history enables constraints on processes such as hydrogen and helium reionization, which left distinct imprints on the thermal evolution of the universe \citep{Hui:1997MNRAS.292...27H,2009ApJ...694..842M,Phoebe:2016MNRAS.460.1885U,2020MNRAS.496.4372U}.
At the same time, the \lya forest is sensitive to the free-streaming scale of warm dark matter (WDM), offering constraints on its particle mass \citep{2005PhRvD..71f3534V,  2013PhRvD..88d3502V, 2017PhRvD..96b3522I, 2020JCAP...04..038P, 2021MNRAS.502.2356G, 2022arXiv220914220V,Rogers:2021a}. Dark matter free-streaming suppresses the formation of structure below a characteristic scale, which can be constrained by precision measurements of the \lya flux power spectrum.
In addition, analyses of the small-scale \lya forest P1D have yielded constraints on other alternative dark matter models, such as fuzzy dark matter (see Ref.~\cite{2021PhRvL.126g1302R}). An important strength of the \lya forest as a cosmological probe is its ability to access non-linear clustering statistics over nearly three orders of magnitude in scale: from the largest scales of $\sim 100$ Mpc probed by cross-sightline correlations (see, e.g., 3D \Lya power spectrum, P3D in Ref.~\cite{Andreu:2018JCAP...01..003F,deBelsunce:2024MNRAS.533.3756D,Naim:2025arXiv250315619K}), to intermediate scales of $\sim 1$ Mpc accessible through the 1D power spectrum in systematics-limited surveys, and $\sim 0.1$ Mpc from small high-resolution datasets.

Large-scale spectroscopic surveys like the SDSS \citep{2019JCAP...07..017C} and DESI \citep{2022AJ....164..207A} have measured the 1D flux power spectrum from the \lya forest. The extended Baryon Oscillation Sky Survey (eBOSS), part of SDSS, mapped the flux power spectrum using 43,751 high signal to noise quasar sightlines \citep{2019JCAP...07..017C}. Despite its scale, eBOSS remained limited by systematic uncertainties, particularly those arising from spectrograph resolution and calibration issues, that dominated its error budget. DESI addresses these limitations with a substantial improvement in spectral resolution, nearly doubling the resolving power of its predecessors. This enhancement allows DESI to probe smaller scales, down to $k \sim 0.035$ s km$^{-1}$ \citep{2024MNRAS.528.3941K,Naim:2025}. The full projected five-year dataset of 700,000 \lya forest quasars from DESI will achieve percent level accuracy even on the smallest scales, substantially improving cosmological constraints and revealing fine details of the IGM thermal history.

A further source of complementary information comes from high-resolution quasar spectra surveys. These include the X-Shooter quasar catalog 100 (XQ100), a blind survey of $100$ quasar spectra, and two catalogues assembled from the telescope archives: the Keck Observatory Database of Ionized Absorption toward Quasars (KODIAQ), and the UVES Spectral Quasar Absorption Database (SQUAD). These surveys provide information on the \lya forest at the smallest physically accessible scales, $k \approx 0.001 - 0.2$ s km$^{-1}$ \citep{2017MNRAS.466.4332I, 2019MNRAS.489.2536D, 2022MNRAS.509.2842K, 2022MNRAS.515..857E,Boera:2019ApJ...872..101B}\footnote{$k \approx 0.001 - 0.2$ s km$^{-1}$ is probed by Ref.~\cite{Boera:2019ApJ...872..101B}.}, and are especially sensitive to the IGM thermal history.
In this work, we present an analysis of the \lya\ P1D from these high-resolution quasar spectroscopic surveys. We reanalyze the XQ100 P1D ($R \sim 5\,000$) \citep{2022MNRAS.509.2423W}
and, for the first time, perform joint inference of cosmological parameters and IGM thermal history using the KODIAQ-SQUAD P1D ($R \sim 38\,000$) \citep{2022MNRAS.509.2842K}, which is the largest high-resolution dataset yet assembled, comprising $\sim 700$ quasar sightlines with $z_\mathrm{QSO} = 2 - 5$.

High-resolution hydrodynamical simulations are required to capture the non-linear small-scale clustering of neutral hydrogen, but they are computationally expensive and face an inherent trade-off between resolution and volume.
To enable cosmological inference, theoretical predictions of the flux power spectrum must achieve percent-level accuracy across a wide range of cosmological and astrophysical parameters.
Currently, all available \lya\ emulators built on hydrodynamical simulations, e.g., Refs.~\citep{Lyssa:2025JCAP...05..099W, Cabayol:2023MNRAS.525.3499C,Bird:2019JCAP...02..050B,Jin:2025MNRAS.536.2277J},
are constrained by this resolution/volume trade-off, which limits their ability to extend to the smallest scales of the P1D and beat down the cosmic variance.

Our work builds on the PRIYA simulation suite \cite{2023simsuite}, a state-of-the-art suite of large-volume, high resolution simulations of the \lya forest which uniquely include patchy hydrogen and helium reionization.
The PRIYA suite \cite{2023simsuite} addresses the resolution/volume challenge by employing a multi-fidelity emulator framework  (see, e.g., Ref.~\cite{2017RSPSA.47360751P,Peherstorfer:2018,2022MNRAS.509.2551H,2022MNRAS.517.3200F,Yang:2025PhRvD.111h3529Y,Diao:2025arXiv250204246D}), integrating low- ($1536^3$) and high-resolution ($3072^3$) hydrodynamical simulations within $(120\, \mathrm{Mpc}/h)^3$ volumes, which is so far the largest volume for a \Lya emulator. This strategy allows PRIYA to span a wide range of cosmological and reionization parameters without the need for ad hoc splicing corrections. Through multi-fidelity emulation, low-resolution outputs are corrected using high-resolution samples, preserving physical accuracy while extending parameter coverage.

In PRIYA, we vary the AGN feedback strength as one of the astrophysical parameters. 
As noted in Ref.~\cite{2023simsuite}, AGN feedback has little impact on the \Lya\ forest flux power spectrum within current survey errors, especially at $z > 2$ (see also Ref.~\cite{Tillman:2025ApJ...980...72T}). 
In this work, however, we take the view that He\,\textsc{ii} reionization is quasar-driven, and therefore we can view the He\,\textsc{ii} reionization effect in PRIYA as a form of AGN (quasar) feedback. 

The structure of this paper is as follows: 
In \S~\ref{sec:simulations}, we describe the PRIYA simulation suite and the construction of the multi-fidelity emulator framework. 
\S~\ref{sec:inference} outlines our inference pipeline, including introductions to KODIAQ-SQUAD and XQ100 P1Ds, high-column density absorber (HCDs) templates, likelihood function and parameter priors.
We present the main results, including posterior constraints from the XQ100 and KODIAQ-SQUAD datasets, in \S~\ref{sec:results}. 
In \S~\ref{sec:discussion}, we discuss the implications of these findings, highlight potential limitations of the current P1D analysis, and suggest directions for future work. 
Finally, \S~\ref{sec:conclusions} summarizes our findings and discusses prospects for next-generation \lya\ forest studies.

\section{Simulation Suite and Emulator}
\label{sec:simulations}


\begin{table}
	\centering
     \begin{tabular}{|c|c|c|c|c|}
		\hline
		Simulation & Box Volume & N$_{\text{gas}}$ & M$_{\text{gas}}$ (M$_{\odot}$ h$^{-1}$)\\
		\hline
		LF & $(120$ Mpc h$^{-1})^3$ & $1536^3$ & $[5.29, 6.98]\times10^6$\\
		HF & $(120$ Mpc h$^{-1})^3$ & $3072^3$ & $[6.73, 7.97]\times10^5$\\
		\hline
	\end{tabular}
    \caption{\label{table:simulations}
    Low-Fidelity (LF) and High-Fidelity (HF) simulation suite details.
    N$_{\text{gas}}$ is the number of gas particles simulated, M$_{\text{gas}}$ is the mass resolution of those particles, which depends on $\Omega_M$.}
\end{table}


\begin{table*}
	\begin{centering}
	  \begin{tabular}{llll}
	  \hline
	  Parameter & Minimum & Maximum & Description \\
		\hline
		$n_P$  &  $0.8$  & $1.05$ & Scalar spectral index \\
		$A_P$  &  $1.2 \times 10^{-9}$  & $2.6 \times 10^{-9}$ & Power amplitude at $k = 0.78 \,\mathrm{Mpc}^{-1}$ \\
		$\Omega_M h^2$ & $0.14$ & $0.146$ & Total matter density \\
		$z_{Hei}$      & $3.5$  & $4.1$  & Start redshift of He~{\sc ii} reionization \\
		$z_{Hef}$      & $2.6$  & $3.2$  & End redshift of He~{\sc ii} reionization \\
		$\alpha_q$     & $1.3$  & $2.5$ & Quasar spectral index during He~{\sc ii} reionization  \\
		$z_{Hi}$        & $6.5$ & $8$   & Median redshift of H~{\sc i} reionization \\
		\hline
		Prior-dominated \\
		$h$    & $0.65$  & $0.75$ & Hubble parameter \\
		$\epsilon_{AGN}$ & $0.03$ & $0.07$ & Thermal efficiency of black hole feedback \\
		\hline
		Post-processing \\
		$\tau_0$ & $0.75$ & $1.25$ & Mean optical depth at $z=3$ in Eq.~\ref{eq:meanflux}.\\
		$d \tau_0$ & $-0.4$ & $0.25$ & $\tau_0$ redshift evolution in Eq.~\ref{eq:meanflux}. \\
		\hline
	  \end{tabular}
	  \caption{Summary of emulator and post-processing parameters. We vary a total of $11$ parameters: $4$ for cosmology, $3$ for the helium reionization model, $1$ for the hydrogen reionization model, $1$ for the strength of AGN feedback and $2$ for the mean optical depth.}
	  \label{tab:emulatorparams}
	  \end{centering}
	\end{table*}
	
We use the PRIYA simulation suite \cite{2023simsuite} to predict the 1D flux power spectrum (or P1D) of the \lya forest as a function of cosmology.
In this section, we briefly recap the simulation suite and the multi-fidelity emulator used to predict the P1D.
Detailed descriptions of the simulation suite and emulator can be found in Refs.~\cite{2023simsuite, 2022MNRAS.517.3200F}.
For reviews of multi-fidelity emulation, see Ref.~\cite{Peherstorfer:2018} for a thorough methodology review and Ref.~\cite{2022MNRAS.509.2551H,Ho:2023MNRAS.526.2903H,Yang:2025PhRvD.111h3529Y,Yang:2025arXiv250707177Y,Yang:2025arXiv250707184Y} for cosmology applications.

The PRIYA suite\footnote{\url{http://priya-sims.readthedocs.io}} is run with the MP-Gadget\footnote{\url{https://github.com/MP-Gadget/MP-Gadget}} N-body and smoothed particle hydrodynamics (SPH) code.
PRIYA suite is designed around the multi-fidelity emulator strategy, and combines simulations with two different resolutions to predict the P1D at high-fidelity through a machine learning model (see Ref.~\cite{2022MNRAS.517.3200F}).
We run two sets of simulations, 60 low-fidelity (LF) and 3 high-fidelity (HF) simulations, with different particle loads, $1536^3$ and $3072^3$, in a $120\,\mathrm{Mpc}/h$ co-moving box, as shown in Table~\ref{table:simulations}.
A multi-fidelity emulator is a machine learning model that interpolates over simulations with multiple fidelity levels. It predicts the output of a simulation at the highest fidelity for arbitrary input cosmological parameters within the parameter limits, just as a traditional emulator with homogeneous simulation input, but with a reduced cost training set. The accuracy of the multi-fidelity emulator is validated against high-fidelity simulation outputs using leave-one-out cross-validation (LOO-CV), as described in Ref.~\cite{2023simsuite}.

\S~\ref{sec:gps} describes and validates our \lya forest 1D flux power spectrum emulator for $k = 0.003 - 0.06 \,\mathrm{s/km}$, which extends the emulator from Ref.~\cite{2024JCAP...07..029F} to model smaller scales.
\S~\ref{sec:parameters} describes the cosmological and astrophysical parameters varied in the emulator, and the parameter prior limits.

\subsection{P1D Emulator for XQ100 and KODIAQ-SQUAD}\label{sec:gps}

\begin{figure}
	\centering
	\includegraphics[width=\columnwidth]{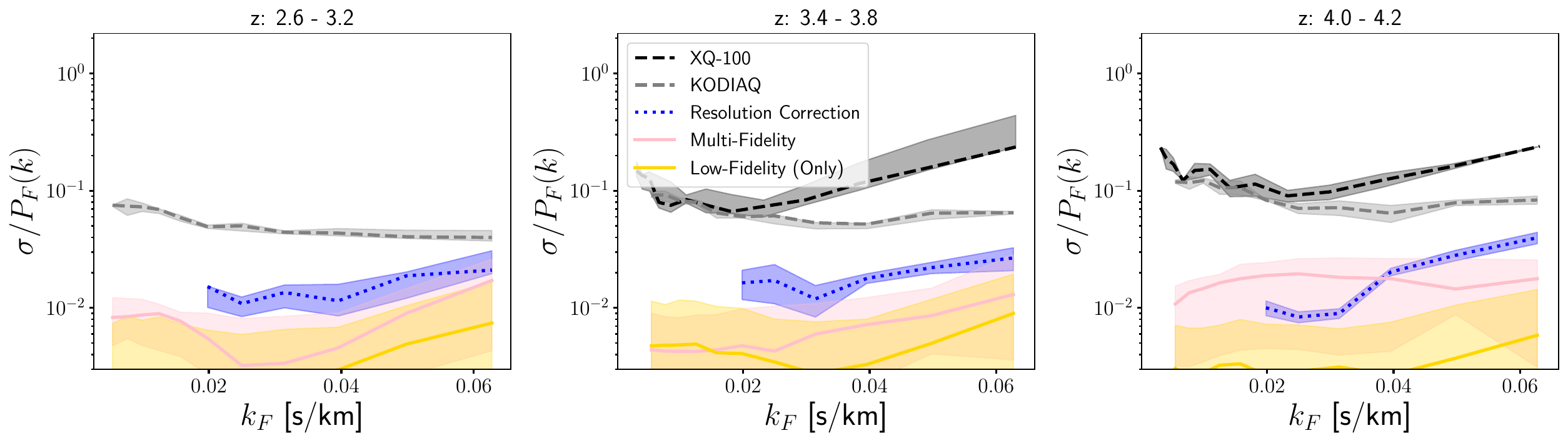}
	\caption{Leave-one-out cross-validation (LOO-CV) accuracy of the multi-fidelity emulator at $z=2.6 - 3.2$, $3.4 - 3.8$, and $4.0 - 4.2$.
	Pink lines (multi-fidelity) show the LOO-CV error against HF simulations ( median and 68\% percentiles), while yellow lines (LF only) show the LOO-CV of LF emulator against LF simulations.
	We overplot the square root of the diagonal elements of the covariance matrix of KODIAQ-SQUAD (gray dotted) and XQ100 (black dashed) for comparison.
	Blue lines show the convergence error estimated in Ref.~\cite{2023simsuite} between HF and the extreme high-resolution simulation ($2\times 768^3$ particles in a $15\,\mathrm{Mpc}/h$ box).}
	\label{fig:emulator_error}
\end{figure}

In Ref.~\cite{2024JCAP...07..029F}, we trained a multi-fidelity emulator model which predicts the P1D at $\sim 1\%$ accuracy for the scales measured by eBOSS, $k = 0.001 - 0.02 \,\mathrm{s/km}$.
In this work, we extend the emulator to predict the P1D on the smaller scales measured by XQ100 and KODIAQ-SQUAD, $k = 0.003 - 0.06 \,\mathrm{s/km}$. Our emulator construction, including multi-fidelity corrections and hyperparameters, is identical to Ref.~\cite{2024JCAP...07..029F}.

Following Ref.~\cite{2023simsuite}, we validate our emulator using leave-one-out-cross-validation (LOO-CV) with each high fidelity simulation left out in turn.
The LOO-CV results are shown in Figure~\ref{fig:emulator_error} for three redshift ranges, $z = 2.6 - 3.2$, $3.4 - 3.8$, and $4.0 - 4.2$.
Note that we only have 3 HF simulations, so the LOO-CV would significantly over-estimate the HF interpolation error because we are discarding 1/3 of the HF training set.
We find that the emulator interpolates the P1D at $k = 0.01 - 0.06 \,\mathrm{s/km}$ at $\sim 1 \%$ accuracy (median).
At higher redshifts, $z = 4 - 4.2$, the interpolation error is slightly larger, $\sim 2 \%$, as the \lya forest probes smaller physical scales at higher redshifts \cite{2009MNRAS.398L..26B}.

The interpolation error is not the only source of uncertainty in the emulator, as the emulator is trained on simulations with a finite resolution.
Ref.~\cite{2023simsuite} checked the resolution convergence and found it to be $\sim 1\%$ for $k < 0.05\,\mathrm{s/km}$ and $\sim 3\%$ for $k = 0.05 - 0.06\,\,\mathrm{s/km}$, which is substantially smaller than the statistical error of $10\%$ for KODIAQ-SQUAD and $\sim 20\%$ for XQ100. Smaller scales are less well resolved, with on the order of $4-10\%$ convergence error, and for this reason we conservatively exclude these scales from our analysis. We attempt to improve resolution convergence by applying a cosmology-independent correction function in the emulator. Specifically, we multiply by P1D for $k = 0.02 - 0.06 \,\mathrm{s/km}$ by the ratio between the two highest resolution simulations run by Ref.~\cite{2023simsuite}. These are, respectively, $2\times 512^3$ particles in a $15\,\mathrm{Mpc}/h$ box (thus a resolution equal to our high fidelity tier), and $2\times 768^3$ particles in a $15\,\mathrm{Mpc}/h$ box.

\subsection{Cosmological \& Astrophysical Parameters}\label{sec:parameters}

The cosmological and astrophysical parameters varied in the emulator are identical to Ref.~\cite{2024JCAP...07..029F}, and are summarised in Table~\ref{tab:emulatorparams}.\footnote{We have constructed an online widget to visualize how the P1D depends on emulator parameters, available at \url{https://jibancat.github.io/priya-widget/}}
Our emulator contains two parameters related to the primordial (the pre-transfer function) power spectrum
are varied: the scalar spectral index, $n_P$, and the power amplitude at $k = 0.78 \,\mathrm{Mpc}^{-1}$, $A_P$:
\begin{equation}
    P(k) = A_P \left(\frac{k}{0.78\, \mathrm{Mpc}^{-1}}\right)^{n_P - 1}\,.
\end{equation}
The pivot scale $0.78 \,\mathrm{Mpc}^{-1}$ is chosen to match the scales probed by the \lya forest. For eBOSS, this pivot scale minimises the degeneracy between $n_P$ and $A_P$, although we shall see that some degeneracy exists for the high-resolution P1D we examine here, which probes smaller scales (see Figure~\ref{fig:1pvar_ap_aLLS}).
We also vary the Hubble parameter $h$ and the total matter density $\Omega_M h^2$, although these have small ($1-4\%$) effects on the \lya forest and are not strongly constrained by the high resolution measurements. The dominant effect of $h$ is to slightly change the mapping between simulated (Mpc/h) and observed (km/s) units.

There are three Helium reionization parameters, which control the thermal history of the IGM.
The He~{\sc ii} reionization model follows Ref.~\cite{2020MNRAS.496.4372U}: $z_{Hei}$ and $z_{Hef}$ are the redshifts for the start and end of He~{\sc ii} reionization. $\alpha_q$ is the quasar spectral index, which scales the peak temperature during He~{\sc ii} reionization, with
a \textit{lower} $\alpha_q$ value corresponding to a \text{higher} heating rate.
$z_\mathrm{Hi}$ is the midpoint redshift of H~{\sc i} reionization, defined as the redshift at which
$50\%$ of the hydrogen is ionized.
Note that our emulator includes scenarios where hydrogen reionization completes at $z < 6$, as supported by recent quasar dark gap measurements \cite{2022ApJ...932...76Z}; for $z_{Hi} = 6.5$, hydrogen is $95\%$ ionized at $z = 5.5$.



Finally, we have two additional parameters for the \Lya~effective optical depth, which measures the average amount of the \Lya~forest absorption as a function of redshift and is set by the unknown UV photon background density. Higher redshifts have a larger effective optical depth, corresponding to a more neutral IGM. We parameterize the mean flux $\bar{\mathcal{F}} = \exp(-\tau^{\text{eff}}_{\text{H~{\sc i}}})$ as variations from the power law redshift evolution model of Ref.~\cite{2007MNRAS.382.1657K}.
\begin{align}
	\begin{split}
		\tau^{\text{Kim}}_{\text{H~{\sc i}}}(z) &= 0.0023 \times (\tau_{z=3})\times (1+z)^{3.65}\,, \\
		\tau^{\text{eff}}_{\text{H~{\sc i}}}(z) &= \tau_0 \left(\frac{\tau^{\text{Kim}}_{\text{H~{\sc i}}}(z)}{\tau^{\text{Kim}}_{\text{H~{\sc i}}}(3)}\right)^{d\tau_0} \tau^{\text{Kim}}_{\text{H~{\sc i}}}(z)\,\\
		&= \tau_0 \left(\frac{1+z}{4}\right)^{3.65+d\tau_0} \tau^{\text{Kim}}_{\text{H~{\sc i}}}(z)\,.
	\end{split}
 \label{eq:meanflux}
\end{align}
In practice, we rescale the optical depths in our simulated spectra by a constant factor in post-processing to match the desired $\bar{\mathcal{F}}$. We sample $10$ linearly spaced values of the mean flux, independent of redshift, to cover the range of $\tau_0$ and $d\tau_0$ at the extreme ($z=2.2$ and $z=4.6$) redshifts.
The final set of P1D training set for the emulator is thus ten times the number of simulations, or $600$ LF, and $30$ HF simulated flux power spectra.


Our parameterization in Eq.~\ref{eq:meanflux} is based on Ref.~\cite{2007MNRAS.382.1657K}, which is measured from 18 high resolution, high signal-to-noise VLT/UVES QSO spectra, and we allow the mean flux to vary (with a pivot point at $z=3$) by $\tau_0$ and the redshift evolution by $d\tau_0$ with a large enough prior.
This allows us to capture a wide variety of the mean flux evolutions.\footnote{We show the mean flux prior in Fig.~\ref{fig:meanflux-inferred} alongside the meanflux measurement from Ref.~\cite{2007MNRAS.382.1657K} and Ref.~\cite{2013MNRAS.430.2067B}.}

\section{Inference Scheme and Likelihood Function}\label{sec:inference}

We perform the likelihood analysis of the XQ100 and KODIAQ-SQUAD P1D using the emulator described in \S~\ref{sec:simulations}.
The inference scheme is as follows:
\begin{enumerate}
    \item Predict the flux power spectrum (P1D) from the emulator for a set of input parameters (see Table~\ref{tab:emulatorparams}).
    \item Calculate a likelihood comparing these predictions to the XQ100 or KODIAQ-SQUAD P1D.
    \item Run Markov Chain Monte Carlo (MCMC) chains using Cobaya \cite{2021JCAP...05..057T, 2019ascl.soft10019T} to compute posterior parameter constraints.
\end{enumerate}
Step 1 is straightforward, as the emulator trained on the PRIYA simulation suite can predict the P1D at the desired redshifts and scales.
For Step 2, we directly take the covariance matrix from the XQ100 or KODIAQ-SQUAD data, and compute the multi-variate normal likelihood using Equation~\ref{eq:likelihood}.
With the likelihood in hand, we run MCMC chains using Cobaya.

\S~\ref{sec:theoryerror} describes the covariance matrix used in the likelihood calculation.
Likelihood function and prior are detailed in \S~\ref{sec:likelihood} and \S~\ref{sec:priors}, respectively.
A validation of our MCMC inference using simulated data is presented in \S~\ref{sec:simdat}.
\S~\ref{sec:fpsdata} discusses the flux power spectrum data from XQ100 and KODIAQ-SQUAD.
Finally, \S~\ref{sec:dlalimit} describes the high-column density absorbers (HCDs) template used for marginalizing out the HCD contamination.

\subsection{Covariance Matrix}
\label{sec:theoryerror}

Below we describe the covariance matrix used in the likelihood calculation.
\begin{equation}
    \boldsymbol{K} = \boldsymbol{K}_\mathrm{FPS} 
    + \boldsymbol{\sigma}_{CV} \cdot \boldsymbol{\sigma}_{CV}^T \,.
    \label{eq:covariance}
\end{equation}
Here $\boldsymbol{K}_\mathrm{FPS}$ is the covariance matrix from either XQ100 data or KODIAQ-SQUAD data.
$\boldsymbol{\sigma}_{CV}$ is residual sample variance from the finite box size.
We include an estimate of sample variance using the leave-one-out errors discussed in Ref.~\cite{2023simsuite}, a technique made possible by the inclusion of $h$ in our simulation suite.
Gadget uses Mpc/$h$ units, and the conversion of wavenumbers from $h$/Mpc to s/km changes the Fourier modes between bins depending on the value of $h$, which acting as a proxy for the sample variance from different initial phases.
We approximate $\boldsymbol{\sigma}_{CV}$ with the averaged variance of the leave-one-out errors using the low fidelity simulations.
Details of this approximation are discussed in Ref.~\cite{2023simsuite}.

\subsection{Likelihood}\label{sec:likelihood}

The likelihood function is a multivariate normal likelihood summed over all redshifts and, for the flux power, all wavenumber bins:
\begin{equation}
    \mathrm{log}\mathcal{L} = -\frac{1}{2} \sum_{z=z_\mathrm{min}}^{z=z_\mathrm{max}} \left(\left(\boldsymbol{P}_F^{\mathrm{diff}}\right)^\top \cdot \boldsymbol{K}^{-1} \cdot \boldsymbol{P}_F^{\mathrm{diff}} + \log\left( \mathrm{det}(\boldsymbol{K})\right)\right)\,.
    \label{eq:likelihood}
\end{equation}
Here $\boldsymbol{P}_F^{\mathrm{diff}} = \boldsymbol{P}_F^{\mathrm{sim}} - \boldsymbol{P}_F^{\mathrm{obs}}$ is the vector difference between the simulation prediction and the observation.
We chose to compute the $\boldsymbol{P}_F^{\mathrm{sim}}$ on the bins of the observed data.
The covariance matrix, $\boldsymbol{K}$, is described in Equation~\ref{eq:covariance}.
The minimum and maximum redshifts are set to $z=3.4$ and $z=4.2$ for the XQ100 data ($z=2.6$ and $z=4.2$ for the KODIAQ-SQUAD data).
We use the same Cobaya MCMC sampler as in Ref.~\cite{2024JCAP...07..029F}, which uses the Metropolis method discussed in Ref.~\cite{2013PhRvD..87j3529L}, and a Gaussian + exponential proposal distribution that dynamically learns the proposal covariance.


\subsection{Priors}
\label{sec:priors}

The parameter priors are shown in Table~\ref{tab:emulatorparams}.
We use an informative Gaussian prior on the AGN feedback parameter, $\epsilon_{AGN}$, with $\mu = 0.05$ and $\sigma = 0.005$, as it is poorly constrained by the data and has minimal correlations with other parameters.
We have shown in Ref.~\cite{2023simsuite} that the AGN feedback parameter has minimal effect on the \Lya forest 1D flux power spectrum for the redshift range of interest.
The $\epsilon_{AGN}$ changes the P1D at the level $2\%$ at $k \sim 0.06\,\mathrm{s/km}$ rather than the sub-percent level for $k < 0.02\,\mathrm{s/km}$.
However, this is still small compared to the statistical $\sim 10\%$ uncertainty of KODIAQ-SQUAD at $k \sim 0.06\,\mathrm{s/km}$.
We experimented with using a uniform prior on $\epsilon_{AGN}$. The mode shifts to $\epsilon_{AGN} \sim 0.03$, but there is negligible change to the other parameters.


We also place a weak Gaussian prior on the Hubble parameter, $h$, with $\mu = 0.70$ and $\sigma = 0.015$, as it is weakly constrained by the P1D.
For all other parameters we use uniform priors within the parameter limits.
\add{In particular, we use uniform (linear) priors on the primordial power spectrum parameters, $A_P \sim \mathrm{Uniform}(1.2 \times 10^{-9}, 2.6 \times 10^{-9})$ and $n_P \sim \mathrm{Uniform}(0.8, 1.05)$.}
\add{As a robustness test, we also reran the baseline XQ100 and KODIAQ-SQUAD chains with a log-uniform prior on $A_P$. The posterior shifts are small, with the largest shift being $0.2\sigma$ (see Appendix~\ref{sec:ap-prior-robustness}).}

\subsection{Inference Using Simulation Data from PRIYA}\label{sec:simdat}

\begin{figure}
    \includegraphics[width=0.49\columnwidth]{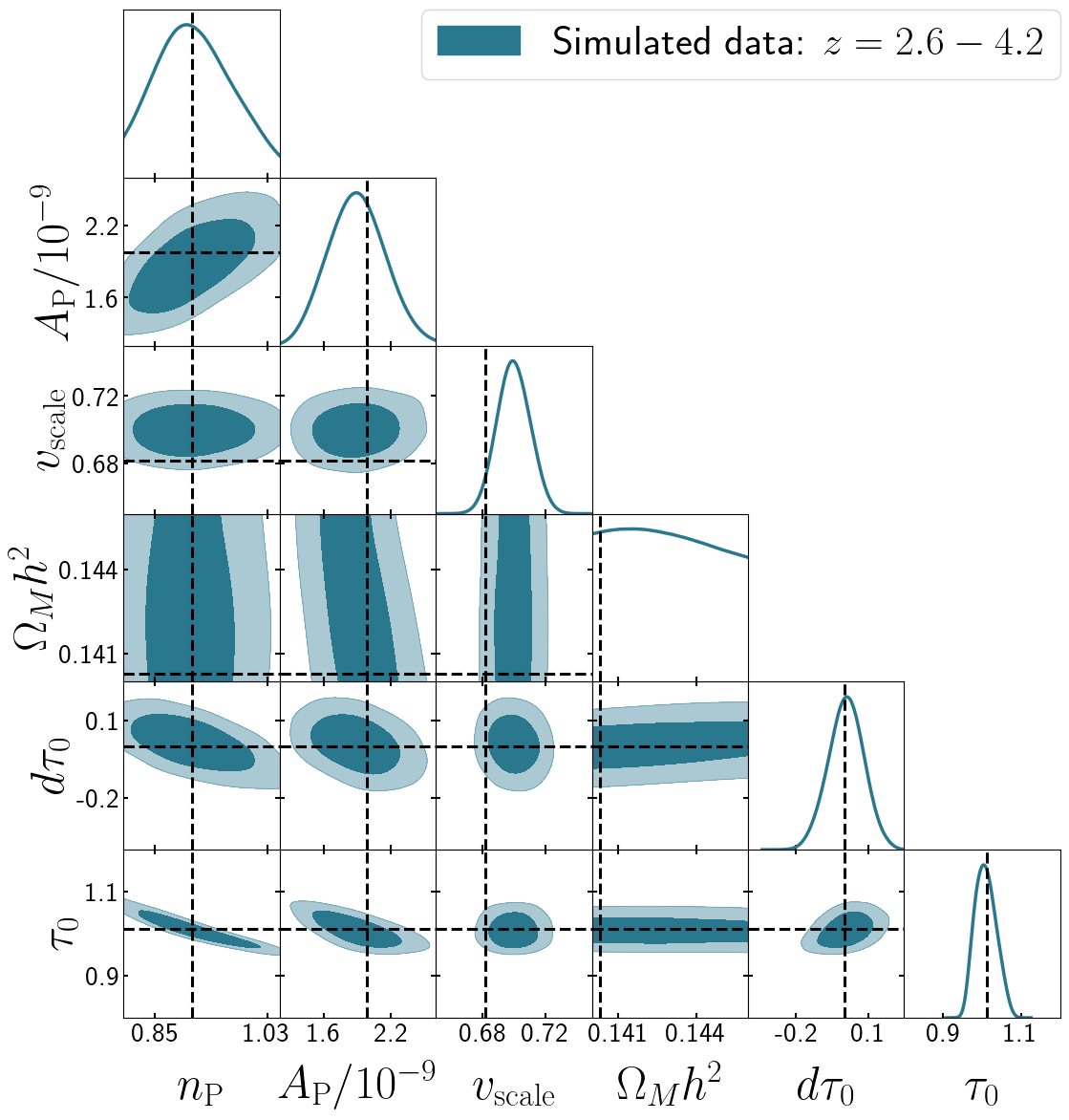}
    \includegraphics[width=0.49\columnwidth]{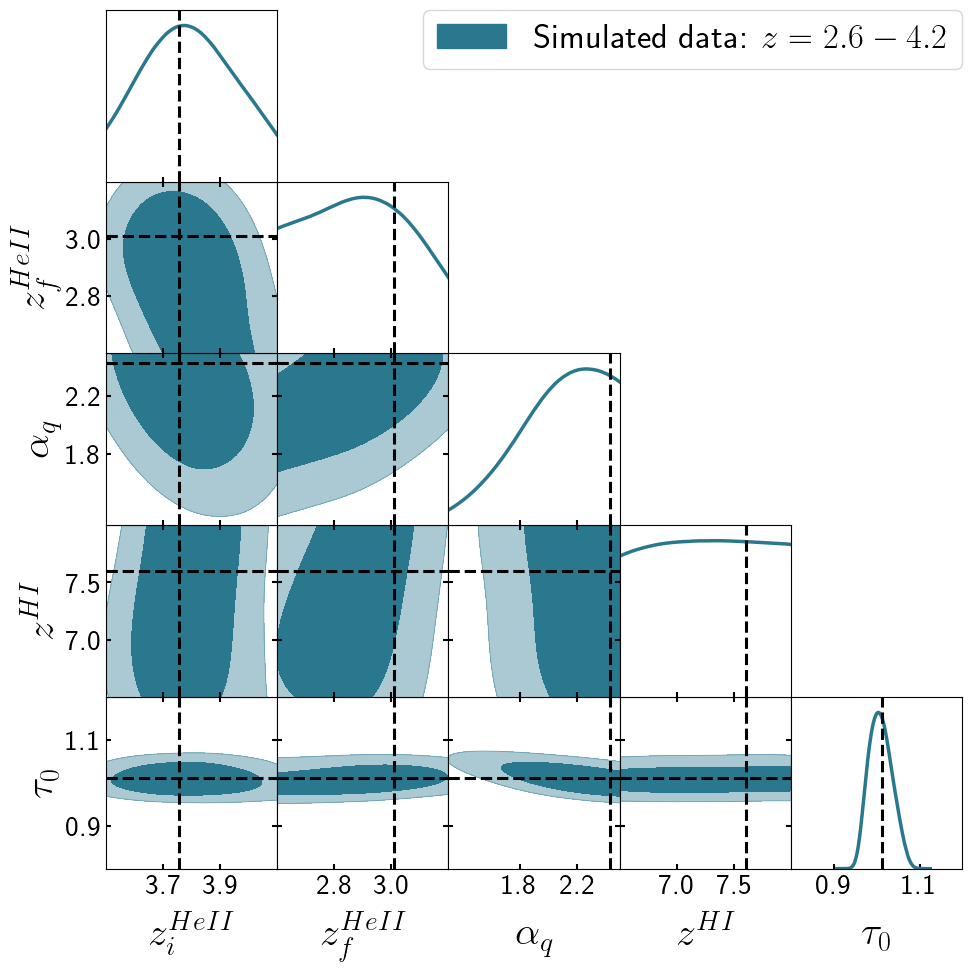}
    \caption{Posterior contours from our likelihood function and emulator, using simulated data and the KODIAQ-SQUAD covariance matrix for validation, restricted to $z = 2.6-4.2$.
    The left panel shows the cosmological parameters, and the right panel shows the reionization parameters.
    The dashed black lines indicate the correct input values of the parameters.
    Contours show the $1-\sigma$ and $2-\sigma$ confidence regions.
    }
    \label{fig:cornerplot_simdat}
\end{figure}

Figure~\ref{fig:cornerplot_simdat} shows a validation of our inference scheme, using the simulation outputs in place of the observational data.
Dashed black lines indicate the correct input parameters, and contours show the posterior outputs of the likelihood.
We have a similar validation in Ref.~\cite{2024JCAP...07..029F}, but here we use the covariance matrix and bins from KODIAQ-SQUAD instead of eBOSS.

With the simulated data as validation, we can test whether the likelihood function recovers the correct mode of the parameters, and we can also verify which parameters are not well constrained by this likelihood function.
From Figure~\ref{fig:cornerplot_simdat}, we see the likelihood function recovers the correct mode of $n_P, A_P, z^\mathrm{HeII}_i, z^\mathrm{HeII}_f, \tau_0, d\tau_0$, and $\alpha_q$.
All parameters are falling within the $1-\sigma$ range of the correct input values.

There are some parameters that are not well constrained by the likelihood function, such as $v_\mathrm{scale}$, which is the Hubble parameter $h$, and $\Omega_M h^2$.
As we mentioned in Ref.~\cite{2023simsuite}, the MP-Gadget simulation is run with comoving units Mpc/$h$, so the $h$ does not directly affect the gravitational evolution in our simulations.
The primary effect of $h$ in this case is the change of comoving unit to velocity units (km/s) in P1D, that is, shifting the Fourier modes between bins.
This effect is dominated by the residuals of the sample variance due to our finite box size.
Even with this residual, the P1D is only weakly sensitive to $h$ at $ < 2\%$ effect within the prior range.

$z_{HI}$, the mid-point of the redshift of hydrogen reionization, is also not well constrained by the likelihood function.
This is because the P1D is not very sensitive to the redshift of hydrogen reionization, as it is higher redshift ($z \gtrsim 5$) than the redshift range of the P1D data we use ($z  = 2.6 - 4.2$).


\subsection{Flux Power Spectrum: XQ100 \& KODIAQ-SQUAD}
\label{sec:fpsdata}

\begin{figure}
    \centering
    \includegraphics[width=\columnwidth]{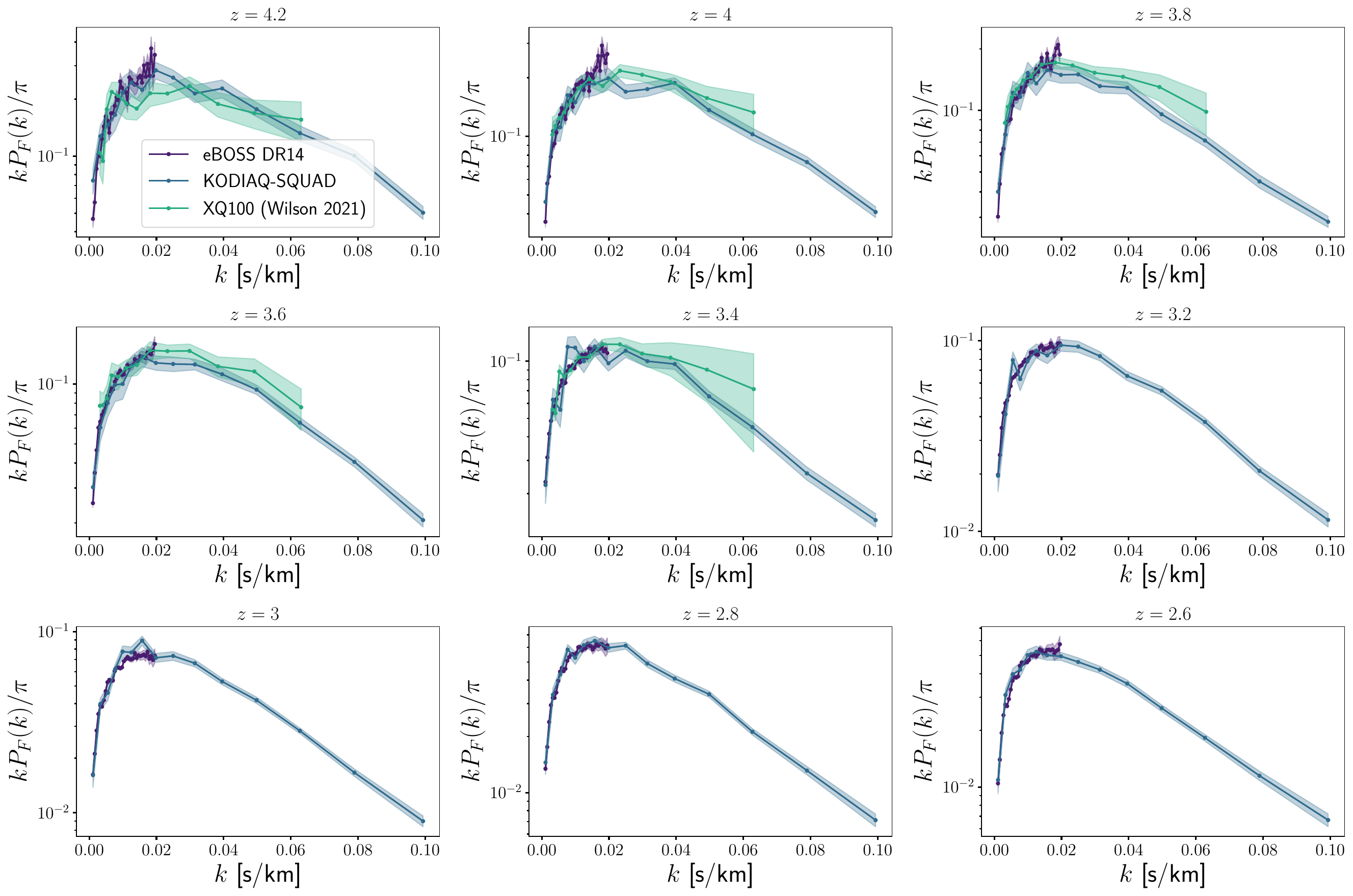}
    \caption{The flux power spectrum data from XQ100 (green) and KODIAQ-SQUAD (blue).
    We show eBOSS DR14 (purple) data for comparison.
    The XQ100 data is re-analyzed from Ref.~\cite{2022MNRAS.509.2423W}, and the KODIAQ-SQUAD data is from the optimal estimator in Ref.~\cite{2022MNRAS.509.2842K}.
    The old XQ100 data from Ref.~\cite{2017MNRAS.466.4332I} is  not used in the analysis.
    The XQ100 data is limited to $z=3.4-4.2$ and $k=0.003-0.065 \,\mathrm{s/km}$, while the KODIAQ-SQUAD data is limited to $z=2.6-4.2$ and $k=0.0055-0.065 \,\mathrm{s/km}$ in our analysis.
    The KODIAQ-SQUAD P1D span a wider range of $k$ than XQ100 with a tighter 1$\sigma$ errorbar.}
    \label{fig:fpsdata}
\end{figure}

Figure~\ref{fig:fpsdata} shows the P1D data used in this work. We use measurements from three high-resolution and high SNR quasar surveys: XQ100 \cite{Lopez:2016A&A...594A..91L}, KODIAQ \cite{Lehner:2014ApJ...788..119L,OMeara:2015AJ....150..111O,OMeara:2017AJ....154..114O}, and SQUAD \cite{Murphy:2019MNRAS.482.3458M}. The XQ100 survey comprises 100 quasars at $3.5 < z_\mathrm{QSO} < 4.5$ \cite{Lopez:2016A&A...594A..91L}, observed with the X-Shooter spectrograph on the ESO Very Large Telescope (VLT) \cite{Vernet:2011A&A...536A.105V}, with a resolving power of $R \sim 4\,000-7\,000$. We adopt the re-analysis by Ref.~\cite{2022MNRAS.509.2423W}, which introduces a refined and improved covariance calculation using Ly$\beta$ regions. We restrict the usable redshift range to $z = 3.4-4.2$, following the re-analysis by Ref.~\cite{2022MNRAS.509.2423W}.

The KODIAQ-SQUAD dataset combines archival spectra from Keck/HIRES \cite{Vogt:1994SPIE.2198..362V} and VLT/UVES \cite{Dekker:2000SPIE.4008..534D}, both offering higher resolving power ($R \gtrsim 36{\,}000-40{\,}000$) over the redshift range $z = 2.0-5.0$ with 767 quasars in total. The joint measurement of P1D in Ref.~\cite{2022MNRAS.509.2842K} uses an optimal quadratic estimator, which is robust to masking and spectrograph inhomogeneities. For this work, we use only the KODIAQ and SQUAD subsamples from Ref.~\cite{2022MNRAS.509.2842K} (excluding XQ100 to avoid redundancy).


We have also made some decisions on the data cuts (redshift and $k$ ranges).
The $k$-range and redshift range of the XQ100 data is limited to $k=0.003 - 0.064 \,\mathrm{s/km}$ and $z=3.4-4.2$, which are largely unchanged from the conservative data cuts already implemented by Ref.~\cite{2022MNRAS.509.2423W}.

We limit the KODIAQ-SQUAD P1D to $k=0.0055 - 0.065 \,\mathrm{s/km}$ and $z=2.6-4.2$.
It was suggested in Ref.~\cite{2022MNRAS.509.2842K} to only use $0.004 < k < 0.1 \,\mathrm{s/km}$ as a conservative cut.
We remove additional bins, so that the lowest $k$ bin is $k=0.0055 \,\mathrm{s/km}$.
This is to avoid potential cosmic variance issues due to the small sample size, and because eBOSS already measures the low-$k$ P1D well. Our focus here is on the small-scale P1D.
The maximum $k$ is set to $0.065 \,\mathrm{s/km}$, due to the limited resolution of our simulation suite rather than the data itself.
Scales above $0.065 \,\mathrm{s/km}$ have $4 - 10\%$ resolution convergence error in our HF simulation suite at some redshift bins, which is comparable to the error of the KODIAQ-SQUAD P1D.

We also limit the analysed redshift range to $z=2.6 - 4.2$, excluding data at high and low redshifts, as it is known that these ranges are difficult to handle data-wise.
There have been discussions on the disagreement between eBOSS and other P1D measurements such as DESI DR1 and KODIAQ-SQUAD in bins with $z < 2.6$ (see, e.g., Ref.~\cite{Naim:2025, 2024JCAP...07..029F}).
These low-redshift bins are from quasar sightlines with shorter wavelength segments, which are more challenging for DLA finders.
\add{In particular, Ref.~\cite{Naim:2025} suggests this low-$z$ disagreement may stem from automated DLA finders mistakenly classifying sub-DLAs as DLAs, contaminating the DLA mask. We have verified that including the $z=2.2$--$2.4$ bins shifts the cosmological parameters $A_P$ and $n_P$ by $\leq 0.2\sigma$ while the mean-flux parameters $\tau_0$ and $\delta\tau_0$ change significantly; we therefore conservatively omit these bins, since the omission only mildly affects cosmology while reducing data-level tension between surveys.}
At high redshift $z > 4.2$, the sample size is small, which makes the covariance difficult to estimate, and implies limited impact on the overall cosmological inference in any case.


The KODIAQ-SQUAD P1D is a re-reduction of all available Keck and UVES spectra across a wide range of target selection strategies over 300+ PIs. Ref.~\cite{2022MNRAS.509.2842K} noted that KODIAQ targets O~{\sc iv} systems while UVES targets known DLAs, which introduce a selection bias toward higher HCD contamination.
We will discuss the impact of this selection bias on our results in \S~\ref{sec:results} and \S~\ref{sec:discussion}.

\subsection{Template for DLAs and LLSs}
\label{sec:dlalimit}

Both the XQ100 and KODIAQ-SQUAD spectra have high resolution and SNR, enabling observers to visually identify and mask regions with prominent damped \lya\ absorbers (DLAs), which are neutral hydrogen (HI) systems with column densities $N_\mathrm{HI} > 10^{20.3}\,\mathrm{cm}^{-2}$.
However, sub-DLAs, which have column densities $10^{19.0} < N_\mathrm{HI} < 10^{20.3}\,\mathrm{cm}^{-2}$, are easier to miss. Lyman limit systems (LLSs), which have column densities $10^{17.2} < N_\mathrm{HI} < 10^{19.0}\,\mathrm{cm}^{-2}$ cannot be reliably identified on an absorber-by-absorber basis. To maintain consistency across datasets, we include the full four-parameter HCD template in the likelihood: $\alpha_{\mathrm{LLS}}$, $\alpha_{\mathrm{subDLA}}$, $\alpha_{\mathrm{small-DLA}}$, and $\alpha_{\mathrm{large-DLA}}$.\footnote{We sometimes use the shorter names $\alpha_{lls}$, $\alpha_{sub}$, $\alpha_{sdla}$, and $\alpha_{ldla}$. These notations are interchangeable throughout this work.}
We use the template from Ref.~\cite{2018MNRAS.474.3032R}, which is based on simulated absorber populations in the Illustris simulation (which has a $106.5\,\mathrm{Mpc}$ box).

This differs from our modeling choice in Ref.~\cite{2024JCAP...07..029F}, where we adopted a coarser two-parameter scheme separating systems above and below the DLA column density threshold (constant ratios for $\alpha_\mathrm{LLS}/\alpha_\mathrm{subDLA} = 1$ and $\alpha_\mathrm{small-DLA}/\alpha_\mathrm{large-DLA} = 1$). Since KODIAQ-SQUAD especially probes significantly smaller scales than eBOSS DR14, the contribution from LLSs and sub-DLAs becomes more relevant than that from  DLAs ($N_\mathrm{HI} > 10^{20.3}\,\mathrm{cm}^{-2}$). It is therefore more appropriate to explicitly separate the LLS and sub-DLA terms, while retaining small- and large-DLA parameters to verify that masking was sufficiently effective.

The PRIYA simulations contain a realistic population of high-column-density systems. We follow the same masking procedure as in Ref.~\cite{2019JCAP...07..017C}, masking all DLAs with $N_\mathrm{HI} \gtrsim 10^{20}\,\mathrm{cm}^{-2}$. Consequently, we do not expect DLAs to contribute significantly to the model P1D, and set the priors on $\alpha_{\mathrm{small-DLA}}$ and $\alpha_{\mathrm{large-DLA}}$ to be positive, accounting for potential residual contamination in the data. For sub-DLAs, which are only partially masked in PRIYA, we also expect $\alpha_{\mathrm{subDLA}}$ to be positive.
For LLSs, we only allow $\alpha_{\mathrm{LLS}}$ to take positive values, which corresponds to adding extra LLSs from the Illustris-based template on top of the existing PRIYA prediction. We avoid negative values, since Ref.~\cite{2018MNRAS.474.3032R} only models the difference between the P1D with HCDs and the HCD-free P1D, and does not account for removing absorbers from an existing population.

Importantly, since KODIAQ-SQUAD reaches scales where LLS contamination becomes significant, the pixel-level suppression from unmasked systems, i.e., saturated absorption that washes out small-scale structure, is no longer negligible. While Ref.~\cite{2018MNRAS.474.3032R} suggests that the mean effect of HCD contamination can be absorbed into the mean flux calibration (effectively shifting the P1D vertically), we find the amplitude of this shift to be nontrivial. In KODIAQ-SQUAD, it can alter the inferred meanflux-redshift relation. We caution that this implies the inferred $\tau_0$ from the survey may include a LLS component.


\section{Results}
\label{sec:results}

This section presents the MCMC posterior computed from applying the PRIYA emulator to XQ100 P1D and KODIAQ-SQUAD P1D measurements.We limit the redshift range of XQ100 to $z = 3.4-4.2$, and KODIAQ-SQUAD to $z = 2.6-4.2$, as discussed in Section~\ref{sec:fpsdata}. 
For reference, we also include the PRIYA baseline MCMC chains from eBOSS DR14 \citep{2024JCAP...07..029F}, which use $z = 2.6-4.6$.
All results incorporate data-related nuisance parameters, including contamination from HCDs (DLAs, sub-DLAs, LLSs) via the template model of Ref.~\cite{2018MNRAS.474.3032R}.

The purpose of our analysis is to examine how the inferred posteriors respond across datasets with differing spectral resolution of \Lya forest sightlines.
In particular, we are interested in whether high-resolution data, while less statistically powerful, reveals qualitatively different features in the posterior or better sensitivity to thermal parameters.

We begin by presenting the cross-parameter correlations in \S~\ref{sec:correlations}, which provides some physical intuition before jumping into the parameter constraints.
Then, in \S~\ref{sec:xq100-eboss}, we compare the XQ100 P1D to the eBOSS P1D, focusing on cosmological parameters $(A_P, n_P)$ in \S~\ref{sec:cosmo} and astrophysical parameters, including the IGM thermal history, in \S~\ref{sec:astroparams}.  Next, in \S~\ref{sec:kodiaq-interpretation}, we analyze the KODIAQ-SQUAD P1D, with the cosmological and astrophysical results detailed in \S~\ref{sec:cosmo_ks} and \S\ref{sec:astroparams_ks}, respectively.
We then examine the potential LLS contamination due to KODIAQ-SQUAD's selection bias in \S~\ref{sec:separate_z}.
Finally, we discuss the mean flux model in \S~\ref{sec:meanflux}.
We do not include maximum a posteriori (MAP) predictions in the main results, but they can be found in Appendix~\ref{sec:map}.

\subsection{Parameter Correlation Matrix}
\label{sec:correlations}

\begin{figure}
    \centering
    \includegraphics[width=\columnwidth]{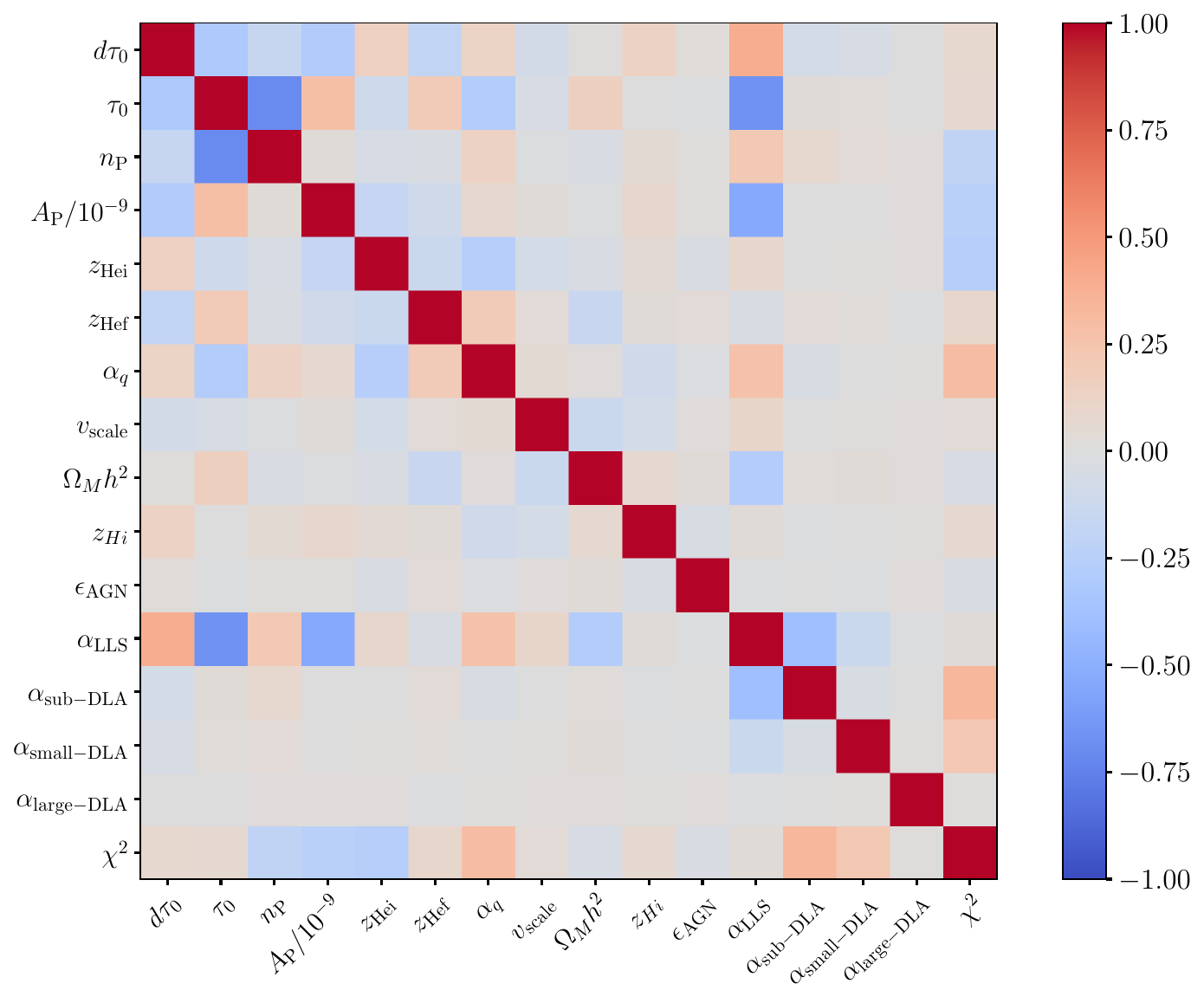}
    \caption{Correlation matrix computed from the MCMC chains for KODIAQ-SQUAD at $z = 2.6 - 4.2$.
    Positive correlations are shown in red, negative in blue, and zero correlation in white (or gray).
    We retain the upper triangle for visual clarity, even though the matrix is symmetric.
    $\chi^2$ refers to the $\chi^2$ values in the MCMC chains, i.e., the log-likelihood of the model prediction given the data.
    }
    \label{fig:corr_matrix}
\end{figure}

In this section, we examine how the cosmological and astrophysical parameters in PRIYA covary at the small scales probed by the P1D from high-resolution measurements, a regime that behaves quite differently from our previous eBOSS DR14 analysis (Ref.~\cite{2024JCAP...07..029F}).
Figure~\ref{fig:corr_matrix} shows the correlation matrix for all parameters varied in the MCMC chains for the KODIAQ-SQUAD P1D ($z = 2.6-4.2$). We use the KODIAQ-SQUAD matrix here, as the redshift range of XQ100 is a subset of KODIAQ-SQUAD, and the correlations are qualitatively similar.

\begin{figure}
    \centering
    \includegraphics[width=0.495\columnwidth]{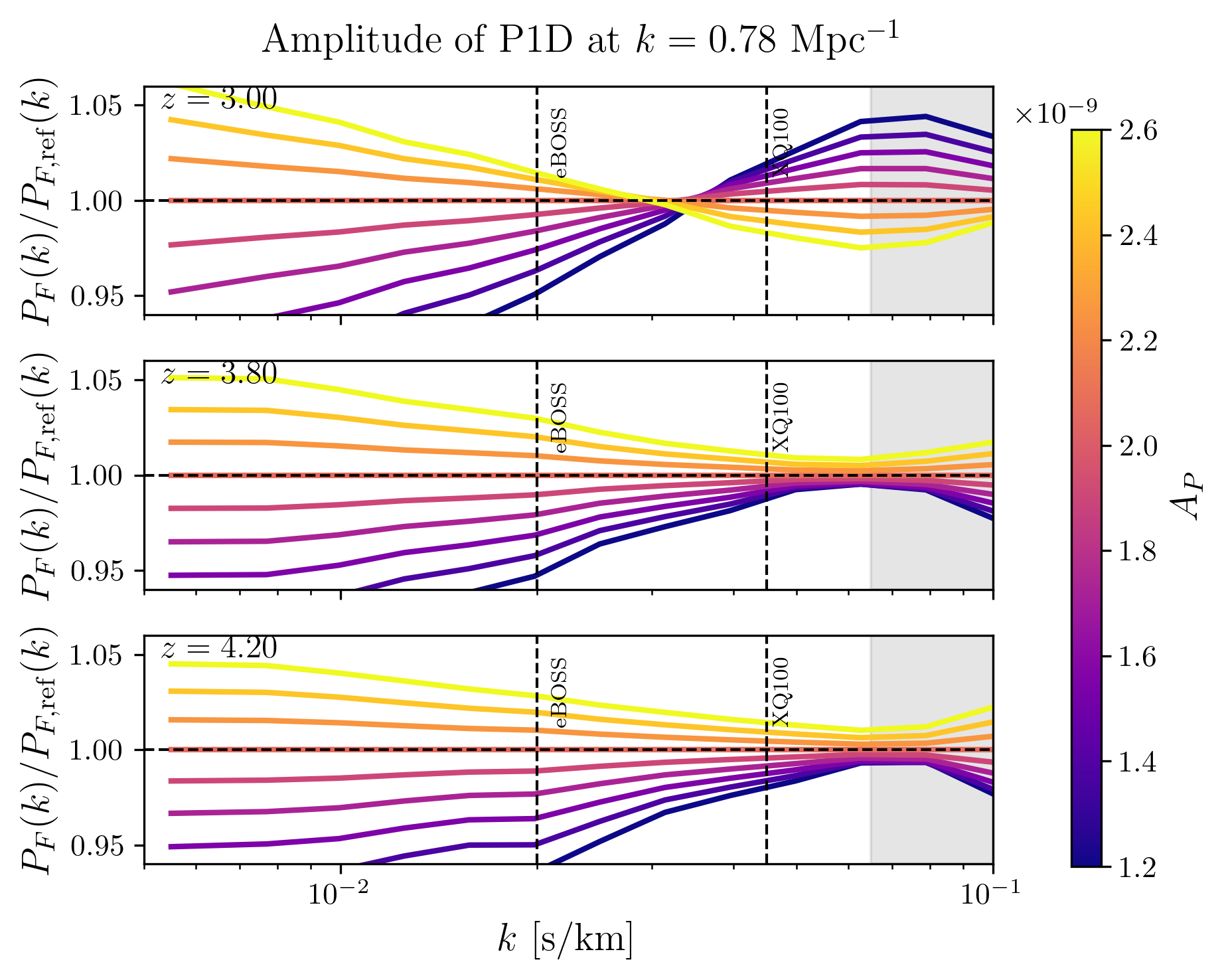}
    \includegraphics[width=0.495\columnwidth]{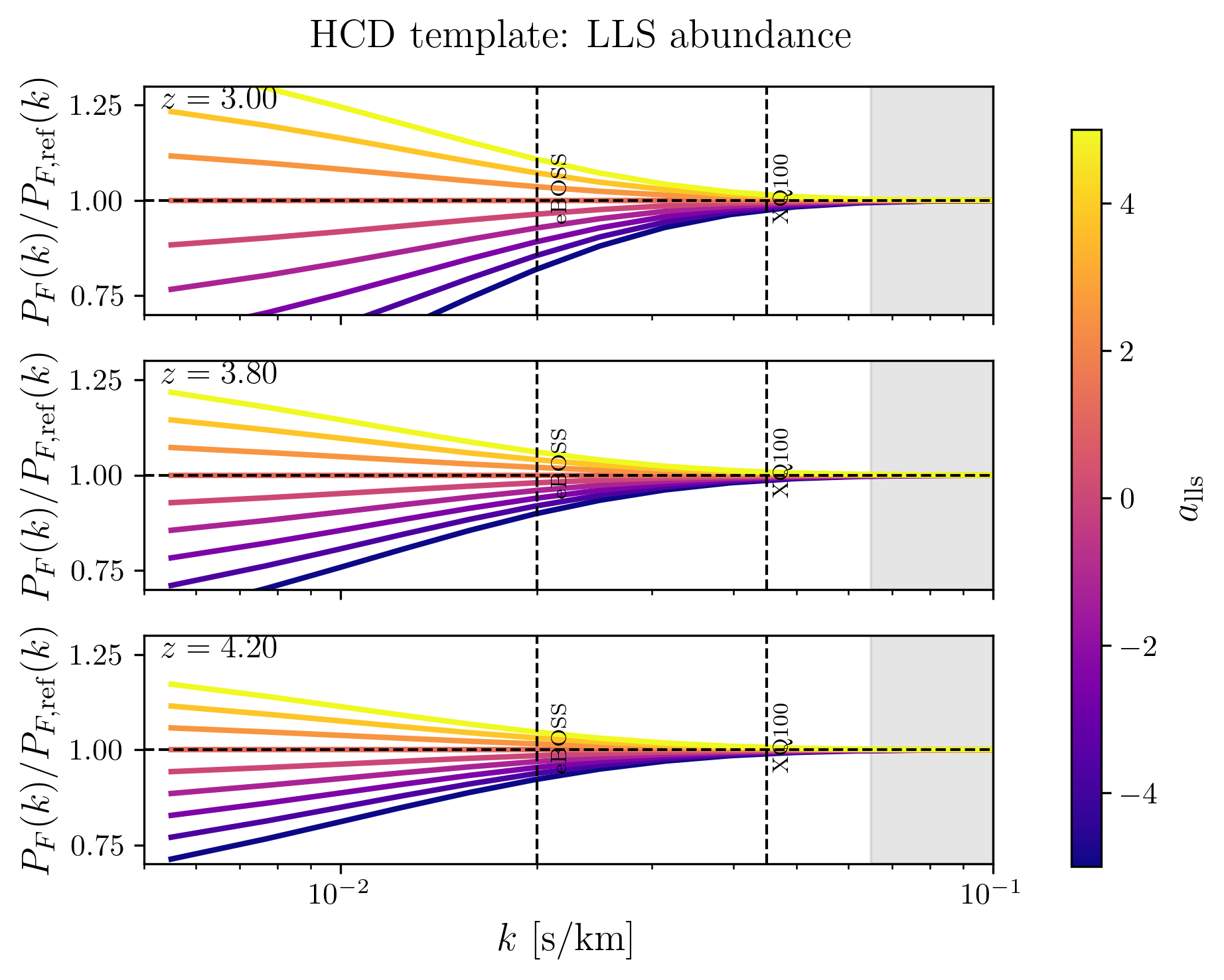}
    \caption{One-parameter variation of the PRIYA emulator, with all other parameters fixed to the eBOSS baseline chains from Ref.~\cite{2024JCAP...07..029F}.
    The left panel shows the variation of $A_P$ (the amplitude of the P1D), and the right panel shows the variation of $a_{\rm LLS}$ (the LLS contribution).
    Note that the impact of LLS variation is significantly larger than that of $A_P$.
    We also label the scales relevant for eBOSS and XQ100: $k = 0.02\,\mathrm{s/km}$ and $k = 0.045\,\mathrm{s/km}$, respectively, which correspond to the maximum sensitive $k$-values for each dataset.}
    \label{fig:1pvar_ap_aLLS}
\end{figure}

Starting with $A_P$: there is a strong negative correlation between $A_P$ and $a_{\rm LLS}$. 
As shown in Figure~\ref{fig:1pvar_ap_aLLS}, the one-parameter variations of $A_P$ (left panel) and $a_{\rm LLS}$ (right panel) produce similar responses in the P1D. 
The large-scale boost from LLSs comes from the damping wings of their absorption profiles. 
By contrast, increasing $A_P$ raises the P1D amplitude at $k < 0.02\,\mathrm{s/km}$, but the effect weakens on smaller scales because of peculiar velocity smoothing, causing a tilt rather than a uniform boost.

Because the HCD template does not include the saturated absorption suppression expected from LLSs, increasing $a_{\rm LLS}$ must be compensated by lowering $\tau_0$ to keep the small-scale shape of the P1D consistent. 
This explains the negative correlation between $\tau_0$ and $a_{\rm LLS}$. 
Physically, an increase in $A_P$ leads to a larger amount of overdense regions in the IGM, with strong correlated absorption.
LLS arise from overdense, biased and partially shielded, regions of the IGM and have minimal damping wings.
Their absorption signature thus strongly resembles the correlated overdensities that indicate increased $A_P$.

In terms of $\chi^2$,
interestingly, despite these strong parameter degeneracies, particularly for $\tau_0$, and $d\tau_0$, the LLS amplitude $a_{\rm LLS}$ shows only a weak correlation with $\chi^2$. 
This suggests that these degeneracies are largely unconstrained by the P1D alone.

\begin{figure}
    \centering
    \includegraphics[width=0.495\columnwidth]{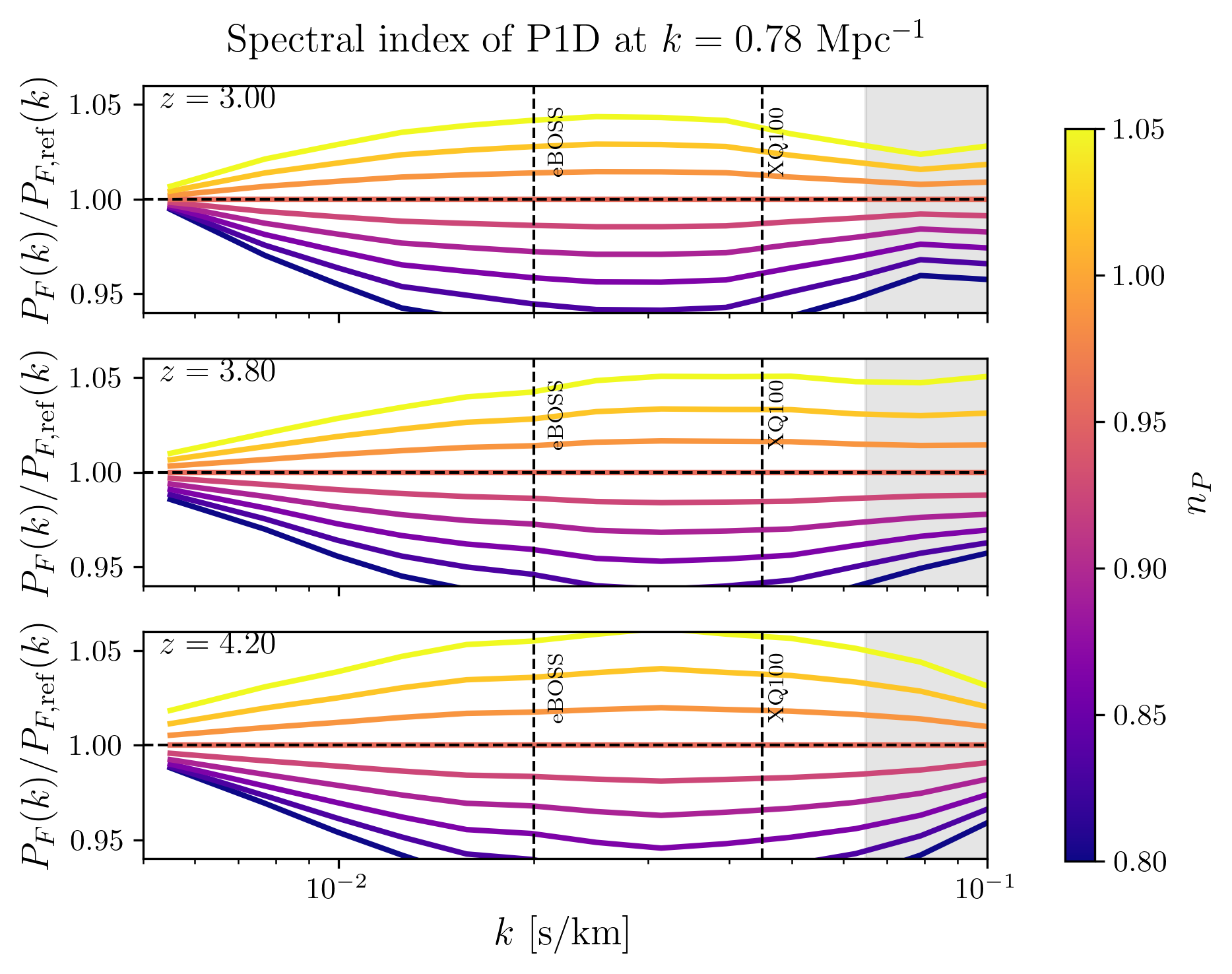}
    \includegraphics[width=0.495\columnwidth]{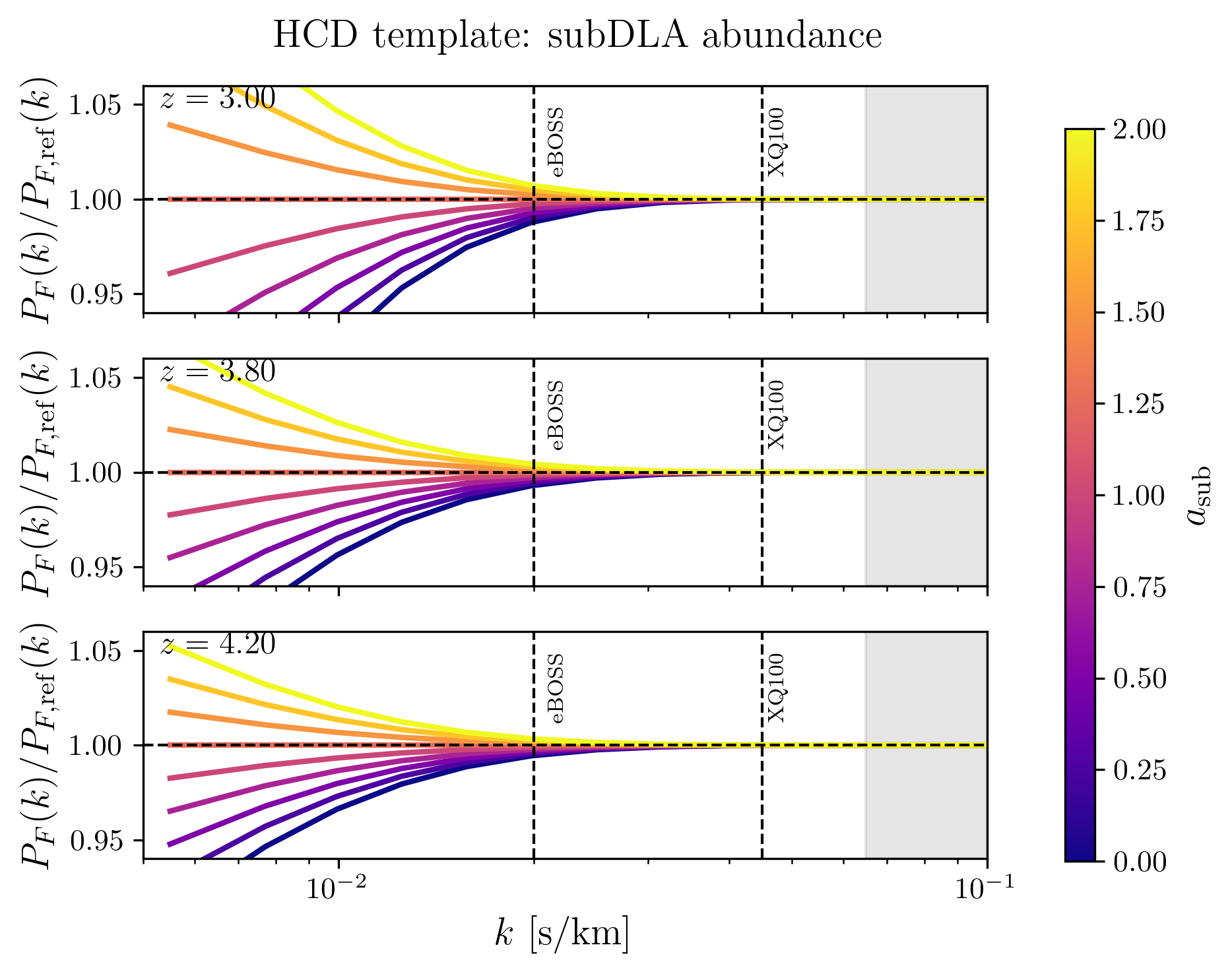}
    \caption{One-parameter variation of the PRIYA emulator, with all other parameters fixed to the eBOSS baseline chains from Ref.~\cite{2024JCAP...07..029F}.
    The left panel shows the variation of $n_P$ (slope of the P1D), and the right panel shows the variation of $a_{\rm sub}$ (sub-DLA contribution).}
    \label{fig:1pvar_ns_asub}
\end{figure}

Other than $A_P$, the parameters most decisive for the overall log-evidence are $a_{\rm sub}$, $\alpha_q$ (along with HeII reionization parameters in general), and $n_P$.
Figure~\ref{fig:1pvar_ns_asub} offers a clue as to why. The parameter $n_P$ modifies the P1D slope at $k \sim 0.001\,\mathrm{s/km}$ and gradually bends it into an arc shape over the range $k \sim 0.005-0.06\,\mathrm{s/km}$. This is due to peculiar velocity effects smoothing out the slope change at $k > 0.02\,\mathrm{s/km}$. In the eBOSS range, $n_P$ affects the slope; in the overlapping eBOSS-XQ100 range, it mostly changes the normalization; beyond XQ100 range ($k \sim 0.045\,\mathrm{s/km}$), it again alters the slope, but in the opposite direction.
This is a fairly unique feature that other parameters cannot easily reproduce, which explains its correlation with $\chi^2$.

The right panel of Figure~\ref{fig:1pvar_ns_asub} shows the variation of $a_{\rm sub}$, the sub-DLA contribution. Unlike LLSs, sub-DLAs primarily affect the largest scales in the P1D relevant to XQ100 and KODIAQ-SQUAD, with a pivot scale around $k \sim 0.02\,\mathrm{s/km}$.
This also provides a unique variation that other parameters cannot reproduce.


\begin{figure}
    \centering
    \includegraphics[width=0.32\columnwidth]{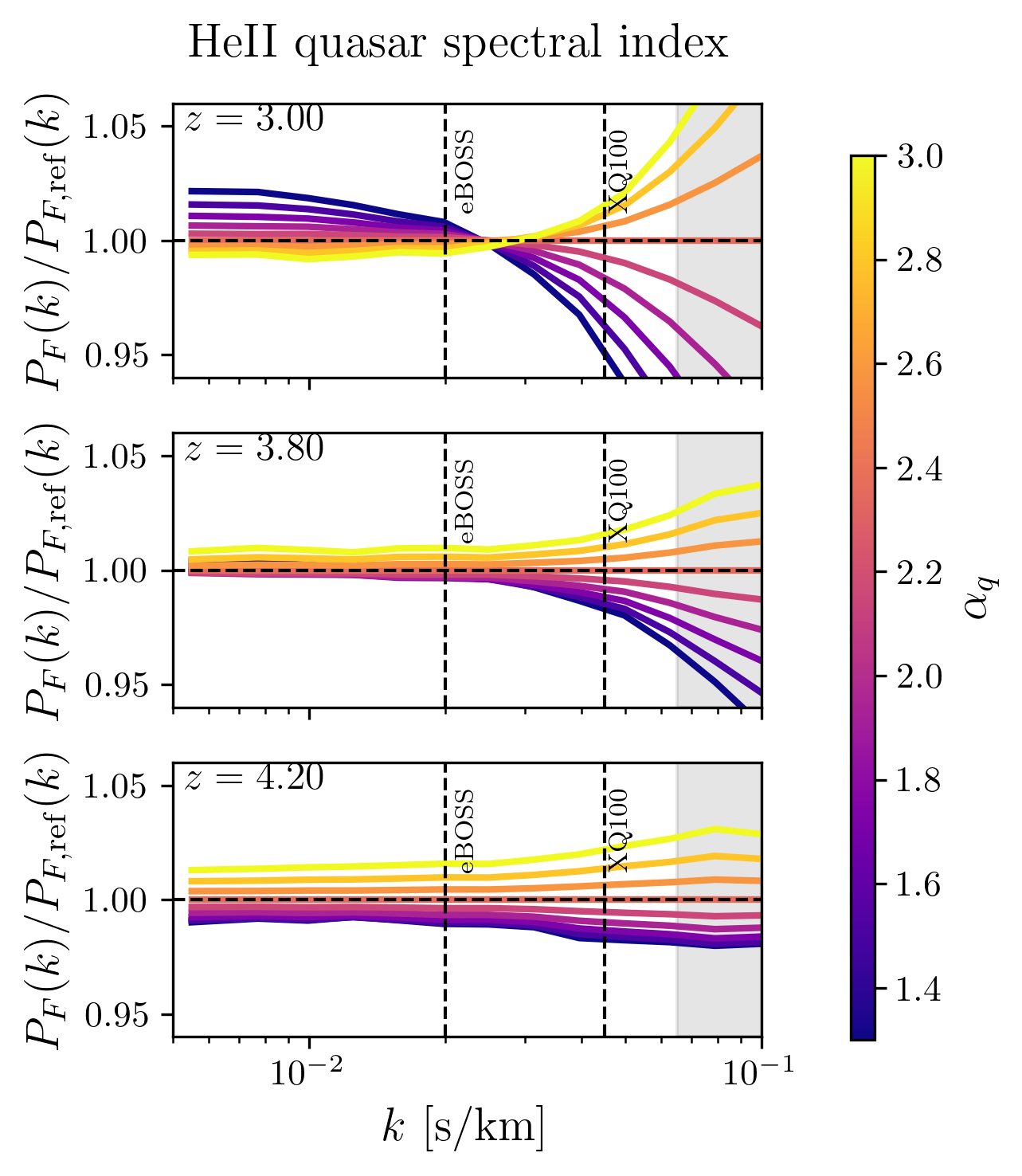}
    \includegraphics[width=0.32\columnwidth]{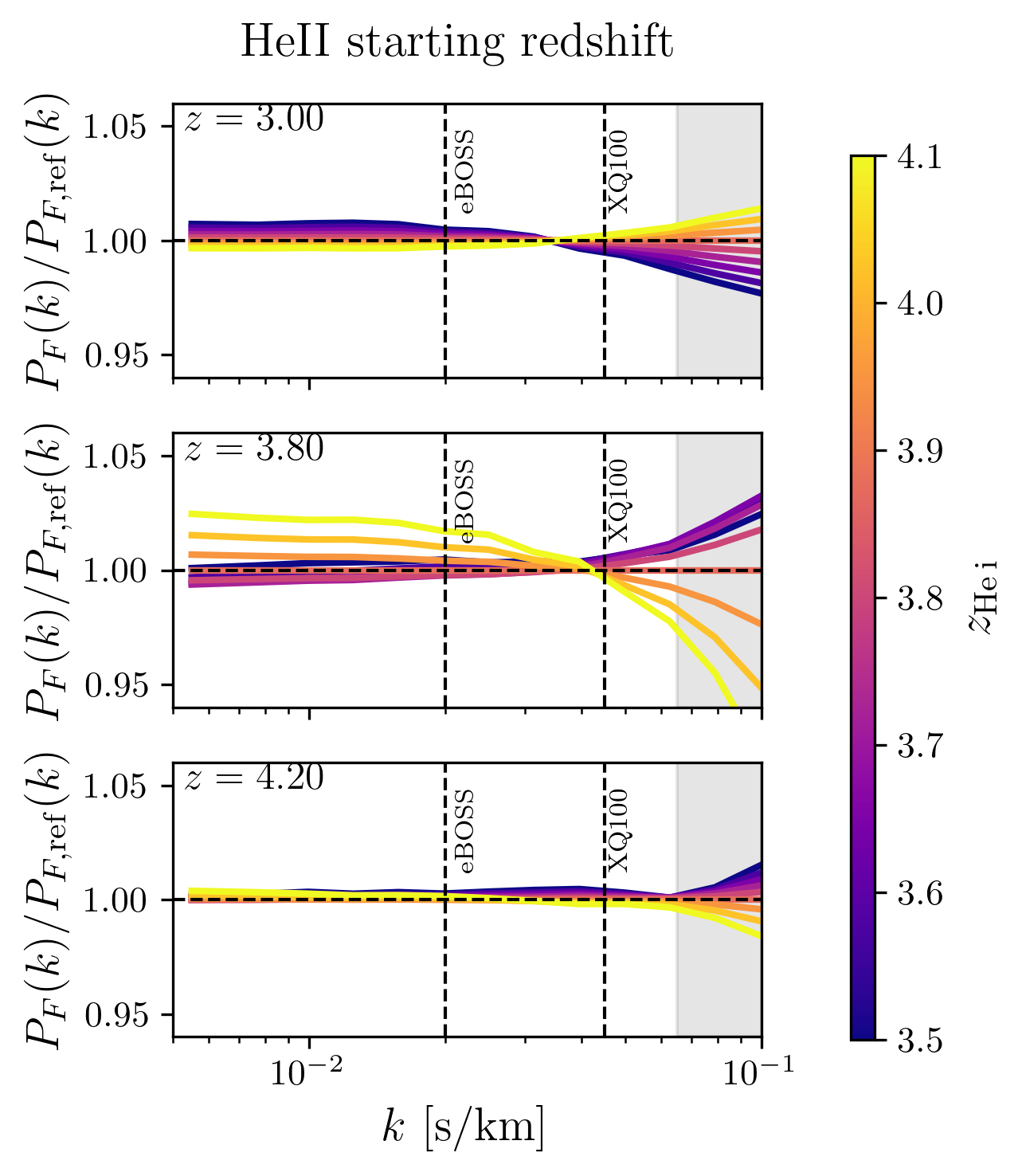}
    \includegraphics[width=0.32\columnwidth]{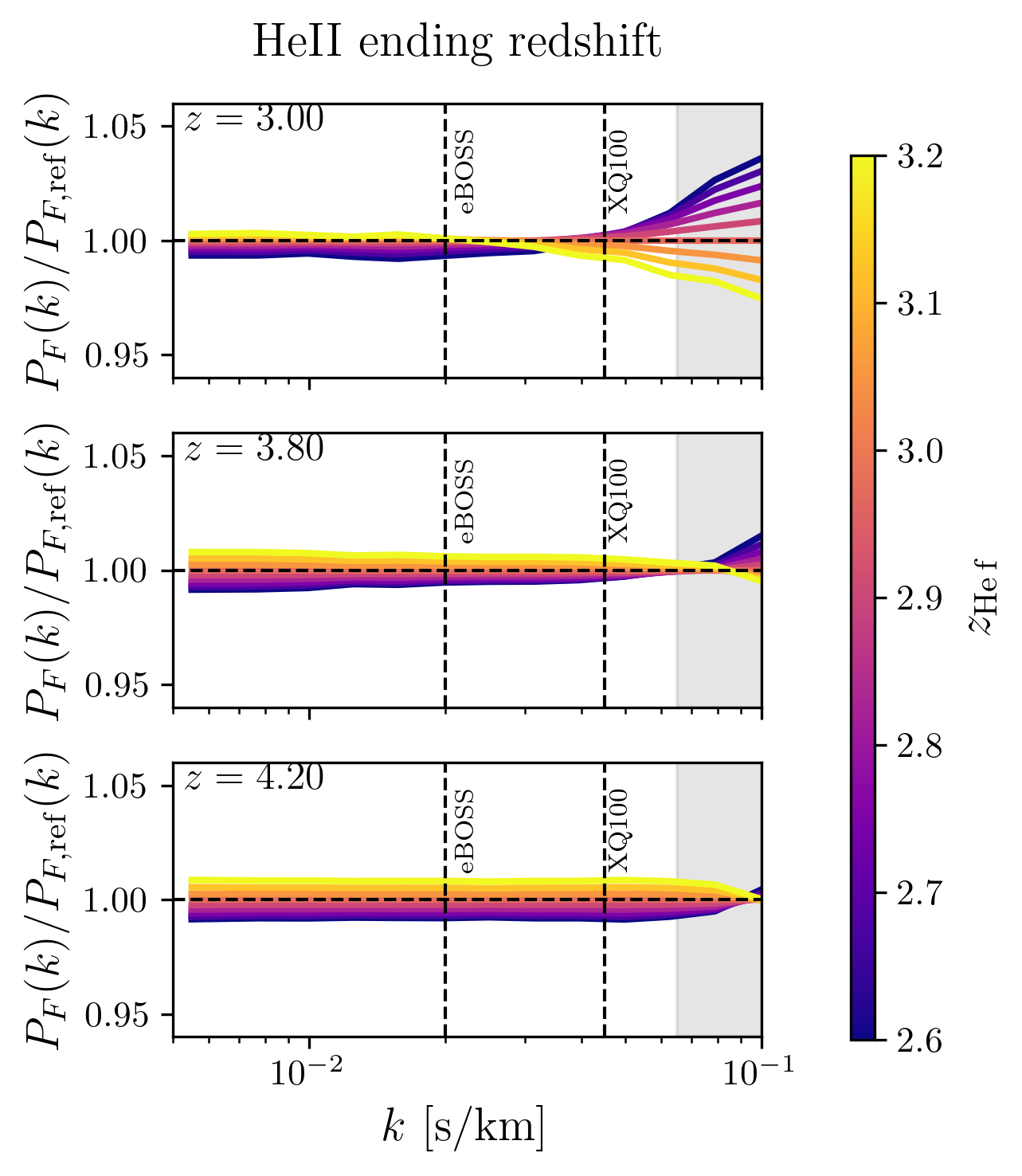}
    \caption{One-parameter variation of the PRIYA emulator, with all other parameters fixed to the eBOSS baseline chains in Ref.~\cite{2024JCAP...07..029F}.
    The left panel shows the variation of $\alpha_q$ (HeII reionization heating rate), the middle panel shows the variation of $z^{\mathrm{HeII}}_i$ (HeII reionization starting redshift), and the right panel shows the variation of $z^{\mathrm{HeII}}_f$ (HeII reionization ending redshift).}
    \label{fig:1pvar_heii}
\end{figure}

Figure~\ref{fig:1pvar_heii} shows the one-parameter variations of the HeII reionization parameters: $\alpha_q$, $z^{\mathrm{HeII}}_i$, and $z^{\mathrm{HeII}}_f$.
These parameters exhibit very interesting and highly nonlinear behavior in terms of redshift evolution.

The fiducial values are approximately $\alpha_q \sim 1.3$, $z^{\mathrm{HeII}}_i \sim 4.12$, and $z^{\mathrm{HeII}}_f \sim 2.63$.
A value of $\alpha_q \sim 1.3$ corresponds to a fairly large heating rate, which suppresses small-scale neutral hydrogen clustering beyond $k > 0.03\,\mathrm{s/km}$ after the onset of HeII reionization at $z^{\mathrm{HeII}}_i$.
This redshift evolution is a unique feature of the HeII reionization parameters, so it is not surprising that they are strongly correlated with $\chi^2$.

\subsection{Consistency between XQ100 and eBOSS}
\label{sec:xq100-eboss}

We first compare the XQ100 results to the PRIYA baseline analysis of eBOSS DR14 from Ref.~\cite{2024JCAP...07..029F}, which uses the P1D over $z = 2.6 - 4.6$ and incorporates external priors on the IGM thermal history, ultimately derived from KODIAQ DR2 Ref.~\cite{2021MNRAS.506.4389G}. Note that the IGM temperature measurements of Ref.~\cite{2021MNRAS.506.4389G} are derived assuming a fixed Planck cosmology.
The purpose of this comparison is to determine the thermal history constraints available from the XQ100 P1D alone, without using external datasets or cosmological priors.

\subsubsection{Cosmological Parameters: $A_P$, $n_P$}
\label{sec:cosmo}

\begin{figure}
    \centering
    \includegraphics[width=0.75\columnwidth]{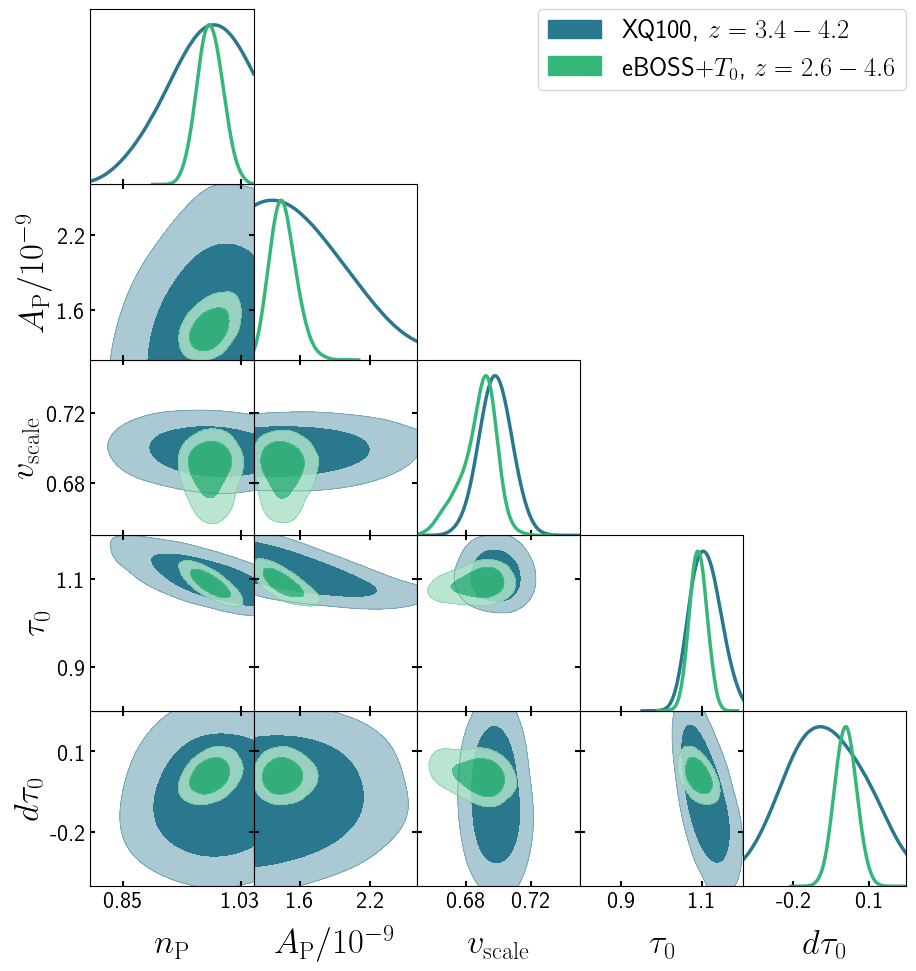}
    \caption{Posterior distributions for $n_P$, $A_P$, and $v_\mathrm{scale}$ from XQ100 (blue) and eBOSS (green) P1Ds.
    The XQ100 posteriors are broadly consistent with eBOSS, despite broader uncertainties due to smaller sample size and narrower redshift range.
    $v_\mathrm{scale}$ (the Hubble parameter) is already priored out to $v_\mathrm{scale} \sim \mathcal{N}(\mu=0.7, \sigma=0.15)$ in the MCMC chains, as it is not sensitive to the P1D as suggested in Ref.~\cite{2024JCAP...07..029F}.
    }
    \label{fig:corner_cosmo}
\end{figure}

Figure~\ref{fig:corner_cosmo} shows the marginalized posteriors for three cosmological parameters: the primordial amplitude $A_P$, the spectral index $n_P$ (pivoted at $k = 0.78\,\mathrm{Mpc}^{-1}$), and the velocity scale $v_\mathrm{scale}$.

The XQ100 posteriors for $A_P$ and $n_P$ remain consistent with eBOSS baseline within $1\sigma$. This is despite a narrower redshift range ($z = 3.4 - 4.2$), higher spectral resolution ($R \sim 5{\,}000$ versus eBOSS's $R \sim 2{\,}000$), and much larger statistical uncertainty. Although the constraints are broader due to increased cosmic variance, the $n_P$ posterior mode aligns well with eBOSS, supporting recent results from PRIYA and Lyssa (see Refs.~\cite{2024JCAP...07..029F,Lyssa:2025JCAP...05..099W}) that the tilt of the linear power spectrum inferred from the \lya\ forest is consistent with Planck. Interestingly, the $A_P$ posterior from XQ100 also peaks near the PRIYA eBOSS value, albeit with a slight shift toward lower linear power amplitude ($A_P \sim 1.2 \times 10^{-9}$) compared to eBOSS ($A_P \sim 1.46 \times 10^{-9}$).

Note that XQ100 probes somewhat smaller physical scales ($k \sim 3\,h\,\mathrm{Mpc}^{-1}$) than eBOSS ($k \sim 1\,h\,\mathrm{Mpc}^{-1}$). The agreement in $A_P$ across these datasets suggests that the small-scale cosmology from a high-resolution quasar survey like XQ100 and the large-volume eBOSS survey are at least consistent with each other, despite probing different scales.
As we will see in \S~\ref{sec:reduced_likelihood}, the $1\sigma$ constraints from XQ100 on the amplitude and slope of the linear theory power spectrum parameters, $\Delta_\mathrm{L}^2$ and $n_\mathrm{eff}$, are in fact consistent with both eBOSS and Planck, with the posterior mode slightly closer to eBOSS. The mean optical depth parameters $\tau_0$ and $d\tau_0$ remain consistent with eBOSS, as we shall discuss further in \S \ref{sec:meanflux}.

\subsubsection{Astrophysical Parameters: Thermal History and mean IGM temperature}
\label{sec:astroparams}

\begin{figure}
    \centering
    \includegraphics[width=0.49\columnwidth]{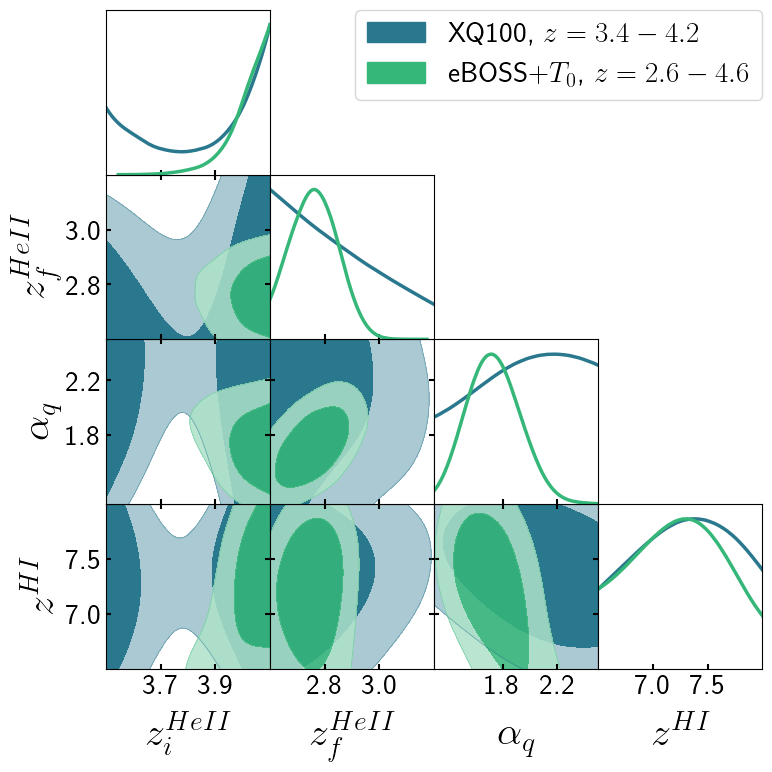}
    \includegraphics[width=0.49\columnwidth]{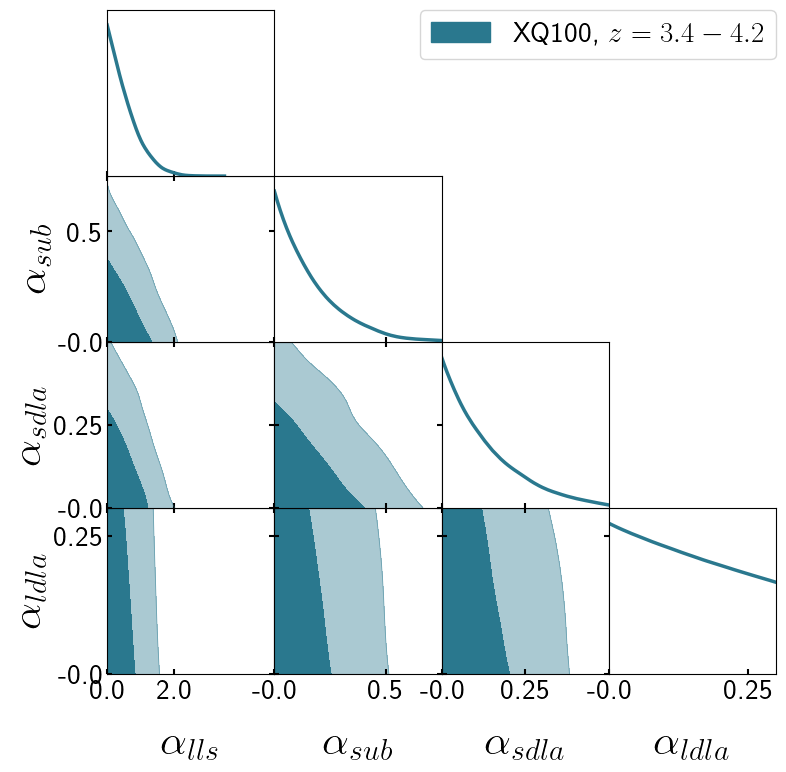}    
    \caption{Posterior distributions for thermal history parameters (left) and DLA/LLS contamination parameters (right) for XQ100 (blue) and a combination of eBOSS and IGM mean temperature measurements (eBOSS+$T_0$, green).
    Note that eBOSS+$T_0$ constraints come predominantly from the IGM mean temperature prior (from Ref.~\citep{2021MNRAS.506.4389G}; for details see Ref.~\cite{2024JCAP...07..029F}), while XQ100 constraints come from the XQ100 P1D alone.
    }
    \label{fig:corner_astro}
\end{figure}
\begin{figure}
    \centering
    \includegraphics[width=0.75\columnwidth]{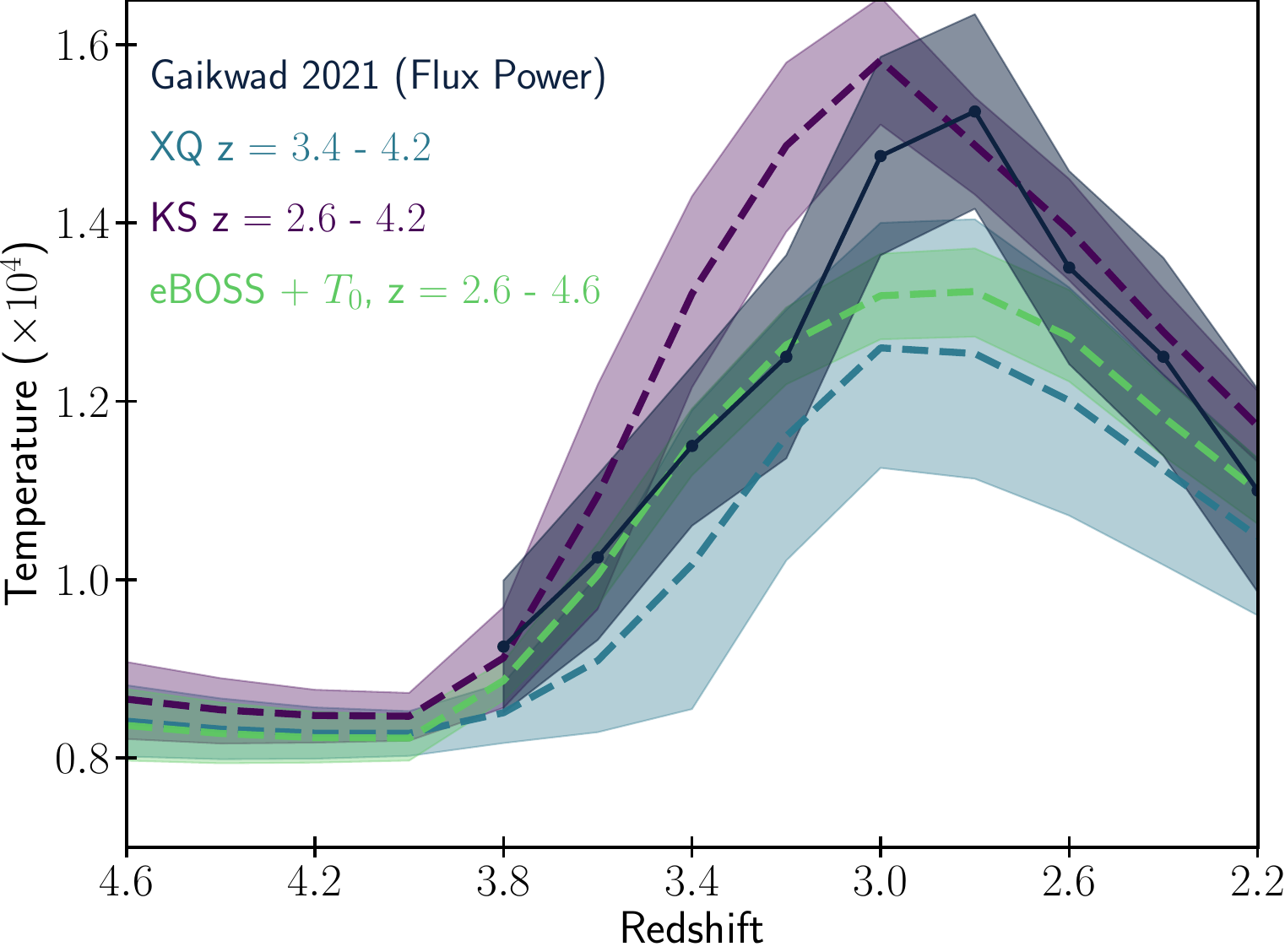}
    \caption{Mean IGM temperature posterior predictive distributions (shaded: 68\% confidence interval) using the P1D from XQ100 (blue) and KODIAQ-SQUAD (purple), compared to the eBOSS+$T_0$ DR14 baseline (green). Note that eBOSS+$T_0$ includes data from an independent analysis of KODIAQ DR2.
    }
    \label{fig:meanT_pred}
\end{figure}

Figure~\ref{fig:corner_astro} presents the marginalized posteriors for the astrophysical parameters. The left panel focuses on thermal and ionization history parameters ($z^{\mathrm{HeII}}_i$, $z^{\mathrm{HeII}}_f$, $\alpha_q$, $z^{\mathrm{HI}}$), while the right panel shows the four nuisance parameters associated with residual DLA and LLS contamination, based on the HCD templates from Ref.~\cite{2018MNRAS.474.3032R}.

Starting with the thermal parameters, both XQ100 and eBOSS+$T_0$ prefer an extended He~{\sc ii} reionization history, with the starting redshift $z^{\mathrm{HeII}}_i \gtrsim 3.9$ and ending redshift $z^{\mathrm{HeII}}_f \lesssim 3.0$. The XQ100 posterior on $z^{\mathrm{HeII}}_i$ is bimodal, with a secondary peak around $z^{\mathrm{HeII}}_i \sim 3.6$,
though we will see later in Figure~\ref{fig:meanT_pred} that this bimodality does not significantly affect the predicted IGM mean temperature.
The spectral slope $\alpha_q$, which controls the quasar heating rate, is consistent across datasets within $1\sigma$, with XQ100 favoring slightly weaker heating (larger $\alpha_q$ implies less heating). As expected, $z^{\mathrm{HeII}}_f$ is only weakly constrained by XQ100 alone, since the ending redshift ($z \sim 2.8$) lies outside the redshift range probed.

Note that in Ref.~\cite{2024JCAP...07..029F}, we showed that the eBOSS P1D alone \textit{cannot} constrain thermal parameters without an external measurement of the IGM mean temperature. 
Figure~\ref{fig:meanT_pred} compares the predicted IGM mean temperature from the XQ100 MCMC chains to the eBOSS baseline. 
Despite \textit{not} using any external temperature data, the XQ100 P1D prediction is consistent with the eBOSS baseline (which includes external data) within $1\sigma$, suggesting that the small-scale P1D alone encodes substantial information about the thermal history. 

The HI reionization redshift $z^{\mathrm{HI}}$ remains poorly constrained from P1D, as anticipated, since its imprint on the P1D is limited to residual memory effects in the IGM's temperature-density relation at high redshift. All chains remain within the prior range $z^{\mathrm{HI}} \sim 6.5-8$.

The right panel of Figure~\ref{fig:corner_astro} shows the four HCD nuisance parameters: $\alpha_{\mathrm{LLS}}$, $\alpha_{\mathrm{subDLA}}$, $\alpha_{\mathrm{small-DLA}}$, and $\alpha_{\mathrm{large-DLA}}$. As described in \S~\ref{sec:dlalimit}, we marginalize over these to account for residual contamination from unmasked HCDs.
All HCD parameters are consistent with zero within their prior bounds, indicating no significant excess contamination.
Notably, $\alpha_{\mathrm{large-DLA}}$ is unconstrained, implying the XQ100 P1D is insensitive to large DLAs. $\alpha_{\mathrm{small-DLA}}$ is also near zero, as small DLAs have already been masked in both data and simulation. Recall that $\alpha_{\mathrm{sub-DLA},{\rm LLS}} = 0$ implies exact agreement between the PRIYA model and the real Universe observed by XQ100. This suggests that the LLS and sub-DLA populations in PRIYA are broadly consistent with those inferred from XQ100.


\subsection{KODIAQ-SQUAD: Interpretation from PRIYA}
\label{sec:kodiaq-interpretation}

We present the first cosmological interpretation of the KODIAQ-SQUAD P1D using the PRIYA suite.
KODIAQ-SQUAD probes smaller scales even than XQ100. These scales are uniquely sensitive to the scale-dependent effects of Lyman Limit Systems, which have not been fully explored in earlier cosmological analyses.

As mentioned in \S~\ref{sec:fpsdata}, the KODIAQ-SQUAD is biased due to the selection of quasar sightlines that preferentially target known DLAs and O~{\sc iv} absorbers, though Ref.~\cite{2022MNRAS.509.2842K} has masked all identified DLAs carefully, sub-DLAs and LLSs remain in the data.
In this section, we will assess how this selection bias impacts the inferred cosmology and whether the HCD template model can adequately account for this bias.

As before, we start with the cosmological parameters in \S~\ref{sec:cosmo_ks}, then move to the astrophysical parameters in \S~\ref{sec:astroparams_ks}. We further examine the LLS bias and HCD template mismatch in \S~\ref{sec:separate_z}.

\subsubsection{Cosmological parameters: $A_P$, $n_P$}
\label{sec:cosmo_ks}

\begin{figure}
    \centering
    \includegraphics[width=0.75\columnwidth]{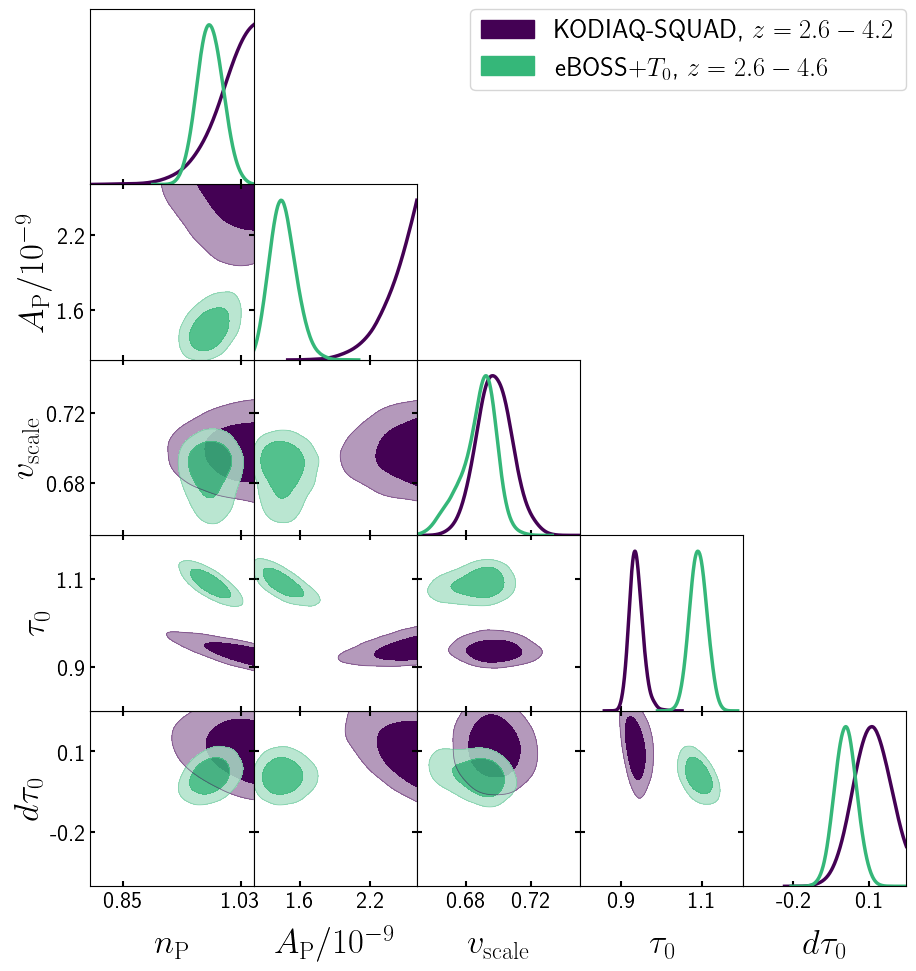}
    \caption{Constraints for $n_P$, $A_P$, and $v_\mathrm{scale}$ from  KODIAQ-SQUAD (purple) and eBOSS (green) P1Ds.
    }
    \label{fig:corner_cosmo_ks}
\end{figure}

Figure~\ref{fig:corner_cosmo_ks} shows the marginalized posteriors from KODIAQ-SQUAD for the primordial amplitude $A_P$, spectral index $n_P$, and velocity scale $v_\mathrm{scale}$. We compare these results to the PRIYA baseline analysis of eBOSS DR14 from Ref.~\cite{2024JCAP...07..029F}.
Note that both eBOSS and KODIAQ-SQUAD span a similarly wide redshift range and include measurements down to $z = 2.6$.\footnote{As discussed in \S~\ref{sec:fpsdata}, we omit the $z = 2.2 - 2.4$ bins due to known inconsistencies across surveys in this range, based on the analyses in Refs.~\cite{2024JCAP...07..029F,Lyssa:2025JCAP...05..099W}. We defer further investigation of $z = 2.2 - 2.4$  to future work following the release of DESI DR1.}

Somewhat unexpectedly, the KODIAQ-SQUAD posteriors for both $A_P$ and $n_P$ hit the upper boundary of the prior range, in clear tension with the posteriors from both eBOSS and XQ100 (as well as Planck \cite{2020A&A...641A...6P}).
One might initially expect that the elevated $A_P$ inferred from KODIAQ-SQUAD simply boosts the overall P1D amplitude, as would be the case for the linear matter power spectrum. However, this intuition breaks down at small scales. Figure~\ref{fig:1pvar_ap_aLLS} shows the one-parameter variation of the P1D with respect to $A_P$ and $\alpha_{\mathrm{LLS}}$, with all other parameters fixed to the eBOSS baseline chain values from Ref.~\cite{2024JCAP...07..029F}.
In the left panel, we examine how varying $A_P$ impacts the P1D. Rather than uniformly increasing the P1D, increasing $A_P$ introduces a tilt: it boosts power at large scales ($k \lesssim 0.02\,\mathrm{s/km}$) but suppresses it at smaller scales. This counterintuitive behavior arises because the P1D becomes increasingly dominated by peculiar velocity effects at high-$k$, causing the response to $A_P$ to change sign (see Appendix~\ref{sec:peculiarvel}).

Since the KODIAQ-SQUAD P1D has unusually small uncertainty at $k > 0.045\,\mathrm{s/km}$, it is more affected by this scale-dependent response. In other words, the high $A_P$ inferred from KODIAQ-SQUAD is actually compensating for suppressed small-scale power and a steepened P1D slope, rather than signaling a higher amplitude in the flux power spectrum.

The leading suspicion for this behavior is selection effects. The KODIAQ-SQUAD spectra are selected from the literature, with a bias toward DLAs or metal absorbers. Since these are correlated with high density regions, they also select for an over-abundance of LLS, which we will show in \S~\ref{sec:separate_z} produces a similar scale-dependent suppression to that seen from varying $A_P$ (see again Figure~\ref{fig:1pvar_ap_aLLS}). In addition, the correlation with higher density regions of the IGM may imply a biased region of the cosmological sky and lead to residual systematics that are difficult to fully marginalize out.



\subsubsection{Astrophysical Parameters: Thermal History}
\label{sec:astroparams_ks}

\begin{figure}
    \centering
    \includegraphics[width=0.49\columnwidth]{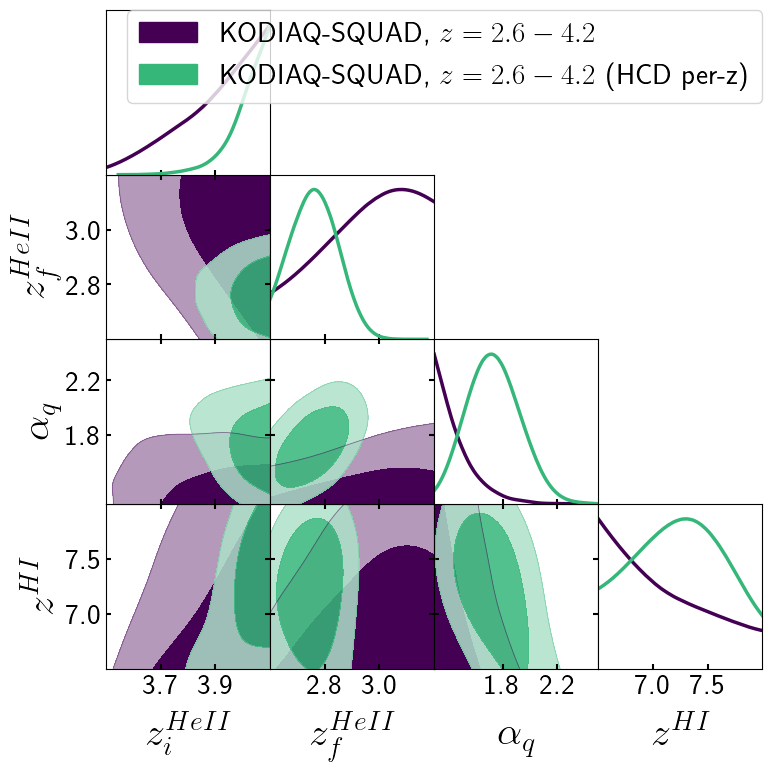}
    \includegraphics[width=0.49\columnwidth]{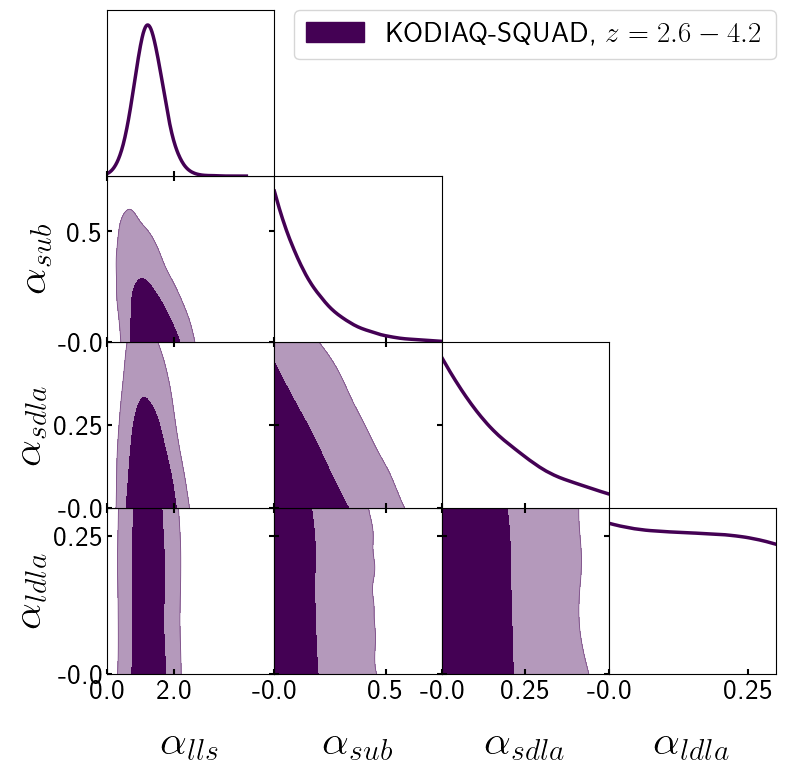}    
    \caption{Posterior distributions for thermal history parameters (left) and DLA/LLS contamination parameters (right) for KODIAQ-SQUAD (purple), XQ100 (blue), and eBOSS with the IGM mean temperature prior (eBOSS+$T_0$, green).
    eBOSS constraints are tight for thermal history because it uses the IGM mean temperature prior (from Ref.~\citep{2021MNRAS.506.4389G}; details see Ref.~\cite{2024JCAP...07..029F}), while KODIAQ-SQUAD is P1D only.
    }
    \label{fig:corner_astro_ks}
\end{figure}

Figure~\ref{fig:corner_astro_ks} shows the posteriors for the astrophysical parameters in our emulator, focusing on the thermal and ionization history ($z^{\mathrm{HeII}}_i$, $z^{\mathrm{HeII}}_f$, $\alpha_q$, $z^{\mathrm{HI}}$; left) and the four nuisance parameters associated with residual DLA and LLS contamination (right), based on the HCD template from Ref.~\cite{2018MNRAS.474.3032R}, which we will discuss in \S~\ref{sec:separate_z}.

Starting with the thermal parameters, we find that KODIAQ-SQUAD prefers an extended HeII reionization history, with the start of HeII reionization $z^{\mathrm{HeII}}_i \gtrsim 3.9$.
The end of HeII reionization prefers a slightly higher $z^{\mathrm{HeII}}_f \gtrsim 3.0$ value, shifting the best-fit peak temperature to a higher redshift than XQ100 and eBOSS.
The posterior on $\alpha_q$ shows a preference for stronger heating rates, though the posterior hits the lower bound of the prior range.


\subsubsection{LLS Abundance is not Consistent with HCD Template}
\label{sec:separate_z}

\begin{figure}
    \centering
    \includegraphics[width=0.75\columnwidth]{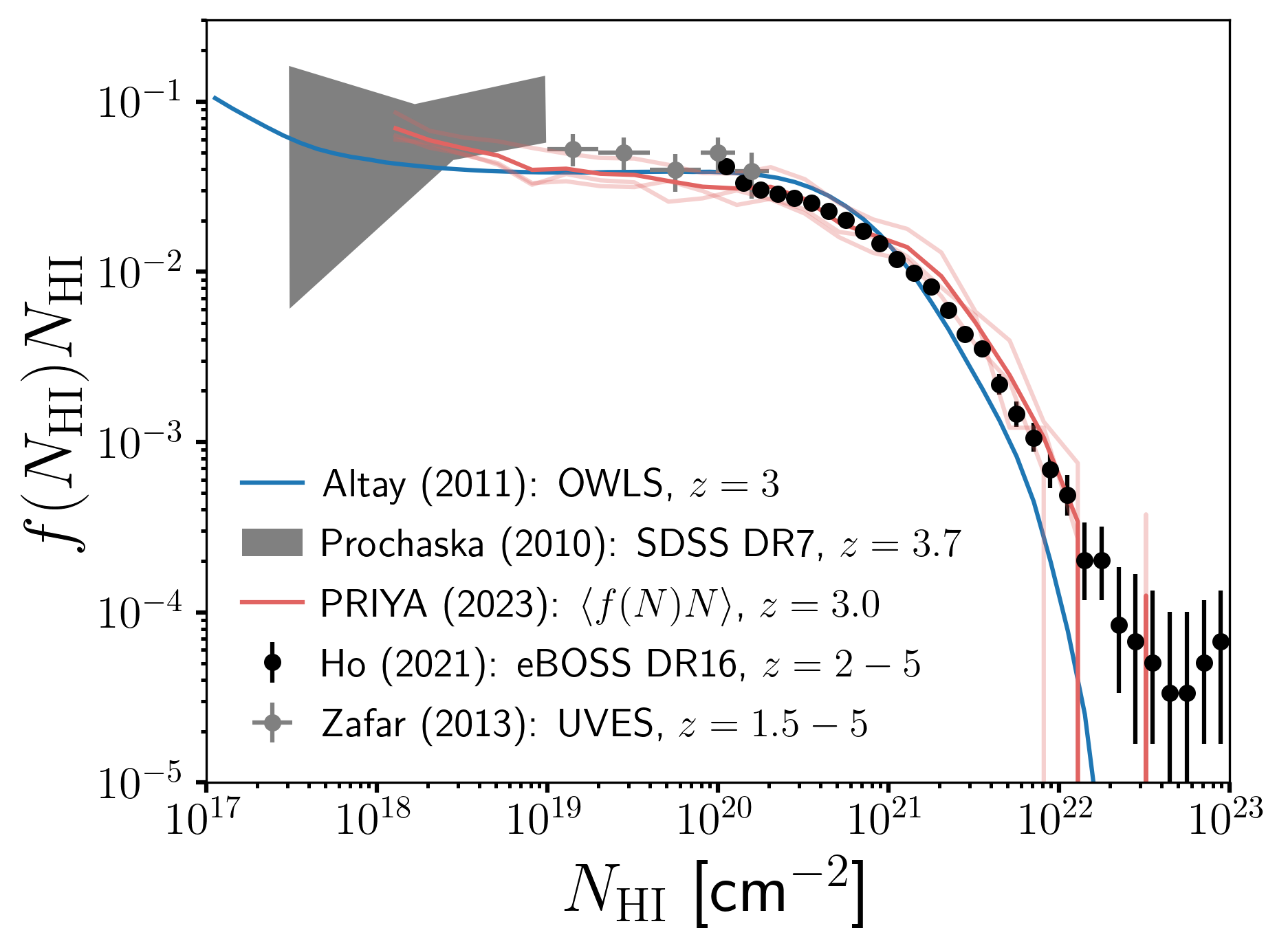}
    \caption{Column density distribution function (CDDF) moments (red solid line), $f(N_{\rm HI}) N_{\rm HI}$, as a function of neutral hydrogen column density. We show the Ho 2021 measurement from eBOSS DR16 quasars (Ref.~\cite{2021MNRAS.507..704H}, $z=2-5$), Altay (2011) simulations using OWLS at $z=3$ (Ref.~\cite{Altay:2011ApJ...737L..37A}), and observational constraints from Zafar 2013 (Ref.~\cite{Zafar:2013A&A...556A.141Z}, UVES) and Prochaska 2010 (Ref.~\cite{Prochaska:2010ApJ...718..392P}, SDSS DR7). The light red lines represent outputs from three independent simulations of the PRIYA suite at $z=3.0$, with the thick line denoting the mean across realizations. All data are scaled by $N_{\rm HI}$ to highlight the first moment relevant for $\Omega_{\rm DLA}$ calculations.
    }
    \label{fig:priya_cddf}
\end{figure}

The HCD contamination is shown in the right panel of Figure~\ref{fig:corner_astro_ks}.
KODIAQ-SQUAD favors a substantially higher abundance of LLSs than the PRIYA simulations, with $\alpha_{\mathrm{LLS}} \sim 2$, at high confidence.
A higher $\alpha_{\mathrm{LLS}}$ tilts the P1D downward at small scales, in a way that mimics the effect of increasing $A_P$ (see Figure~\ref{fig:1pvar_ap_aLLS}).
$\alpha_{\mathrm{LLS}} \sim 2$ implies that KODIAQ-SQUAD is detecting triple the number of LLS present in PRIYA, i.e., that $2\times$ the LLSs from the Illustris-based HCD template \citep{2018MNRAS.474.3032R} should be added to the LLS catalogue already in PRIYA.


There are two possible explanations. First, the PRIYA simulations may not accurately model the LLS population. Second, selection effects or systematics in the KODIAQ-SQUAD data may cause it to have a higher LLS abundance than a randomly selected part of the Universe, potentially with a redshift evolution not encoded in the LLS template we use.

We begin with the first possibility. Ref.~\cite{2023simsuite} notes that the PRIYA includes galaxy physics models and a high enough resolution for a realistic population of DLAs with $N_\mathrm{HI} > 10^{20}\,\mathrm{cm}^{-2}$.
As the simulation contains a realistic model of the IGM, of DLAs and of galaxies, it seems a priori likely that it would also contain a LLS population in reasonable agreement with observations. However, LLS present unique modeling challenges, as they trace lower-density circumgalactic gas, are physically small \cite{2019ApJ...885...61R}, and tend to lie further from halos than DLAs, so that accurate modeling requires capturing the CGM-IGM interface.
Furthermore, they lie in a transition region between ionized gas in photoionization equilibrium with the UVB, and the self-shielded neutral gas which produces DLAs.
Their neutral fraction thus depends sensitively on incident ionizing radiation and is sensitive to radiative transfer effects such as shadowing, as emphasized by previous simulation efforts (e.g., Refs.~\cite{Kohler:2007ApJ...655..685K,Fan:2024ApJ...963...45F,Georgiev:2025MNRAS.536.3689G,McQuinn:2011ApJ...743...82M,DAloisio.2020ApJ...898..149D}).
Although PRIYA implements inhomogeneous reionization and reproduces global quantities like the IGM temperature and DLA statistics well, it thus remains uncertain whether its LLS population is accurate. However, for P1D analyses, what matters most is that the total number density and the clustering of LLSs is statistically correct, as the P1D is averaged over sightlines.

We therefore examine the column density distribution function (CDDF) of DLAs/LLSs in PRIYA, shown in Figure~\ref{fig:priya_cddf}.
The CDDF is defined as the number of absorbers per unit redshift per unit column density and represents the abundance of DLAs and LLSs.
The PRIYA CDDF is actually in good agreement with previous sub-DLA measurements (e.g., Ref.~\cite{Zafar:2013A&A...556A.141Z}).
It also agrees with earlier OWLS simulation (Ref.~\cite{Altay:2011ApJ...737L..37A}), and agrees with SDSS DR7 measurements (Ref.~\cite{Prochaska:2010ApJ...718..392P}) in the LLS regime ($N_\mathrm{HI} \sim 10^{17.2} - 10^{19}\,\mathrm{cm}^{-2}$). In addition, PRIYA matches the observed DLA abundance from eBOSS DR16 (CDDF in DLA GP, Ref.~\citep{2021MNRAS.507..704H}).
At the very least, it is unlikely that the LLS abundance in PRIYA is low by a factor of $\sim 3$, as would be required to explain the high confidence observations.

\begin{figure}
    \centering
    \includegraphics[width=0.60\columnwidth]{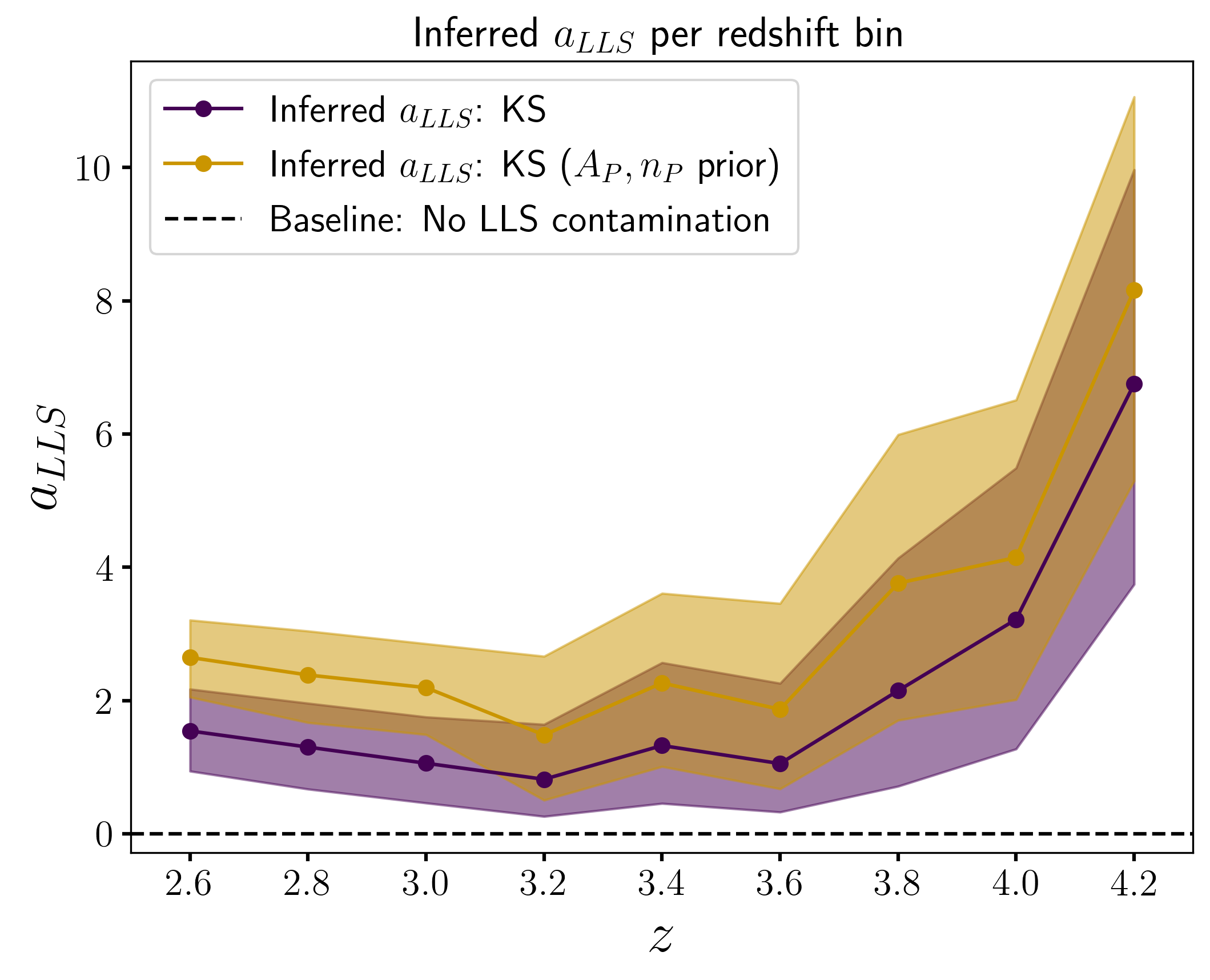}
    \caption{
    The inferred $\alpha_{\mathrm{LLS}}$ for each redshift bin, with or without the eBOSS $A_P,n_P$ priors.}
    \label{fig:separate_z_corner_astro}
\end{figure}

The second possibility is that the KODIAQ-SQUAD selection function is biased toward a higher LLS abundance. It is known \cite{2017AJ....154..114O, 2022MNRAS.509.2842K} that KODIAQ-SQUAD contains multiple spectra deliberately selected to contain large DLAs, and so a correlation between LLS and DLAs would imply that the selection function is also biased towards LLS. To test this, we impose eBOSS priors on $A_P$ and $n_P$, which helps isolate the $A_P$ constraint from small-scale astrophysical effects.
The results, shown as yellow contours in Figure~\ref{fig:separate_z_corner_astro}, indicate that with the eBOSS priors, the inferred $\alpha_{\mathrm{LLS}}$ shifts even higher, reaching $\sim 3$. This shift effectively compensates for the extra P1D slope tilt that would otherwise be absorbed by $A_P$. Interestingly, imposing the $(A_P, n_P)$ priors has a negligible impact on the goodness of fit: the reduced $\chi^2$ changes only slightly, from $\chi^2/\mathrm{dof} = 1.58$ to $1.57$. This suggests that the apparent inconsistency of $(A_P, n_P)$ with XQ100/eBOSS is just the MCMC sampler struggling to break the $A_P-\alpha_\mathrm{LLS}$ degeneracy, and that the KODIAQ-SQUAD P1D is still roughly consistent with the eBOSS $A_P$.

To further examine the complexity of the redshift-dependent LLS bias, we allow $\alpha_{\mathrm{LLS}}$ to vary independently in each redshift bin in the MCMC sampler. Figure~\ref{fig:separate_z_corner_astro} shows the inferred values of $\alpha_{\mathrm{LLS}}$ for each redshift bin, with and without the eBOSS $(A_P, n_P)$ priors. The redshift evolution turns out to be notably non-trivial: $\alpha_{\mathrm{LLS}} \sim 3$ at $z = 2.6$, $\sim 2$ at $z = 3.2$, and rising to $\sim 8$ at $z = 4.2$. This redshift trend helps explain why $A_P$ appears biasedly high in the MCMC chains: KODIAQ-SQUAD's selection function is skewed toward higher LLS abundance, and that bias is both non-trivial and redshift-dependent. When we marginalize over the HCD template (which assumes a different redshift evolution), the mismatch ends up being absorbed by a larger $A_P$.

This result suggests that in order to extract cosmological information from the KODIAQ-SQUAD P1D, one must either correct for the selection bias or mask LLSs in quasar spectra directly. However, both are challenging. Due to their higher abundance and relatively (compared to DLAs) subtle damping-wing features, there is currently no robust way to identify LLSs, apart from an early-stage random forest-based classifier (Ref.~\cite{Fumagalli.2020MNRAS.498.1951F}). Even if such a tool is applied, the KODIAQ-SQUAD selection function is still biased toward regions with high neutral hydrogen clustering, since targeting quasars with known DLAs will tend to trace overdense environments near halos.

\subsection{Mean Flux and Inferred Optical Depth}
\label{sec:meanflux}

\begin{figure}
    \centering
    \includegraphics[width=0.8\columnwidth]{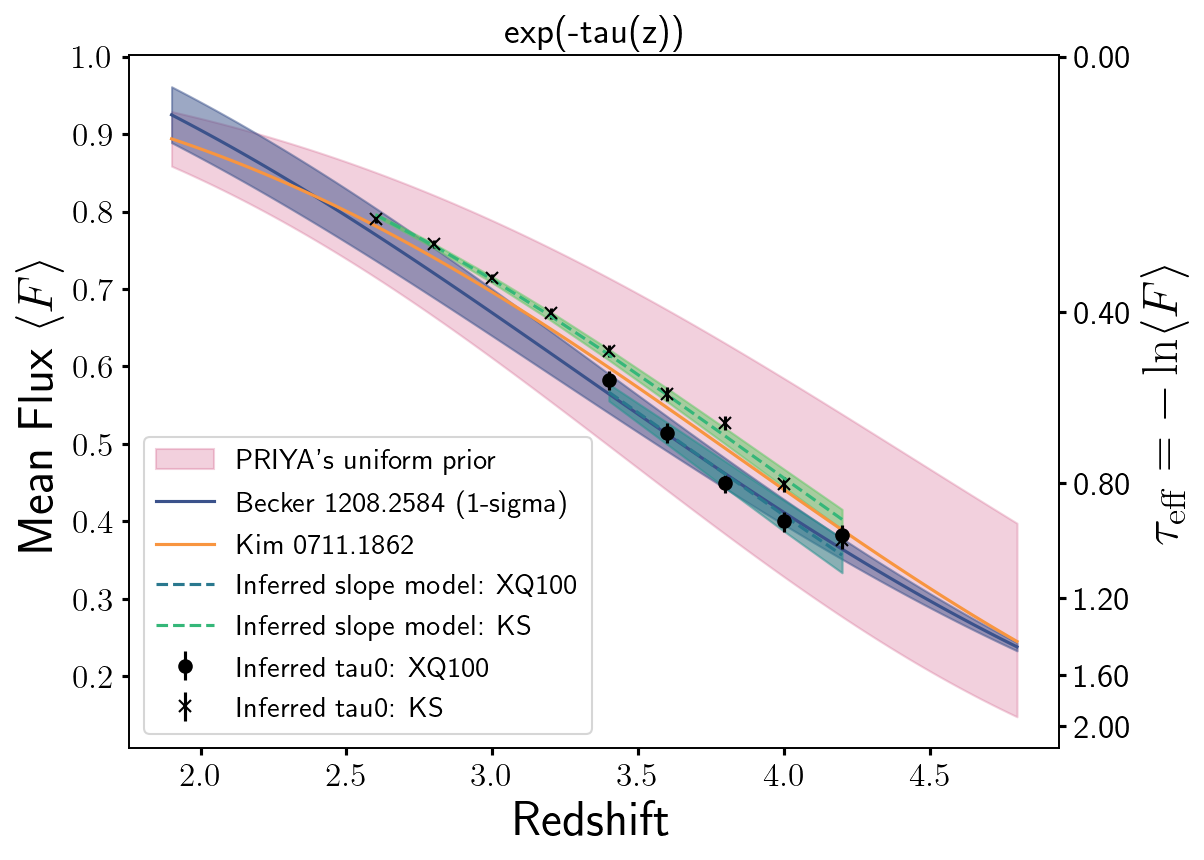}
    \caption{Inferred meanflux relation for XQ100 (blue dashed) and KODIAQ-SQUAD (green dashed). The blue-shaded region shows the measurement from Ref.~\cite{2013MNRAS.430.2067B} (Becker13), and the orange from Ref.~\cite{2007MNRAS.382.1657K} (Kim07). Dashed lines indicate the inferred power-law model parameters ($\tau_0$, $d\tau_0$), while dots (XQ100) and crosses (KODIAQ-SQUAD) show the results from fits assuming $\tau_0$ is independent for each redshift bin. Even though PRIYA does not impose an informed prior on the meanflux parameters (allowing them to be freely inferred from the data) we still find the inferred meanflux models to be broadly consistent with either Kim07 or Becker13.
	Any normalization shift in the $\tau_0$ is actually correlated with HCD parameters and should not interpret as physically meaningful. We will discuss this in \S~\ref{sec:dlalimit}.
    }
    \label{fig:meanflux-inferred}
\end{figure}

Figure~\ref{fig:meanflux-inferred} shows that the inferred mean flux models are broadly consistent with either Ref.~\cite{2007MNRAS.382.1657K} or Ref.~\cite{2013MNRAS.430.2067B} However, there is a normalization offset for KODIAQ-SQUAD.

This is not surprising. As discussed in \S~\ref{sec:dlalimit}, pixel-level contamination from DLAs and LLSs is \textit{not} explicitly modeled in the HCD template. As a result, the inferred $\tau_0$ should be interpreted as an effective quantity that absorbs both the mean flux and residual contamination from unmasked HCDs. In other words, the HCD template accounts only for \textit{damping wings}, not the \textit{saturated absorption troughs} that suppress small-scale clustering in the \lya\ forest and lower the overall P1D normalization. This suppression is indirectly captured by the $\tau_0$ posterior.

To test the validity of the power-law slope assumption in PRIYA, we also ran separate MCMC chains for XQ100 and KODIAQ-SQUAD, allowing $\tau_0$ to vary independently in each redshift bin. These results are shown in Figure~\ref{fig:meanflux-inferred} as dots and crosses. The independently inferred $\tau_0$ values are broadly consistent with the global slope model (dashed lines; Equation~\ref{eq:meanflux}), except for a mild discrepancy at $z = 4.2$. We do not assign strong physical meaning to this deviation: $\tau_0$ is a nuisance parameter entangled with both HCD contamination and $n_P$. At $z > 4$, quasar spectra become increasingly noisy and the sample size drops. As a result, $\tau_0$ normalization shifts in the P1D are more easily absorbed by $\tau_0$ than by parameters like $n_P$ or $\alpha_{\rm LLS}$, which are more tightly constrained by lower-redshift data. Therefore, when jointly fitting multiple redshift bins via the per-bin $\tau_0$ model, normalization shifts at high redshift are more likely to be absorbed into the per-$z$ $\tau_0$ rather than $n_P$ or $\alpha_{\rm LLS}$ (which are shared across redshift bins), explaining the offset at $z = 4.2$ seen in Figure~\ref{fig:meanflux-inferred}.




\section{Discussion}
\label{sec:discussion}
The results presented above illustrate several important points, regarding the small-scale Lyman-alpha forest P1D and its sensitivity to astrophysical/cosmological parameters, particularly the role of HCDs and the thermal history of the IGM through HeII reionization.

\begin{enumerate}
    
    \item \textbf{XQ100 likelihood agrees with eBOSS DR14:} In Ref.~\cite{2024JCAP...07..029F}, we derived cosmological parameters using an eBOSS DR14 likelihood, using external IGM temperature measurements, $T_0$, from Ref.~\cite{2021MNRAS.506.4389G} (based on KODIAQ), to calibrate the thermal history and constrain our inhomogeneous He\,\textsc{ii} reionization model. However, Ref.~\cite{2024JCAP...07..029F} was not able to assess the consistency between the eBOSS and external IGM temperature likelihoods.
    In this work, we find that the high-resolution P1D from XQ100 alone yields a thermal history consistent with the joint eBOSS DR14 and external IGM temperature (FPS + $T_0$) baseline. The posterior cosmological parameters are also in very good agreement, albeit with larger uncertainty in XQ100. This shows that combining the two datasets is a consistent choice and does not artificially bias the cosmological inference. Looking forward, high-resolution P1D datasets like XQ100 can be jointly analyzed with eBOSS or DESI to self-consistently constrain the thermal history, removing the need for external $T_0$ measurements altogether.

    
    \item \textbf{Amplitude}: The power spectrum amplitude parameter, $A_P$, tilts the P1D slope rather than uniformly rescaling the amplitude (see Figure~\ref{fig:1pvar_ap_aLLS}). It is strongly degenerate with the LLS contamination parameter, $a_{\rm LLS}$ on the scales probed by XQ100.

    \item \textbf{Slope}: The small-scale P1D ($k > 0.02\,\mathrm{s/km}$) is more sensitive to the slope parameter, $n_P$, than to the amplitude parameter, $A_P$, due to $n_P$ modifying the shape of the P1D in a distinct way.

    \item \textbf{HCD Contamination}: All unmasked HCDs, such as LLSs and sub-DLAs, have a significant impact on the scales ($k = 0.005 - 0.10\,\mathrm{s/km}$) relevant to high-resolution P1D measurements.
    The LLS contamination in KODIAQ-SQUAD is substantially larger than expected in the HCD template from simulations (Ref.~\cite{2018MNRAS.474.3032R}), especially at high redshift ($z = 3.8 - 4.2$). This appears due to a redshift-dependent selection effect, which biases the KODIAQ-SQUAD selection function towards DLAs and thus LLS and overdense regions of gas.

    \item \textbf{HeII Reionization}: The HeII reionization parameters, i.e., $\alpha_q$ (the heating rate), $z^{\mathrm{HeII}}_i$ (the start redshift), and $z^{\mathrm{HeII}}_f$ (the end redshift), strongly affect the total log-evidence of the fit. These parameters modify the small-scale ($k > 0.03\,\mathrm{s/km}$) P1D in a highly nonlinear and redshift-dependent way. Thus, any P1D inference at these small scales must account for HeII reionization.

    \item \textbf{Subtlety of $\tau_0$ with HCD Templates}: The effective mean flux parameter $\tau_0$ is not a direct physical observable but instead absorbs residual contamination from HCD systematics, particularly at high $k$. Therefore, it should not be interpreted in isolation, nor should it be placed with a Gaussian prior unless co-varied with $a_{\rm HCD}$.

\end{enumerate}

\subsection{Thermal history and cosmology are scale-separate}
\label{sec:ks-cosmo-separate}

\begin{figure}
    \centering
    \includegraphics[width=0.49\columnwidth]{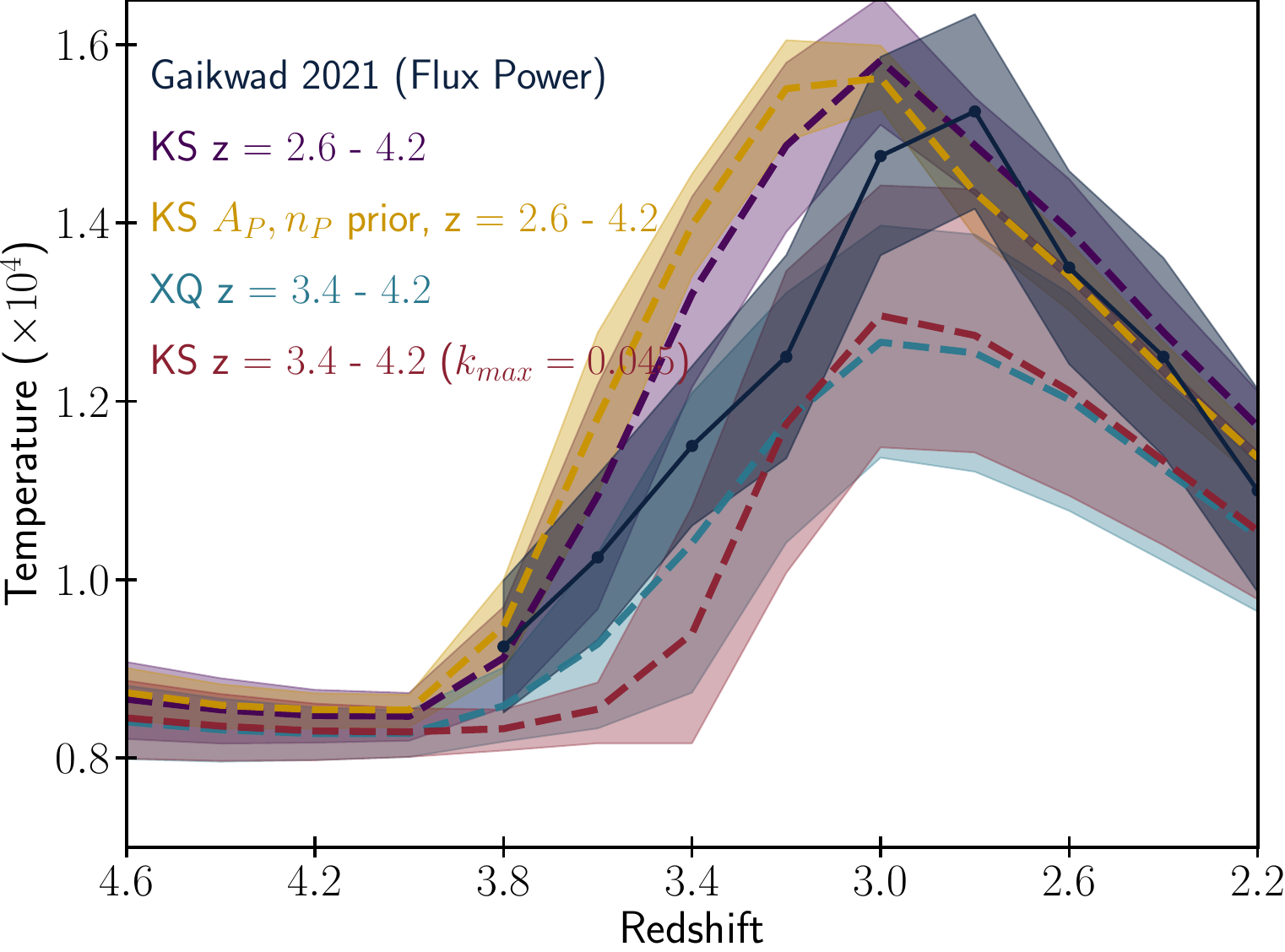}
    \includegraphics[width=0.49\columnwidth]{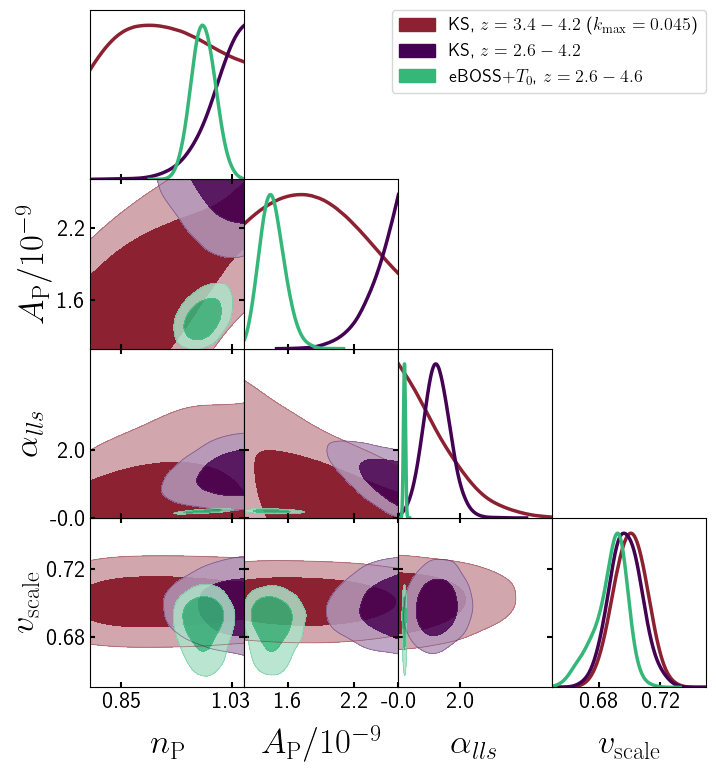}    
    \caption{
        Left: Posterior predictions of the IGM mean temperature, $T_0$, from the MCMC chains shown on the left. Additional MCMC chains include the XQ100 (blue) and a reduced KODIAQ-SQUAD redshift range $z = 3.4 - 4.2$ (red).
        Right: Comparison of the KODIAQ-SQUAD posteriors at full redshift ($z = 2.6-4.2$, purple) and restricted redshift ($z = 3.4-4.2$, red).
    }
    \label{fig:corner_and_meanT_ks}
\end{figure}

In \S~\ref{sec:separate_z}, we showed that the KODIAQ-SQUAD P1D is biased toward denser regions of the IGM, and that the degeneracy between HCD contamination and $A_P$ is difficult to disentangle given KODIAQ-SQUAD's complex selection function. We also found that marginalizing over cosmology using eBOSS priors breaks the $A_P$-$\alpha_{\rm LLS}$ degeneracy, while leaving the reduced $\chi^2$ nearly unchanged. This suggests that the cosmological prior serves to disentangle that specific degeneracy, without significantly altering the thermal parameter constraints. If so, it implies that the cosmological and thermal information in the P1D are largely scale-separated: thermal history is encoded in the small-scale modes of P1D, while cosmological constraints are driven by large-scale modes.

In Figure~\ref{fig:corner_and_meanT_ks} (left), we compare the inferred thermal histories with (yellow) and without (purple) the cosmological priors. The thermal posteriors remain nearly unchanged, with both chains yielding consistent $T_0(z)$ evolution. The only noticeable shift occurs in $\alpha_{\rm LLS}$, which rises to $\sim 3$ when the cosmology is priored out. This supports the idea that thermal constraints are not significantly affected by cosmological information fixed at $k < 0.02\,\mathrm{s/km}$ (eBOSS).

We further test this scale separation by varying the redshift and $k$ range included in the fit. We find that thermal parameters are most sensitive to high-$k$ modes ($k > 0.045\,\mathrm{s/km}$) and to low-redshift data ($z = 2.6-3.2$). As shown in Figure~\ref{fig:corner_and_meanT_ks} (left; red line), using only $z = 3.4-4.2$ and $k < 0.045\,\mathrm{s/km}$ lowers the inferred $T_0$ by nearly $2\sigma$, bringing it into agreement with the XQ100 thermal history. However, either including high-$k$ data at high redshift, or including low-$z$ bins with $k < 0.045\,\mathrm{s/km}$, is sufficient to recover the peak temperature of $T_0 \sim 1.5 \times 10^4\,\mathrm{K}$.

We conclude that thermal information in the P1D is largely separable in scale and redshift from cosmological information. In particular, most of the thermal sensitivity lies in $k \gtrsim 0.045\,\mathrm{s/km}$ and $z \lesssim 3.2$, while cosmological parameters are best constrained at lower $k$. 

\subsection{KS consistent with XQ100 when $z = 3.4-4.2$ and $k < 0.045\,\mathrm{s/km}$}
\label{sec:ks-cosmo-xq}

We now examine the $A_P$ degeneracy in this restricted chain, shown in the right panel of Figure~\ref{fig:corner_and_meanT_ks}. Interestingly, under this redshift and $k$ cut, the KODIAQ-SQUAD posterior (yellow) becomes consistent with the eBOSS baseline cosmology (green), yielding
$
A_P \sim 1.85^{+0.24}_{-0.63} \times 10^{-9},
$
though with large uncertainty. This agreement arises because the $k < 0.045\,\mathrm{s/km}$ cut removes part of the covariance between $A_P$ and $\alpha_{\rm LLS}$.

As a result, the LLS contamination in these reduced chains is only weakly constrained, with an upper bound of $\alpha_{\rm LLS} < 1.49$, and no strong posterior mode near $\alpha_{\rm LLS} \sim 2$ as seen in the full redshift and scale range.

\add{We also checked that applying only the $k<0.045\,\mathrm{s/km}$ cut, without any redshift restriction, already moves $(n_P,\,A_P)$ by about $0.5\sigma$ off the upper prior boundary (Figure~\ref{fig:kcut_appendix}, Appendix~\ref{sec:kcut-appendix}). Most of the recovery therefore comes from dropping the small-scale modes.}

Overall, these tests show that by restricting to high redshift and removing some small-scale modes, we can mitigate the some LLS contribution to the P1D and obtain cosmological constraints that are less sensitive to the biased selection function of KODIAQ-SQUAD. However, residual contamination likely persists, and it remains difficult to claim that the $A_P$ measurement is fully unbiased, as the quasar sightlines still preferentially probe denser regions of the IGM. There are two connected results: first, because KODIAQ-SQUAD probes a biased sky, this specific measurement of the flux power spectrum on small scales is unreliable for estimating the LLS abundance and cosmological gas clustering that drives the $A_P$ constraints. Second, any future analyses using quasars with an HI-blind selection function to obtain constraints on the small-scale power spectrum should include models for LLS contamination when probing scales $ k > 0.045 \,\mathrm{s/km}$.

\subsection{Reduced likelihood: XQ100}
\label{sec:reduced_likelihood}

\begin{figure}
    \centering
    \includegraphics[width=0.75\columnwidth]{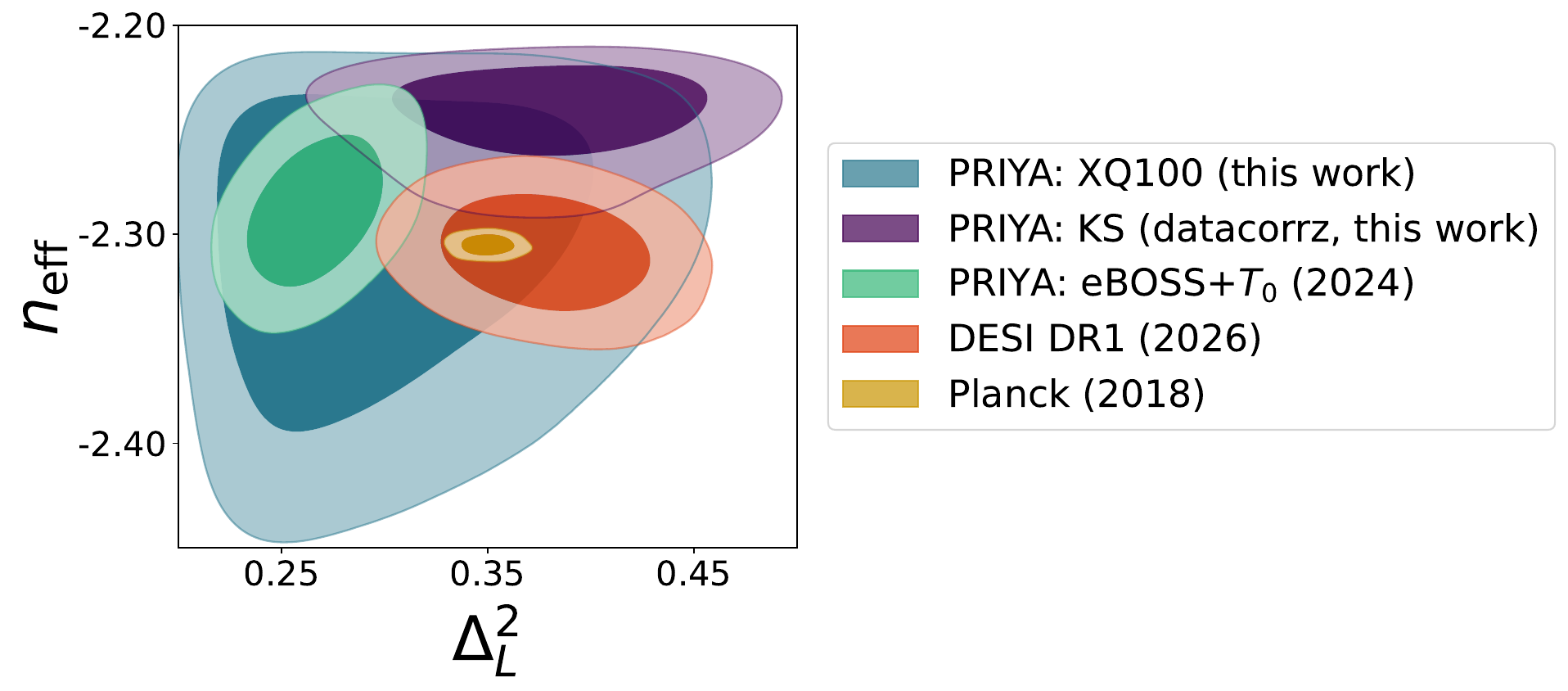}
    \caption{
        Reduced likelihood contours at $k_P = 0.009\,\mathrm{s/km}$, $z=3$:
        eBOSS P1D (green), XQ100 P1D (teal), and Planck 2018 (gold).
        \add{Also shown: KODIAQ-SQUAD with the per-redshift four-component HCD
        correction (\texttt{datacorrz}; purple) and DESI DR1
        P1D~\cite{Chaves-Montero:2026arXiv260121432C} (orange-red). $\Delta_L^2$
        and $n_{\rm eff}$ are derived from the MCMC posteriors via CLASS; see
        \S~\ref{sec:reduced_likelihood} for the choice of \texttt{datacorrz}
        over the global four-component variant \texttt{dc4} (one shared amplitude
        per absorber class across redshift; same as Ref.~\cite{2018MNRAS.474.3032R}) and the impact of prior truncation
        on the apparent precision.}
    }
    \label{fig:redlike_XQ_KS}
\end{figure}

As proposed in Ref.~\cite{Rogers:2025PhRvR...7a2018R}, we can assess the relative constraining power of different surveys by examining their reduced likelihood: the amplitude and slope of the linear theory power spectrum at $k = 1\,\mathrm{Mpc}^{-1}$ and $z = 3$. This choice of pivot scale and redshift is motivated by the fact that the \Lya forest P1D primarily probes Mpc-scale modes around this redshift. Note, however, that this $z = 3$ reference is slightly lower than the redshifts of XQ100, which has a midpoint near $z = 3.8$.

Figure~\ref{fig:redlike_XQ_KS} shows the reduced likelihood contour for XQ100 and \add{KODIAQ-SQUAD using the per-redshift four-component HCD correction (\texttt{datacorrz}; one amplitude per absorber class per redshift bin)} from this work, alongside eBOSS DR14 (FPS + $T_0$)~\add{\cite{2024JCAP...07..029F}}, \add{DESI DR1~\cite{Chaves-Montero:2026arXiv260121432C}}, and Planck 2018. As expected, the XQ100 posterior is broadly consistent with the eBOSS DR14 (FPS + $T_0$) baseline, and the uncertainties are wide enough to overlap with both the Planck baseline. This is consistent with the discussion before: while XQ100 probes smaller scales that are more sensitive to the IGM thermal history, its cosmological constraining power is limited by degeneracies with HCD contamination.
\add{The KODIAQ-SQUAD \texttt{datacorrz} posterior gives $\Delta_L^2 = 0.381 \pm 0.045$, consistent within $1\sigma$ with DESI DR1 ($\Delta_L^2 = 0.379 \pm 0.032$; Table~\ref{tab:posteriors_combined}). In $n_{\rm eff}$, DESI DR1 ($-2.309 \pm 0.019$) is essentially identical to XQ100 ($-2.31^{+0.07}_{-0.03}$) and consistent with Planck; KODIAQ-SQUAD is the outlier at $n_{\rm eff} = -2.24 \pm 0.02$, about $2\sigma$ above the others. Both the KODIAQ-SQUAD and XQ100 contours, however, are sensitive to prior truncation. For KODIAQ-SQUAD, $n_P = 1.032 \pm 0.016$ sits only $\sim 1.2\sigma$ below the upper wall at $n_P = 1.05$, and $A_P = 2.13 \pm 0.24 \times 10^{-9}$ only $\sim 2\sigma$ below upper wall $A_P = 2.6 \times 10^{-9}$, which tightens both $n_{\rm eff}$ and $\Delta_L^2$ on the high side.\footnote{Table~\ref{tab:posteriors_combined} reports this as $n_P > 1.03$, the one-sided 68\% lower bound used for boundary-truncated entries.} XQ100 is affected on the \emph{opposite} side: $A_P = 1.71 \pm 0.34 \times 10^{-9}$ sits only $\sim 1.5\sigma$ above the lower wall at $A_P = 1.2 \times 10^{-9}$, which tightens $\Delta_L^2$ from below. A wider prior in either direction would broaden these contours. We use \texttt{datacorrz} rather than \texttt{dc4} (the global four-component variant: one shared amplitude per absorber class across redshift) because the LLS abundance in KODIAQ-SQUAD is strongly redshift-dependent (\S~\ref{sec:separate_z}, Appendix~\ref{sec:lls-complexity-appendix}); a single global HCD amplitude cannot absorb this $z$-dependent variation, so the residual mismatch leaks into $A_P$ and biases the cosmology. Allowing per-$z$ HCD amplitudes (\texttt{datacorrz}) absorbs the variation directly and gives a less HCD-biased reduced-likelihood contour for KODIAQ-SQUAD.}

\subsection{IGM temperature measurement from Gaikwad 2021}

Figure~\ref{fig:corner_and_meanT_ks} shows four measurements of the thermal history of the IGM from three distinct flux power spectra. We have argued that the KODIAQ-SQUAD dataset is biased. The high peak temperature preferred by that measurement is consistent with a bias towards lower small-scale power, as high IGM temperatures erase small scale power (see Figure~\ref{fig:1pvar_heii}). Of our measurements, we thus prefer the inferred XQ100 thermal history. However, because the minimum redshift is $z=3.4$, it cannot robustly infer the peak IGM temperature at $z\sim 3$. The thermal history measurements from Ref.~\cite{2021MNRAS.506.4389G} uses a similar KODIAQ dataset that we have argued is biased towards DLAs. However, they remove all spectra containing a DLA from the sample, rather than masking only the regions directly affected by damping wings as is done by Ref.~\cite{2022MNRAS.509.2842K}. This will substantially reduce the effect of the selection bias, although LLS are likely still over-represented as the sample will include spectra which attempted to select for DLAs but found only weaker absorbers. Figure~\ref{fig:meanT_pred} shows that in practice the thermal history measurement of Ref.~\cite{2021MNRAS.506.4389G} is consistent with the posterior from XQ100 at the $68\%$ confidence level for $z > 3.0$. This suggests modest effects on the thermal history in the high redshift range most affected by over-sampling LLS.

\section{Conclusions}
\label{sec:conclusions}

In this work, we use the PRIYA simulation suite with a new emulator to analyze the small-scale \lya\ forest flux power spectrum (P1D) from high-resolution quasar datasets, specifically XQ100 (Ref.~\cite{2022MNRAS.509.2423W}) and KODIAQ-SQUAD (Ref.~\cite{2022MNRAS.509.2842K}). The emulator enables simulation-based predictions across a broad parameter space, including cosmology and thermal history, while explicitly modeling the impact of HeII reionization on the P1D.

We find that the XQ100 P1D yields $(A_P, n_P)$ constraints consistent with previous PRIYA results from eBOSS DR14 and Planck, albeit with broader uncertainties. Notably, this is achieved without using external IGM temperature data, demonstrating that the XQ100 P1D alone carries sufficient thermal information. In contrast, the KODIAQ-SQUAD P1D exhibits a biasedly high $\alpha_{\rm LLS}$ that strongly degenerates with $A_P$. However, restricting the analysis to $k < 0.045\,\mathrm{s/km}$ and $z = 3.4-4.2$ significantly reduces sensitivity to LLSs, and $A_P$ becomes consistent with XQ100 and eBOSS.

Two key features of the small-scale P1D are: first, it is highly sensitive to thermal history, particularly the He\,\textsc{ii} heating rate $\alpha_q$ and its timing.  Another is the contamination from LLSs and sub-DLAs, which alter the P1D slope at intermediate and small scales. We find strong degeneracies among LLSs, mean flux, and $A_P$. The LLS abundance inferred from KODIAQ-SQUAD appears biased: the LLS abundance increases with redshift and deviates from the HCD template. At $z = 3.8-4.2$, KODIAQ-SQUAD prefers $\sim6\times$ more LLSs than PRIYA. Given that PRIYA's LLSs match empirical data at $z = 3$, this suggests the excess likely arises from KODIAQ-SQUAD's   selection function, which uses quasars with known DLAs and metal absorbers, both strongly correlated with LLSs.

These findings show the importance of isolating LLS contamination when modeling the P1D at $k > 0.045\,\mathrm{s/km}$. However, masking LLSs on an absorber-by-absorber basis remains infeasible. Even for DLAs, robust masking is difficult at low S/N: the recent DESI DR2 DLA catalog~\cite{Brodzeller:2025arXiv250314740B} achieves 85\% completeness and 80\% purity, but only for $N_{\rm HI} > 10^{20.3}\,\mathrm{cm}^{-2}$, which are much easier to find than LLSs. Thus, the reliability of cosmological inference from high-resolution P1D fully depends on how accurately one can forward-model LLSs in simulations and build a robust HCD template.

Previous work on high-resolution P1D has largely focused on setting bounds on alternative dark matter models (see Refs.~\cite{2017PhRvD..96b3522I,2013PhRvD..88d3502V,2021PhRvL.126g1302R,Irsic:2024PhRvD.109d3511I}). However, these analyses are typically not discussed in the context of HCD contamination (except Ref.~\cite{2022MNRAS.515..857E}, in the context of comparing with cluster cosmology). Ref.~\cite{Rogers:2025PhRvR...7a2018R} also emphasized that the scale-dependent astrophysical uncertainties from HCDs might play a critical role in future \lya DM analysis. In this work, we confirm that LLSs correlate strongly with the cosmological signal of the primordial amplitude in high-resolution P1D. This suggests that, since there is no practical way to mask LLSs, future \lya\ DM analyses will need to be HI-blind and carefully marginalize over LLSs, especially when the DM signal is expected as a suppression of the linear power amplitude. Any residual LLS contamination in the data would show up as an extra suppression of high-$k$ modes in the P1D.

An alternative path forward might be to measure the power spectrum of HCDs directly.
For example, using Voigt profile decomposition rather than sinusoids in Fourier space, or employing wavelet scattering transforms (WST; see~\cite{Bruna:2012arXiv1203.1513B,Anden:2014ITSP...62.4114A,Andreux:2018arXiv181211214A,Tohfa:2024PhRvL.132w1002T}) to isolate LLS abundance from the forest. Our work also motivates gathering samples of high resolution quasar spectra with an HI-blind selection function that extend to lower redshifts than XQ100.

Taken together, this work highlights the value of small-scale \lya\ forest P1D from high-resolution quasar spectra. XQ100 offers reliable thermal constraints while preserving cosmological consistency with eBOSS DR14. With upcoming DESI DR1 P1D measurements~\cite{Naim:2025,Ravoux:2025arXiv250509493R} covering P1D until $k = 0.035\,\mathrm{s/km}$, the XQ100 P1D can serve as powerful constraints on thermal nuisance parameters in a joint cosmological fit.

\acknowledgments
The authors thank Andreu Font-Ribera for discussions on peculiar velocities, and Naim {G{\"o}ksel Kara{\c{c}}ayl{\i}} for helpful discussions on the KODIAQ-SQUAD data.
We thank Bayu Wilson for kindly sharing the data.
MFH thanks Francisco Villaescusa-Navarro and the DREAMS collaboration for useful discussions during the workshop that helped motivate and complete this paper. 
MFH was supported by NASA FINESST grant No. ASTRO20-0022 during the project.
MFH is currently supported by the Leinweber Foundation and DOE grant DE-SC0019193.
MQ was supported by NSF grant AST-2107821.
MQ is supported by the HETDEX collaboration.
SB and YY were supported by NASA-80NSSC21K1840 and NSF AST-2509639.
Computing resources were provided by Frontera LRAC AST21005.
The authors acknowledge the Frontera computing project at the Texas Advanced Computing Center (TACC) for providing HPC and storage resources that have contributed to the research results reported within this paper.
Frontera is made possible by National Science Foundation award OAC-1818253.
URL: \url{http://www.tacc.utexas.edu}. Analysis computations were performed using the resources of the UCR HPCC, which were funded by grants from NSF (MRI-2215705, MRI-1429826) and NIH (1S10OD016290-01A1).

\appendix

\section{Maximum a Posteriori (MAP) Predictions}
\label{sec:map}

\begin{figure}
    \centering
    \includegraphics[width=\columnwidth]{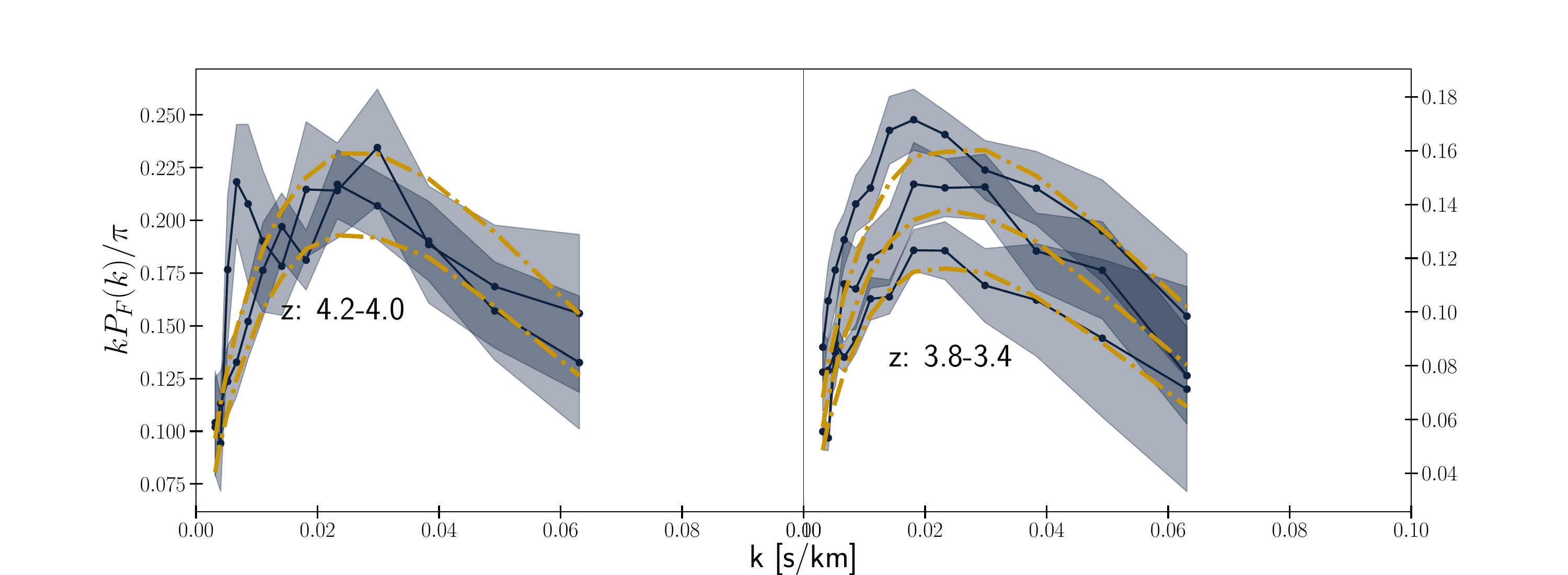}
    \caption{Posterior prediction of the flux power spectrum for XQ100 ($z = 3.4$–$4.2$). Shaded area represents the 1-$\sigma$ confidence interval of the measurement.
}
    \label{fig:fps_xq100}
\end{figure}

\begin{figure}
    \centering
    \includegraphics[width=\columnwidth]{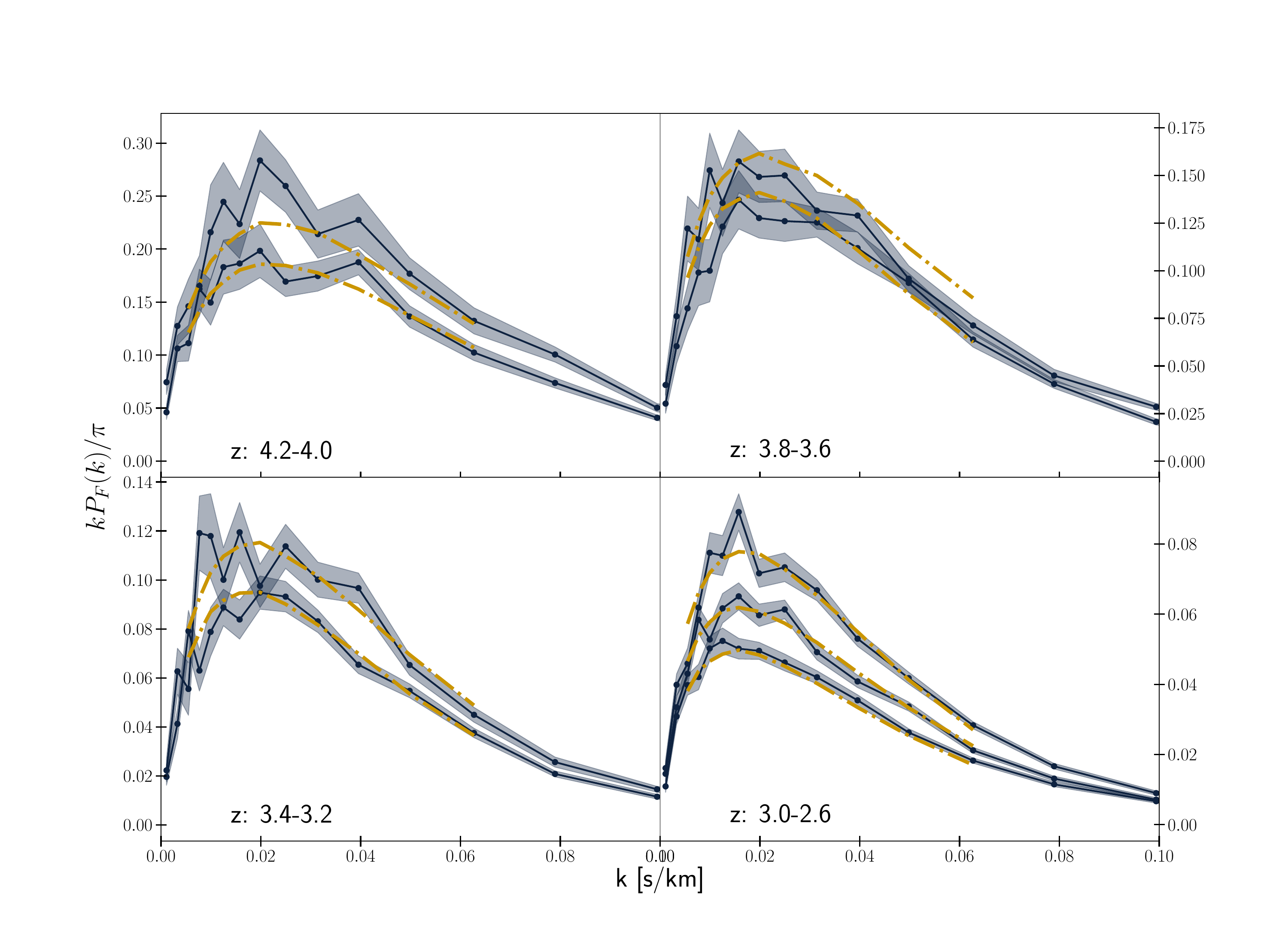}
    \caption{Posterior prediction of the flux power spectrum for KODIAQ-SQUAD ($z = 2.6-4.2$). Dot-dashed lines denote MAP predictions for each redshift bin.}
    \label{fig:fps_kodiaq}
\end{figure}

To visually assess how well the inferred posteriors explain the observed data, we compare the MAP predictions of the flux power spectrum with the actual measurements. Figures~\ref{fig:fps_xq100} and \ref{fig:fps_kodiaq} show the MAP posterior predictions for XQ100 and KODIAQ-SQUAD, respectively.
Notably, even though the raw KODIAQ-SQUAD P1D exhibits stronger small-scale suppression than XQ100, the MAP model reproduces this behavior via a combination of elevated $A_P$ and a higher $\alpha_{\rm LLS}$.
It remains a good fit up to the highest $k = 0.065\,\mathrm{s/km}$.
As seen from the diagonal covariance, the KODIAQ-SQUAD posterior is primarily influenced by the lower redshift bins ($z = 2.6 - 3.2$) rather than the higher ones ($z = 3.4 - 4.2$).
We have seen that the separate redshift bins yield quite different posteriors in \S\ref{sec:ks-cosmo-xq}.

For XQ100, the diagonal covariance is more forgiving, so the MAP predictions comfortably sit within the $1\sigma$ confidence region of the observed P1D in all redshift bins.

KODIAQ-SQUAD, on the other hand, has much smaller covariance at high $k$, allowing the MAP prediction to be more sensitive to small-scale suppression.
However, this is highly degenerate with various other parameters that alter the P1D slope, such as $A_P,\alpha_{\rm LLS},\alpha_q$, which we have discussed in \S\ref{sec:ks-cosmo-separate}, \S\ref{sec:ks-cosmo-xq} and \S\ref{sec:correlations}.

Note that both Figures~\ref{fig:fps_xq100} and \ref{fig:fps_kodiaq} only show the diagonal of the covariance matrix.
Therefore, even if the MAP predictions do not lie within the $1\sigma$ confidence region, this does not imply inconsistency with the data because the off-diagonal terms of the covariance matrix within redshift bins are not being shown.

\section{Table of Cosmological and Astronomical Parameter Constraints}
\label{sec:params-table}

Table~\ref{tab:posteriors_combined} presents the posterior constraints on the model parameters from KODIAQ-SQUAD, XQ100, and eBOSS (FPS + $T_0$). Table~\ref{tab:posteriors_separate_z} shows the posterior constraints on the LLS parameters from KODIAQ-SQUAD in different redshift bins, with $(A_P, n_P)$ fixed to eBOSS prior.

Many parameters are constrained only with upper or lower limits, which is not surprising given the high-dimensional parameter space and the limited sensitivity of the P1D to some of them. As seen in Figure~\ref{fig:corr_matrix}, $\chi^2$ is not very sensitive to many parameters, and several are degenerate with each other.
Nonetheless, this work represents an important step forward in the field of \Lya P1D inference: a self-consistent HeII reionization modeling across a wide range of cosmic evolution ($z = 2.6 - 4.2$), recognizing its importance for small-scale P1D modeling.

\section{Robustness to the $A_P$ Prior}
\label{sec:ap-prior-robustness}

\add{We test the sensitivity of our baseline chains to the prior choice on the power spectrum amplitude $A_P$. In the fiducial analysis, $A_P$ has a uniform (linear) prior within the emulator limits. Since $A_P$ is an amplitude-like parameter, we also run chains with a log-uniform prior on $A_P$. Table~\ref{tab:ap-prior-shifts} summarises the shifts in units of the fiducial posterior standard deviation. The largest shift is on $A_P$ itself ($-0.15\sigma$ for KODIAQ-SQUAD, $-0.20\sigma$ for XQ100); all other parameters move by $\leq 0.13\sigma$, so the prior choice does not change our main conclusions.}


\begin{table}
    \centering
    \def\arraystretch{1.2}
    \begin{tabular}{lcc}
        \hline
        Parameter & KODIAQ-SQUAD $\Delta/\sigma$ & XQ100 $\Delta/\sigma$ \\
        \hline
        $n_P$ & $+0.03$ & $-0.06$ \\
        $A_P$ & $-0.15$ & $-0.20$ \\
        $h$ & $-0.02$ & $-0.03$ \\
        $\Omega_m h^2$ & $+0.02$ & $+0.01$ \\
        $z_{\mathrm{HeII},i}$ & $+0.04$ & $+0.02$ \\
        $z_{\mathrm{HeII},f}$ & $+0.07$ & $+0.02$ \\
        $\alpha_q$ & $+0.06$ & $-0.04$ \\
        $z_{\mathrm{HI}}$ & $0.00$ & -- \\
        $\tau_0$ & $-0.06$ & $+0.12$ \\
        $d\tau_0$ & $+0.03$ & $-0.05$ \\
        $\alpha_{\mathrm{LLS}}$ & $+0.09$ & $+0.12$ \\
        \hline
    \end{tabular}
    \caption{\add{Parameter shifts when replacing the fiducial uniform prior on $A_P$ with a log-uniform prior.
    The sign is defined as log-uniform minus fiducial uniform.
    The shifts are measured in units of the fiducial posterior standard deviation.
    }}
    \label{tab:ap-prior-shifts}
\end{table}

\begin{table}
    \centering
    \def\arraystretch{1.6}
    \resizebox{\textwidth}{!}{%
    \begin{tabular}{| l | c | c | c | c | c |}
    \hline
    \textbf{Parameter}
    & \textbf{KS (dc4)}
    & \textbf{KS (dc4, reduced)}
    & \textbf{KS (datacorrz)$^\dagger$}
    & \textbf{XQ}
    & \textbf{eBOSS (FPS + $T_0$)} \\

    & $z = 2.6$--$4.2$
    & $z = 3.4$--$4.2$
    & $z = 2.6$--$4.2$
    & $z = 3.4$--$4.2$
    & $z = 2.2$--$4.6$ \\

    & $k = 0.0055 - 0.065\,\mathrm{s/km}$
    & $k = 0.0055 - 0.045\,\mathrm{s/km}$
    & $k = 0.0055 - 0.065\,\mathrm{s/km}$
    & $k = 0.003 - 0.065\,\mathrm{s/km}$
    & $k = 0.001 - 0.02\,\mathrm{s/km}$ \\
    \hline

    {\boldmath$d\tau_0$}
    & \add{$0.109^{+0.079}_{-0.067}$}
    & \add{$> 0.0963$}
    & \add{$0.038\pm 0.089$}
    & \add{$-0.08\pm 0.15$}
    & $0.009\pm 0.045$ $\left(^{+0.089}_{-0.087}\right)$ \\

    {\boldmath$\tau_0$}
    & \add{$0.938^{+0.014}_{-0.019}$}
    & \add{$1.038^{+0.047}_{-0.055}$}
    & \add{$0.921\pm 0.013$}
    & \add{$1.107^{+0.035}_{-0.041}$}
    & $1.090\pm 0.022$ $\left(^{+0.043}_{-0.042}\right)$ \\

    {\boldmath$n_\mathrm{P}$}
    & \add{$> 1.00$}
    & $> 0.906$
    & \add{$> 1.03$}
    & \add{$0.966^{+0.079}_{-0.027}$}
    & $0.983\pm 0.020$ $\left(^{+0.040}_{-0.039}\right)$ \\

    {\boldmath$A_\mathrm{P}/10^{-9}$}
    & \add{$> 2.37\cdot 10^{-9}$}
    & \add{$\left(\,1.84^{+0.26}_{-0.58}\,\right)\cdot 10^{-9}$}
    & \add{$\left(2.13^{+0.27}_{-0.24}\right)\cdot 10^{-9}$}
    & \add{$< 1.87\cdot 10^{-9}$}
    & $1.46^{+0.099}_{-0.13}$ $\left(^{+0.22}_{-0.22}\right)$ \\

    {\boldmath$z^{HeII}_i$}
    & \add{$> 3.86$}
    & \add{$< 3.78$}
    & \add{$> 3.81$}
    & ---
    & $> 4.00$ $(> 3.87)$ \\

    {\boldmath$z^{HeII}_f$}
    & \add{$> 2.87$}
    & \add{$< 2.96$}
    & \add{$> 2.94$}
    & $< 2.91$
    & $2.765^{+0.080}_{-0.093}$ $\left(^{+0.14}_{-0.16}\right)$ \\

    {\boldmath$\alpha_{q}$}
    & $< 1.49$
    & \add{$> 2.01$}
    & \add{$< 1.48$}
    & \add{$2.18^{+0.58}_{-0.45}$}
    & $1.74^{+0.18}_{-0.21}$ $\left(^{+0.37}_{-0.38}\right)$ \\

    {\boldmath$v_\mathrm{scale}$}
    & \add{$0.698\pm 0.011$}
    & $0.701\pm 0.011$
    & \add{$0.700^{+0.011}_{-0.012}$}
    & $0.698\pm 0.010$
    & $0.688^{+0.013}_{-0.0074}$ $\left(^{+0.018}_{-0.025}\right)$ \\

    {\boldmath$\Omega_M h^2$}
    & $> 0.143$
    & ---
    & \add{$< 0.144$}
    & $< 0.143$
    & $< 0.143$ ($--$) \\

    {\boldmath$z^{HI}$}
    & \add{$< 7.27$}
    & ---
    & \add{$< 7.28$}
    & \add{$7.26\pm 0.40$}
    & $7.24\pm 0.38$ ($--$) \\

    {\boldmath$\epsilon_{AGN}$}
    & \add{$0.0488\pm 0.0035$}
    & \add{$0.0498\pm 0.0035$}
    & \add{$0.0492\pm 0.0036$}
    & \add{$0.0495\pm 0.0034$}
    & --- \\

    {\boldmath$\alpha_{lls}$}
    & \add{$1.26\pm 0.45$}
    & $< 1.48$
    & per-$z$ (see text)
    & \add{$< 0.674$}
    & --- \\

    {\boldmath$\alpha_{sub}$}
    & \add{$< 0.181$}
    & \add{$< 0.641$}
    & per-$z$ (see text)
    & \add{$< 0.199$}
    & --- \\

    {\boldmath$\alpha_{sdla}$}
    & \add{$< 0.207$}
    & ---
    & ---
    & \add{$< 0.159$}
    & --- \\

    {\boldmath$\alpha_{ldla}$}
    & ---
    & ---
    & ---
    & ---
    & --- \\
    \hline
    \end{tabular}
    }
    \caption{
        Posterior constraints on model parameters from KODIAQ-SQUAD \add{(three HCD model variants and $k$/$z$ cuts)}, XQ100, and eBOSS (FPS + $T_0$ from Ref.~\cite{2024JCAP...07..029F}). We report 68\% confidence intervals or 95\% limits where applicable.
        \add{``dc4'' denotes the global four-component HCD correction (single amplitude per absorber class across all redshifts);
        ``datacorrz'' denotes the per-redshift four-component HCD correction.
        }
    }
    \label{tab:posteriors_combined}
\end{table}

\begin{table}
    \centering
    \def\arraystretch{1.5}
    \begin{tabular}{|l|c|c|}
    \hline
    \textbf{Parameter} & KODIAQ-SQUAD  & KODIAQ-SQUAD  \\
    & $z=2.6$–$4.2$ & $z=3.8$–$4.2$ \\
    & (eBOSS ($A_p, n_P$) prior) & (eBOSS ($A_p, n_P$) prior) \\    
    \hline
    {\boldmath$d\tau_0$}                  & $> 0.197$                             & $> 0.0840$ \\
    {\boldmath$\tau_0$}                  & $0.945\pm 0.013$                      & $0.971\pm 0.022$ \\
    {\boldmath$n_\mathrm{P}$}           & $0.992\pm 0.014$                      & $0.985\pm 0.014$ \\
    {\boldmath$A_\mathrm{P}/10^{-9}$}   & $(1.532\pm 0.078)\cdot 10^{-9}$       & $(1.476\pm 0.079)\cdot 10^{-9}$ \\
    {\boldmath$z^{HeII}_i$}             & $> 4.24$                              & $> 4.14$ \\
    {\boldmath$z^{HeII}_f$}             & $> 3.10$                              & $> 2.62$ \\
    {\boldmath$\alpha_{q}$}             & $< 1.37$                              & $< 2.00$ \\
    {\boldmath$v_\mathrm{scale}$}       & $0.691^{+0.011}_{-0.010}$            & $0.697^{+0.011}_{-0.0098}$ \\
    {\boldmath$\Omega_M h^2$}           & $> 0.145$                             & $> 0.143$ \\
    {\boldmath$z^{HI}$}                 & $< 7.26$                              & --- \\
    {\boldmath$\epsilon_{AGN}$}         & $0.0477\pm 0.0035$                    & $0.0489\pm 0.0035$ \\
    {\boldmath$\alpha_{lls}$}           & $2.79\pm 0.41$                        & $5.0^{+1.9}_{-1.4}$ \\
    {\boldmath$\alpha_{sub}$}           & $< 0.149$                             & $< 1.07$ \\
    {\boldmath$\alpha_{sdla}$}          & $< 0.187$                             & --- \\
    {\boldmath$\alpha_{ldla}$}          & ---                                   & --- \\
    \hline
    \end{tabular}
    \caption{Posterior constraints for KODIAQ-SQUAD in different redshift bins, with $A_P$ and $n_P$ priors applied.}
    \label{tab:posteriors_separate_z}
\end{table}

\section{Comparison of MCMC Chains with Mean Flux Model in Independent Redshift Bins}
\label{sec:per-z-corner}

\begin{figure}
    \centering
    \includegraphics[width=\columnwidth]{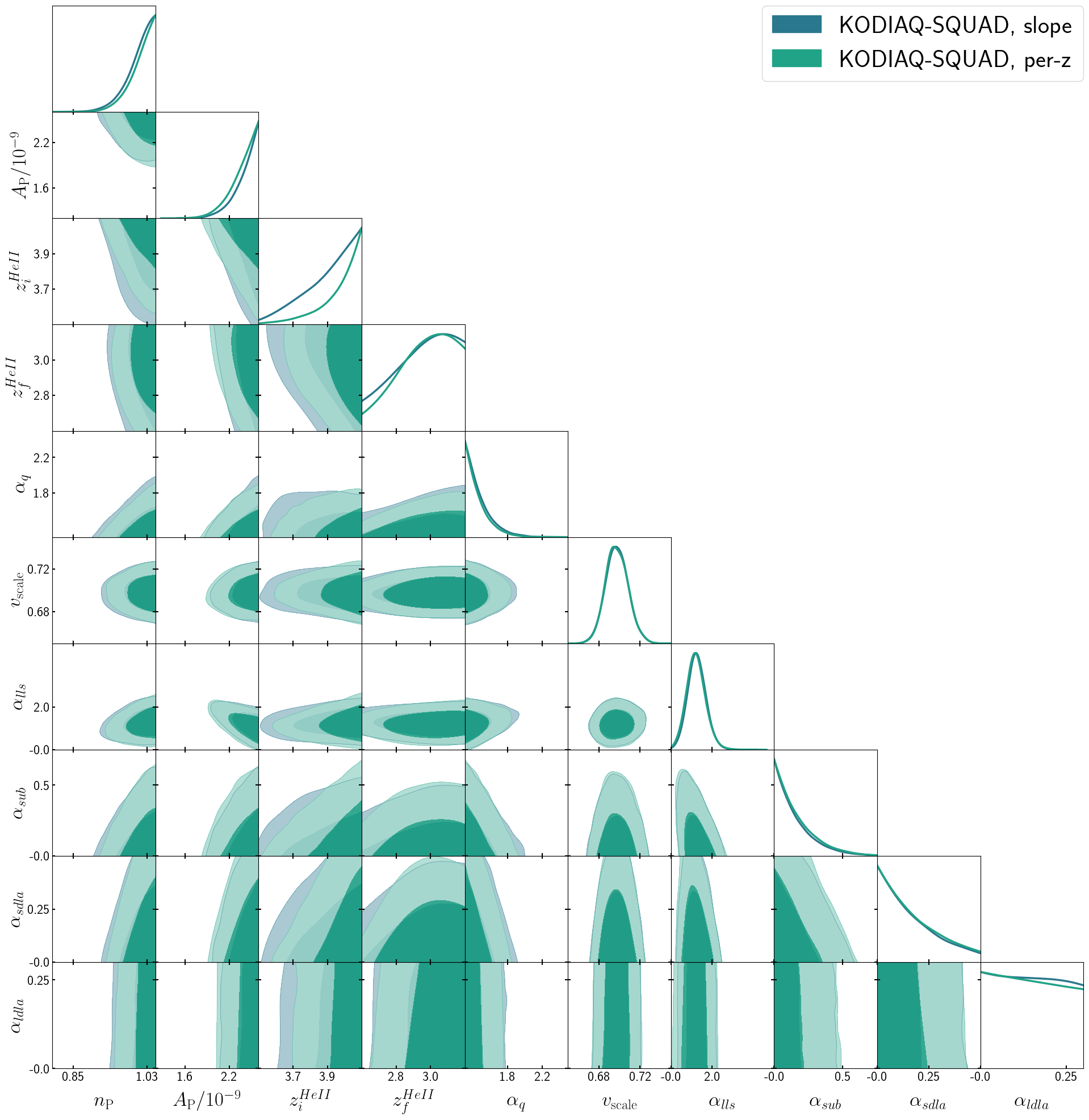}
    \caption{MCMC corner plot for the KODIAQ-SQUAD dataset using a mean flux model with independent redshift bins. This allows for redshift-dependent shifts in the amplitude and shape of the P1D, revealing any hidden degeneracies.}
    \label{fig:per-z-corner-ks}
\end{figure}

\begin{figure}
    \centering
    \includegraphics[width=\columnwidth]{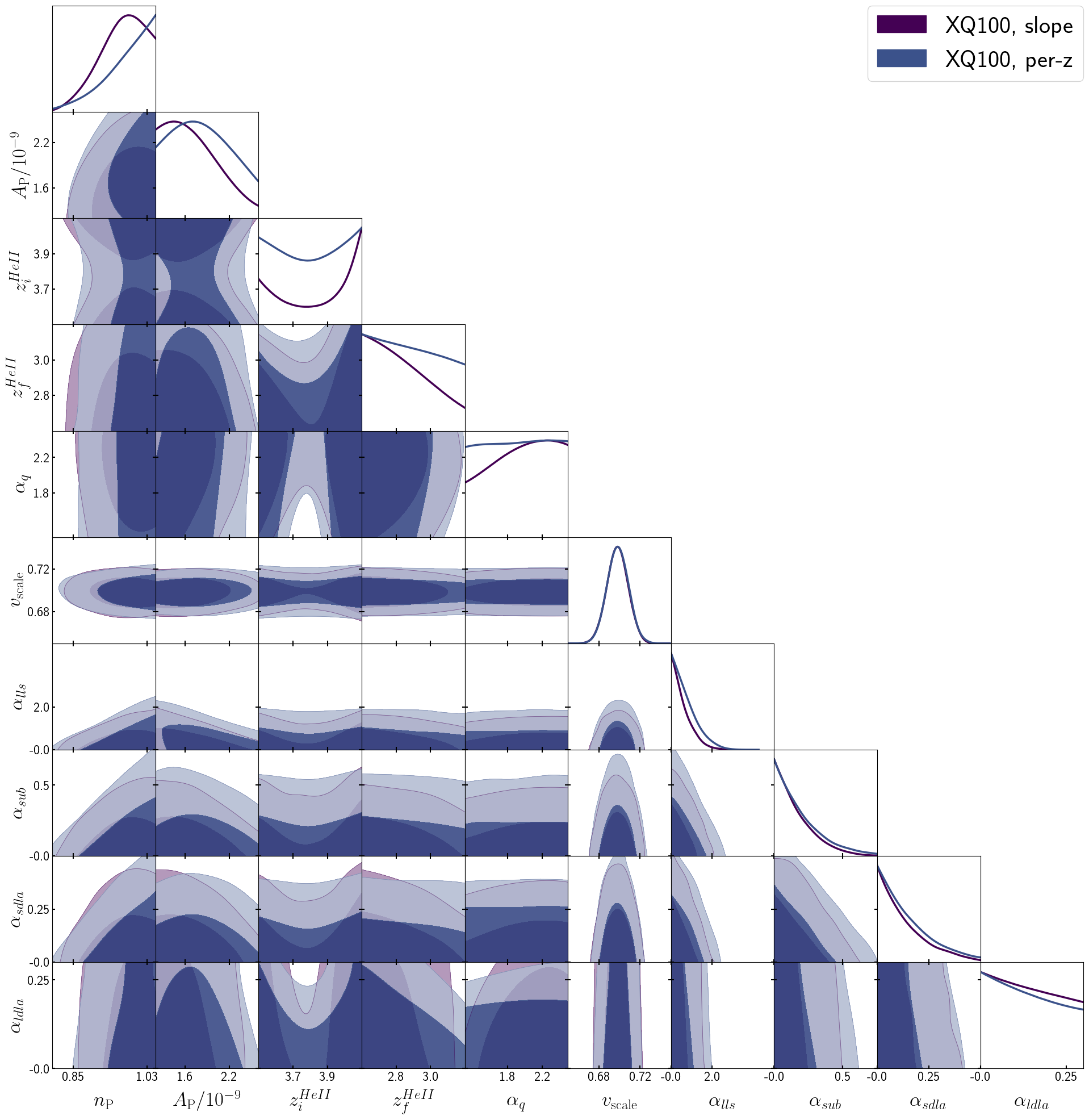}
    \caption{MCMC corner plot for the XQ100 dataset with redshift-independent $\bar{F}(z)$ assumptions relaxed. The results are broadly consistent with the fiducial global model.}
    \label{fig:per-z-corner-xq}
\end{figure}

To examine whether parameter degeneracies or constraints vary with redshift, we perform an alternative fit using a model in which the mean flux is allowed to vary independently in each redshift bin. Figures~\ref{fig:per-z-corner-ks} and~\ref{fig:per-z-corner-xq} show the resulting MCMC corner plots for the KODIAQ-SQUAD and XQ-100 datasets, respectively. This approach serves as a useful cross-check for our fiducial global mean flux model.

Allowing $\bar{F}(z)$ to vary independently can help isolate whether tension in certain parameter directions (e.g., between small-scale amplitude $A_P$ and LLS contributions) is being absorbed by a global $\bar{F}$ assumption. In both datasets, the resulting posteriors are generally consistent with the global model.

\section[Effects of Peculiar Velocities]{Effects of Peculiar Velocities on the Flux Power Spectrum}
\label{sec:peculiarvel}

\begin{figure}
    \centering
    \includegraphics[width=\columnwidth]{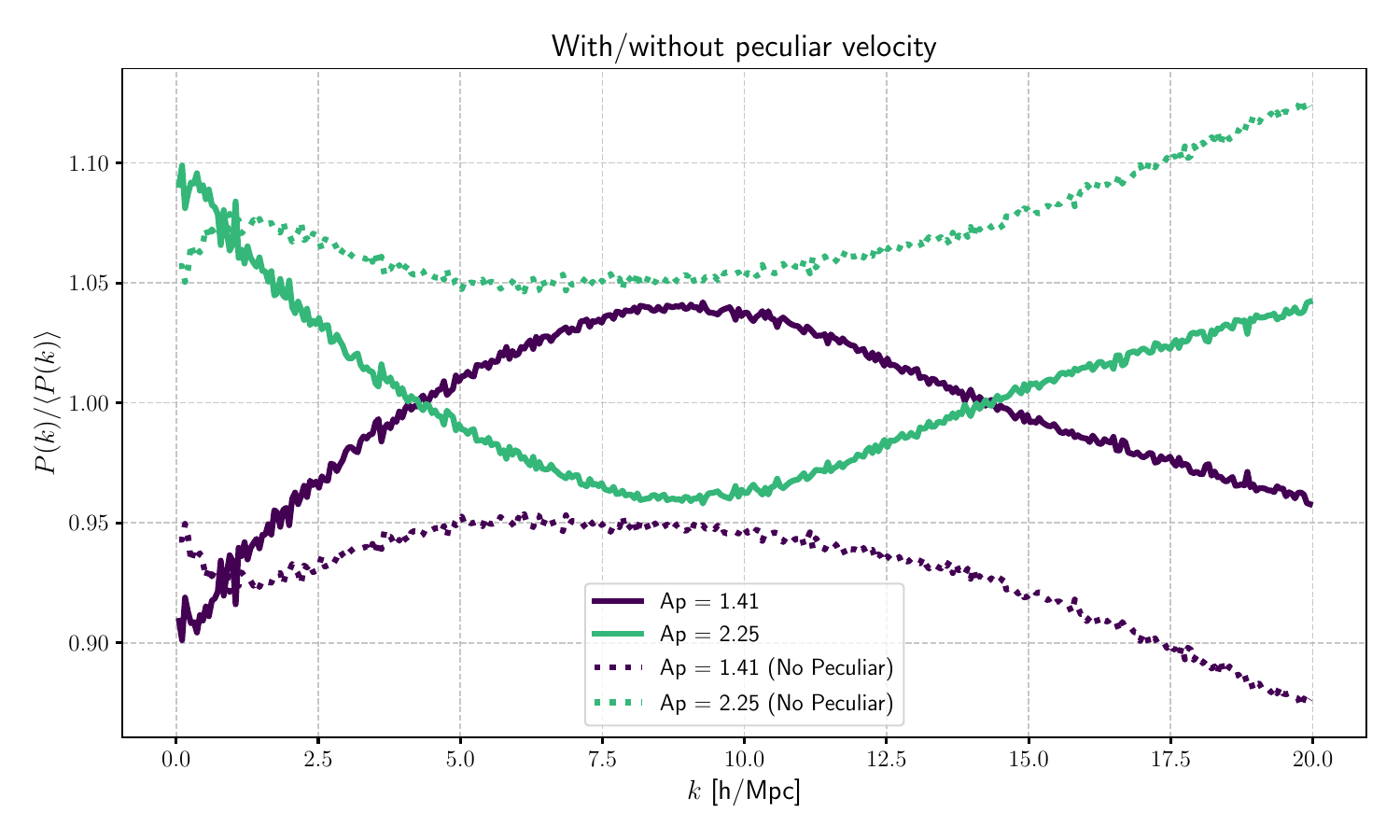}
    \caption{Effect of peculiar velocities on the flux power spectrum. Removing peculiar velocities leads to enhanced small-scale power by eliminating redshift-space broadening.
    }
    \label{fig:peculiar_velocity}
\end{figure}

Figure~\ref{fig:peculiar_velocity} shows the impact of turning off peculiar velocities in simulating \Lya sightlines. As expected, peculiar velocities suppress small-scale P1D by smoothing structure along the line of sight. While not directly varied in our inference, this figure illustrates a well-understood physical effect and helps explain why $A_P$ changes the slope at $k > 0.02\,\mathrm{s/km}$ in the high-$k$ P1D instead of simply boosting the overall amplitude. 

\section{HCD Model Complexity: LLS Degeneracy}
\label{sec:lls-complexity-appendix}

\begin{figure}[ht]
    \centering
    \includegraphics[width=0.92\columnwidth]{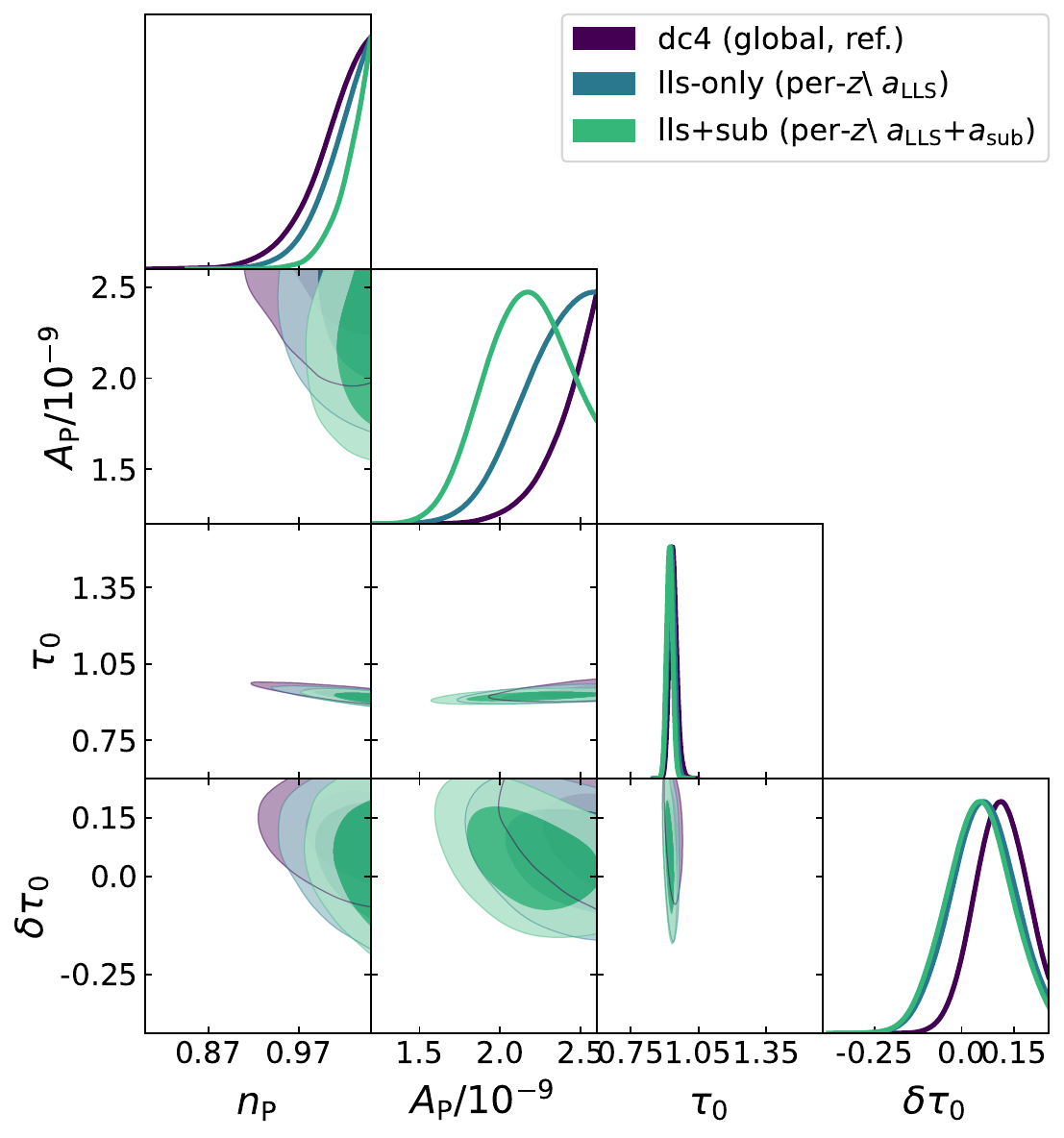}
    \caption{\add{Corner plot of $(n_P,\,A_P,\,\tau_0,\,\delta\tau_0)$ for three HCD model
    variants (KODIAQ-SQUAD $z=2.6$--$4.2$, flat $A_P$ prior): global dc4 (dark purple),
    per-$z$ lls-only (teal), and per-$z$ lls+sub (green).
    The monotone shift of $n_P$ upward and $A_P$ downward traces a single likelihood
    degeneracy between LLS redshift evolution and the primordial power spectrum.
    The mean-flux parameters $\tau_0$ and $\delta\tau_0$ shift along the same degenerate
    direction by up to $\sim 0.9\sigma$, with stable posterior widths across variants.}}
    \label{fig:lls_complexity}
\end{figure}

\add{We compare three progressively flexible HCD correction models for KODIAQ-SQUAD
($z=2.6$--$4.2$, flat $A_P$ prior): the global four-parameter \texttt{dc4} model (based on the four-component HCD template of \cite{2018MNRAS.474.3032R}, with one shared amplitude per absorber class across redshift), the per-$z$
\texttt{lls-only} model (per-redshift $\alpha_{\rm LLS}$, global remaining amplitudes), and the
per-$z$ \texttt{lls+sub} model (per-redshift $\alpha_{\rm LLS}$ and $\alpha_{\rm sub}$).
Figure~\ref{fig:lls_complexity} shows the joint posteriors of $(n_P,\,A_P,\,\tau_0,\,\delta\tau_0)$
for all three converged variants.
As per-$z$ LLS freedom increases, $n_P$ shifts monotonically upward and $A_P$ downward
along a single degenerate direction ($+0.54\sigma$ and $-1.66\sigma$ respectively for \texttt{lls+sub}
relative to \texttt{dc4}); the mean-flux parameters $\tau_0$ and $\delta\tau_0$ shift along the same
degenerate direction by up to $\sim 0.9\sigma$, while their posterior widths remain stable
across variants.
This confirms that the ``high-$A_P$'' result in the baseline analysis is an HCD-model artefact, not a cosmological signal.}

\section{Effect of $k$-range Cut on KODIAQ-SQUAD Posteriors}
\label{sec:kcut-appendix}

\begin{figure}[ht]
    \centering
    \includegraphics[width=0.82\columnwidth]{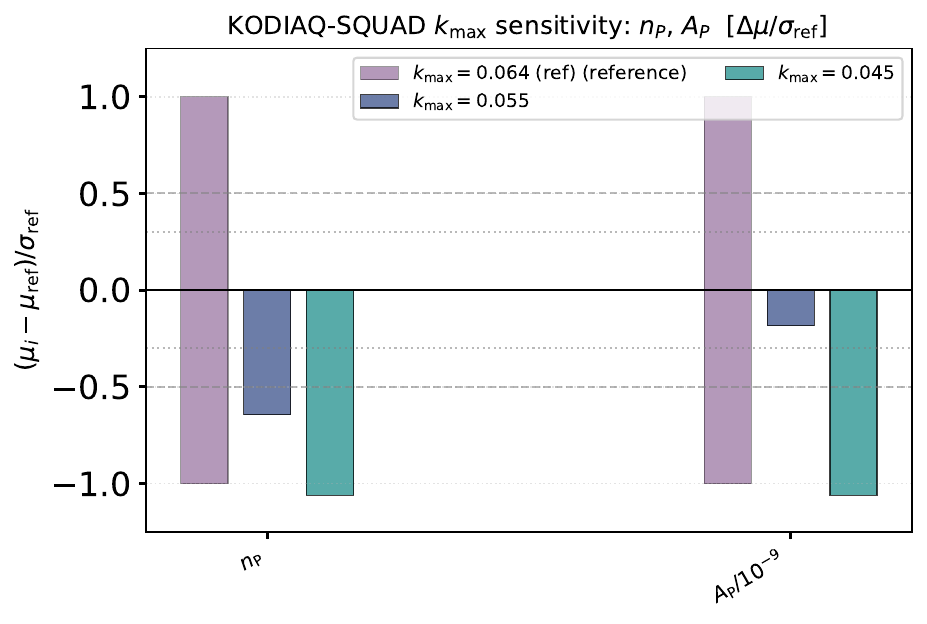}
    \caption{\add{KODIAQ-SQUAD posteriors for $n_P$ and $A_P$ at three values of $k_{\rm max}$.
    Restricting to $k_{\rm max}=0.045\,\mathrm{s/km}$ (without any redshift cut) shifts
    $(n_P,\,A_P)$ by $\sim 0.5\sigma$ away from the upper boundary, confirming that the
    small-scale modes drive the boundary-hitting behavior in the full analysis.}}
    \label{fig:kcut_appendix}
\end{figure}

\add{Figure~\ref{fig:kcut_appendix} shows the KODIAQ-SQUAD posteriors for $n_P$ and $A_P$
when the maximum wavenumber is varied: $k_{\rm max} = 0.064$ (fiducial), $0.055$, and
$0.045\,\mathrm{s/km}$.
Cutting to $k_{\rm max}=0.045\,\mathrm{s/km}$ already shifts both $n_P$ and $A_P$ by
$\sim 0.5\sigma$ away from the prior boundary, \emph{without} requiring a redshift cut.
This demonstrates that the small-scale modes ($k>0.045\,\mathrm{s/km}$) dominate the
LLS-driven boundary-hitting behavior.
The consistency of KODIAQ-SQUAD with XQ100 at $z=3.4$--$4.2$ and $k<0.045\,\mathrm{s/km}$
(discussed in \S\ref{sec:ks-cosmo-xq}) therefore partly reflects the removal of these
high-$k$ modes.}

\bibliographystyle{JHEP.bst}
\bibliography{refs}

\end{document}